\documentclass{aa}              

\usepackage{graphics,psfig,rotate}

\begin{document}


\title{Lumpy Structures in Self-Gravitating Disks}

\author{Daniel Huber, Daniel Pfenniger}

\institute{Geneva Observatory, CH-1290 Sauverny, Switzerland}

\date{Received February 2000 / Accepted May 2001}

\authorrunning{D. Huber, D. Pfenniger}
\titlerunning{Lumpy Structures in Self-Gravitating Disks}

\abstract{
Following Toomre \& Kalnajs (1991), local models of slightly
dissipative self-gravitating disks show how inhomogeneous structures
can be maintained over several galaxy rotations. Their basic physical
ingredients are self-gravity, dissipation and differential rotation.
In order to explore the structures resulting from these processes on
the kpc scale, local simulation of self-gravitating disks are
performed in this paper in 2D as well as in 3D.  The third dimension
becomes a priori important as soon as matter clumping causes a tight
coupling of the 3D equations of motion.  The physically simple and
general framework of the model permits to make conclusions beyond the
here considered scales.  A time dependent affine coordinate system is
used, allowing to calculate the gravitational forces via a
particle-mesh FFT-method, increasing the performance with respect to
previous direct force calculations.\\
Persistent patterns, formed by transient structures, whose intensity
and morphological characteristic depend on the dissipation rate are
obtained and described. Some of our simulations reveal first signs of
mass-size and velocity dispersion-size power-law relations, 
but a clear scale invariant behavior will
require more powerful computer techniques.
}

\maketitle

\section{Introduction}

Classical gravity is scale free, i.e., self-gravitating systems may
form similar structures on different scales.  Indeed, observations of
the interstellar medium, spiral disks and cosmic structures, do 
reveal similar characteristics.  Although the structures in these
systems are lumpy and inhomogeneous, they do not seem yet completely
random.

The observations of molecular clouds reveal hierarchical structures
with power-law behavior over several orders of magnitude in scale
(Larson \cite{Larson81}, Scalo \cite{Scalo85}, Falgarone et
al. \cite{Falga92}, Heithausen et al. \cite{Heit99}).  Larson
(\cite{Larson79}) found first hints that the power-law relation between
velocity dispersion and size is also valid for stellar populations and
that it extends beyond the size of Giant Molecular Clouds. Several
observations confirm that the hierarchical structure of kinematically
cold media is not only present in Milky Way molecular clouds, but is
also found in other systems and on larger scales.  For examples,
Vogelaar \& Wakker (\cite{Vogelaar94}) found perimeter-area correlations
in high-velocity clouds; power-law power spectra of HI emission
were found in the Small and the Large Magellanic Cloud by 
Stanimirovic et al. (\cite{Stanimirovic99}, \cite{Stanimirovic01}), 
and Elmegreen et al. (\cite{Elmegreen00}),
respectively; measurements of the HI distribution in galaxies of the
M81 cluster reveal fractal structures on the galaxy disc scale
(Westpfahl et al. \cite{Westpfahl99}).
 
On cosmic scales, up to about 100 Mpc, matter is also hierarchically
organized.  A common feature of the ISM and the cosmic structure is
that the matter distribution can be characterized by a comparable
fractal dimension. The cosmic and the interstellar fractal dimensions
are, $D_{\rm Galaxies}\approx 2\pm 0.2$ (Sylos Labini et
al. \cite{Sylos98}, Joyce et al. \cite{Joyce99}) and $D_{\rm
ISM}\approx 1.6-2.3$ (Elmegreen \& Falgarone \cite{Elmegreen96},
Combes \cite{Combes98}), respectively.  Thus the precise value of the
fractal dimension does not seem to be universal, but a range between 1
and 2 appears frequent.

All this may suggest that a general scale free factor is mainly
responsible for the matter distribution and the dynamics of cosmic
structures, galactic disks and molecular clouds. 
Only one factor appears to be
dominant over all these scales, namely gravity.

Gravo-thermal experiments on isolated systems show that typically two
possible states are reached asymptotically, a high energy homogeneous
state and a low energy collapsed state, with a halo-core structure
(Lynden-Bell \& Wood \cite{Lynden68}, Hertel \& Thirring
\cite{Hertel71}, Aronson \& Hansen \cite{Aronson72}).

Thus to produce more inhomogeneous structures self-gravitating systems
must be open, such as be subjected to time dependent boundary
conditions.  On cosmic scales the Hubble flow represents a time dependent
boundary condition, and develops lumpy structures.  
On galactic scales down to molecular cloud scales an energy flow,
maintaining the system out of equilibrium, may be sustained by the
shear-flow and small scale dissipation. Indeed, gravitational
instabilities convert directed kinetic energy (shear-flow) into
thermal and turbulent motion (von Weizs\"acker \cite{Weizsacker51}, 
Goldreich \& Lynden-Bell \cite{Goldreich65}). Turbulent motion
may then transport the energy through the scales until it is 
dissipated away by radiation in molecular collisions and shocks.

The lumpy distribution of matter reported by Toomre (\cite{Toomre90})
and Toomre \& Kalnajs (1991, hereafter TK) in local shearing-sheet
experiments of disks reminds us of the ubiquitous inhomogeneous state
of the ISM as well as the flocculent structures of many spirals. The
relevance of these experiments for galaxies is supported by the
recurring spirals found in slightly dissipative complete
self-gravitating disk simulations by Sellwood \& Carlberg
(\cite{Sellwood85}) and many others (e.g., Miller, Prendergast \&
Quirk \cite{Miller70}).  The TK models confirm that purely
self-gravitating systems with time-dependent boundary conditions can
produce very chaotic inhomogeneous structures.

Here, in order to investigate in more detail the matter distribution
produced by self-gravitation, shear and dissipation we perform further
such local shearing-sheet experiments.  To check if the resulting
structures reveal power-law relations we calculate the power-law
indices of the mass-size and the velocity dispersion-size relation. To compare
with earlier models, in particular with those of TK, we start with 2D
simulations and extend then the model to 3 dimensions. This extension
is important because as soon as dense clumps develop in a disk with
horizontal sizes comparable to or smaller than the supposed thickness
of the disk, motion transverse to the plane must be strongly coupled
to the motion in the plane, and the 2D approximation is no longer
valid.

To obtain instructive models it is important not to include too many
ingredients.  We are primarily concerned not with complex physical
objects such as molecular clouds, but with processes.  So our approach
is not to include a maximum of physical ingredients, but just the ones
that appear as the most relevant.  We want to check if
self-gravitation in combination with time-dependent boundary
conditions and a slight dissipation can produce and maintain an
inhomogeneous, lumpy and eventually self-similar structures,
resembling those observed in galactic disks and molecular clouds.

The considered scales are of the order of ${\cal O}(1-10)\, \rm
kpc$. Thus the transition regime between the molecular cloud scale
($\approx 0.05\, \rm kpc$) and the galactic disk scale ($\approx 10\,
\rm kpc$) can be investigated.  However, since the model is
scale-free, we can draw conclusions beyond the here considered scales
and thus eventually contribute to illustrate the scaling laws observed
in sub- or extra-galactic structures.

Preliminary results of our numerical experiments were presented in
Huber \& Pfenniger (\cite{Huber99}, \cite{Huber01a}). 
Since then we continued to
improve our model and to collect more experience, which led to new
insights with respect to the clustering simulation and scaling laws.
In this paper we discuss in detail the model and the results.  
Similar studies have been presented by Semelin (Semelin
\cite{Semelin99}, Semelin \& Combes \cite{Semelin00}).

In the next section, we justify the use of dissipative particles in
order to model the dynamics of self-gravitating gas. The numerical
model in presented in Sect.~3 and a pseudo-code is given in Appendix
A. In Sect.~4 we discuss the methods used to analyze the structures
resulting from our shearing box simulations. The results of the 2D and 
3D simulations are presented in Sect.~5. Finally, Sect.~6 is dedicated
to discussing limitations of the models.

\section{Physical Gas Model}

\subsection{Hierarchical Systems}

For a hierarchical self-similar structure, one can define a
fragmentation efficiency (Scalo \cite{Scalo85}),
\begin{equation}
f=\eta m_{L-1}/m_{L}\;,
\end{equation}
where $m_{L-1}$ and $m_{L}$ are the mean masses of a fragment at level
$L-1$ and $L$, respectively. The factor $\eta$ is the number of
fragments formed at each level. The fragmentation efficiency $f$
indicates how much mass in a clump is concentrated in subclumps. If
$f$ is not very high $(< 95 \% )$, the smallest fragment masses become
negligible after a few levels and a hierarchical description is less
relevant. However, if $f$ is very high, several iteration steps can be
carried out and the bulk mass is still concentrated in the smallest
subclumps. As long as the bulk mass is concentrated in subclumps the
interclump mass can be neglected. For convenience we call the smallest
clumps for which the interclump medium can be neglected basic-clumps.
If the level of the basic clumps is zero, a clump at level $L$ is
formed by $\eta^L$ basic clumps.

Observations of the interstellar medium reveal a highly inhomogeneous
and clumpy structure (see, eg., Dame et al. \cite{Dame00}, Tauber et
al. \cite{Tauber91}, St\"orzer et al. \cite{Storzer00}). Moreover the
structure is for a certain scale range hierarchical. Assuming that the
size of particles is larger or equal to the size of the basic gas
clumps, the structure of the interstellar medium can be described
correctly down to the scale of the particles by the distribution of
these particles.  A particle represents then the lowest resolvable
level, while clumps at higher levels, i.e., at larger scales are
represented by an ensemble of particles.

\subsection{Dissipative Particles}

At the here considered scales, larger than tens of pc,
the description of the dissipative processes taking place in the ISM
is very complex and far from respecting the hypotheses
allowing the full application of Navier-Stokes equations. 
Thus the use of a traditional hydrodynamic code is in 
no way ``better'' than the simpler approach adopted by TK, where 
a simple small drag parameter is all what is introduced as dissipation.

Indeed, we recall the following considerations:\vspace{1ex}

\noindent 1.) Being long range, gravity breaks the fundamental
assumption made in classical thermodynamics that interactions are
short ranged.  In turn, when gravity is sufficiently strong (i.e., the
Jeans' instability threshold is reached), supersonic chaotic
motion is expected, as also systematically reported for 
the interstellar medium.
This means that no local pressure equilibrium is reached at the scales
over which turbulence exists.  Down to the smallest scale at which
supersonic turbulence exists no local thermodynamical equilibrium can
be established, and thus no equation of state can be defined. A basic
assumption allowing to derive the usual Navier-Stokes equations of
fluid dynamics is missing.  Besides, numerous observational evidences
indicate non-thermal cloud clumps. For instance, Beuther et
al. (\cite{Beuther00}) carried out multi-wavelength observations and
compared the line ratios with radiation transport models. They found
that models based on the assumption of a local thermodynamical
equilibrium (LTE) can not reproduce the observed data set. Due to the
lack of a LTE down to smallest scales, thermal physics appears as an
inappropriate tool to represent the statistical state of interstellar
gas. \vspace{1ex}

\noindent 2.) Fundamentally the Navier-Stokes fluid equations describe
a) the local conservation of mass and momentum, with b) additional
constraints such as local smoothness of the quantities subject to
differentiation, c) an equation of state for closing the moment
equations derived from Boltzmann's equation, and d) phenomenological
laws describing viscous forces.  While the mass and momentum
conservation laws are likely to be adapted even for such a clumpy
medium as the ISM, the other constraints do not. In this context the
energy equation is little relevant (is not a constraint) if no control
can be performed on radiative processes, which operate on very short
time-scale in the cold ISM.  Since large but clumpy entities such as
molecular clouds have a mean-free path much larger than their size,
the dynamics of such systems may as well, in the present state of
understanding, be described by semi-collisional, dissipative particles
(Brahic \cite{Brahic77}, Pfenniger \cite{Pfenniger98}).  Casoli \&
Combes (\cite{Casoli82}) studied the formation of giant molecular
clouds through cloud collisions and coalescence in the molecular ring.
They found that the ensemble of clouds never reaches a steady
state. Thus they concluded that clouds are better described by
particles than by a fluid. Another hint that the usual fluid equations
are not better adapted to describe the ISM than sticky particles is
that the rings in barred galaxies are never reproduced by the former
but easily by the latter (e.g., Schwarz \cite{Schwarz84}). \vspace{1ex}

\noindent 3.) In the ISM not only the cooling and heating processes
are rapid with respect to global dynamics, but also the energy
reservoir of global dynamics is much larger than the other available
energy reservoirs represented by gas pressure, stellar radiation,
cosmic rays, or magnetic fields (Pfenniger \cite{Pfenniger96}). 
The virial theorem, expressing a balance of negative and
positive energy reservoirs, is a useful tool to order the
importance of respective physical factors according to their
quantitative values.  Since the energy budget at the galactic scale is
dominated by dynamics, to first order the system is well described by
conservative dynamics, and dissipative effects are of second order.
In weakly dissipative systems the stable periodic orbits and fixed
points of the conservative case are transformed into {\it
attractors}, and chaotic orbits typically converge toward {\it strange
attractors\/} with similar chaotic properties. Therefore one can
naturally infer that the exact dissipative force is irrelevant as long
as it remains weak, since the long term behavior is an attractor.  The
dissipative perturbation is weak when during the time-scale of
interest the energy dissipated is small with respect to the total
energy of the system.  Therefore in this regime it is not necessary to
know precisely how energy is dissipated, any weak factor leads to the
same attractors (see Pfenniger \& Norman
\cite{Pfenniger90} for an extended discussion on the topic).

These considerations show that weakly dissipative particles are a
permissible method to study the dynamics of interstellar gas at
sufficiently large scales.  The mass and momentum conservation is
granted by the equations of motion, and the weakly dissipative regime
by a simple linear friction law.

\section{Numerical Model}

In previous studies of shearing sheet disks, the forces of the
self-gravitating particles were computed by direct summation.
Instead, we show that by using a time-dependent affine coordinate
system we can represent the shear-flow in periodic coordinates.
Consequently we can increase the computation performance by
calculating the self-gravitational forces with the popular
FFT-convolution.

\subsection{Principle}
\label{principle}
Here we explain the principle of the local model for the 3D case.
Ignoring all expressions with a $z$, yields the 2D case. 

In a local model of a disk, everything inside a box of a given size is
simulated, and more distant regions in the plane are represented by
replicas of the local box (Toomre \& Kalnajs \cite{Toomre91}, Wisdom
\& Tremaine \cite{Wisdom88}, Salo \cite{Salo95}).  The global galactic
disk attraction made by components such as stars or dark halo not
included in the local box is 1.) cancelled to zeroth order by adopting
a rotating frame, and 2.)  corrected to first order by the linear
terms in the epicycle $\kappa$ and vertical frequency $\nu$ (Binney \&
Tremaine \cite{Binney94}).

In the same spirit as TK, the matter in the box has an undefined
mass composition with a slight dissipation. For a normal spiral each
particle may be considered as a mixture of stellar mass and gas, 
with mean weak dissipation. 

The origin of the particle coordinates in the local box is a reference
point that moves on a circular orbit at distance ${\cal R}_0$ from the
galactic center with the orbital frequency $\Omega_0=\Omega({\cal
R}_0)$. In their model, TK used a rotating Cartesian coordinate
system. The horizontal particle positions are then given by $x={\cal
R}-{\cal R}_0$, $y={\cal R}_0(\theta-\Omega_0 t)$ and the vertical
location by $z$. If $x,y,z\ll {\cal R}_0$, the orbital motion of the
particles is determined by Hill's approximation of Newton's equations
of motion (Hill \cite{Hill1878}). In the present context, they read:
\begin{equation}
\label{eq1}
\begin{array}{ccccccc}
\ddot{x}&-&2\Omega_0\,\dot{y}&=&4\Omega_0 A_0 x& + &F_x\;\;\\
\ddot{y}&+&2\Omega_0\,\dot{x}&=& & &F_y\;\;\\
\ddot{z}& &                  &=&-\nu^2 z&+&F_z\;,
\end{array}
\end{equation}
where $A_0=-{1\over2} {\cal R}_0(d\Omega /d{\cal R})_{{\cal R}_0}$ is
the Oort constant of differential rotation. $F_x, F_y$ and $F_z$ are
local forces due to the self-gravitating particles, that should be
small with respect to the global force field. Like TK and Griv et
al. (\cite{Griv99}) we use these equations also for simulation zones,
where $x,y,z\ll {\cal R}_0$ is not valid for the most part, i.e., for
galactic disc scales, meaning that non-linear higher order effects 
are not taken into account. However, since much of the gravitational
force in any wavy disturbance stems from the nearest particles
(TK, Julian \& Toomre \cite{Julian66}, Toomre \cite{Toomre64}), 
the conclusions of the model should be relevant for galactic disks,
despite the violation of the linearity hypothesis. Indeed,
the swing amplification theory, whose applicability to spirals has
been well established, is based on the same assumptions.


In a Cartesian coordinate system $(x,y,z)$ the positions of the
rectangular boxes (local simulation box and replicas) change with time
due to the differential rotation $d\Omega_0/dx$. The differential
rotation causes a shear-flow, which reads for a flat rotation curve,
$\dot{y}=\Omega_0 x$. Thus a particle at $(x, y, z)$
has images at $(x+nL_x, y-nL_x\Omega_0 t+mL_y, z)$, where $n$ and $m$
are integers (Wisdom \& Tremaine 1988). $L_x$, $L_y$ and $L_z$ are the
sides of the local box.  An initially $(t=0)$ periodic arrangement of
the boxes relative to a fixed Cartesian coordinate system can not be
maintained (see Salo \cite{Salo95}).  As a consequence the forces of
the self-gravitating particles have been determined in previous
simulations by direct summation with upper and 
lower cut-offs, meaning
that the computation time for $N$ particles is proportional to $N^2$.

However we can improve the performance by computing the forces 
in the Fourier space with the convolution method and the FFT algorithm
(Press et al. \cite{Press86}).  
Thereby the potential computation time is reduced to be
proportional to $N_c{\rm log}(N_c)$, where $N_c$ is the number of
cells, taken here as proportional to the number of particles. The FFT
approach requires a system spatially isolated and/or periodic at all
times. Here the system, representing the local dynamics of a disk, is
isolated in $z$-direction. In the $x-y-$ plane the system is periodic,  
but only on affine coordinate systems whose pitch angles change periodically. 

\begin{figure*}[htb]
\centerline{
\psfig{file=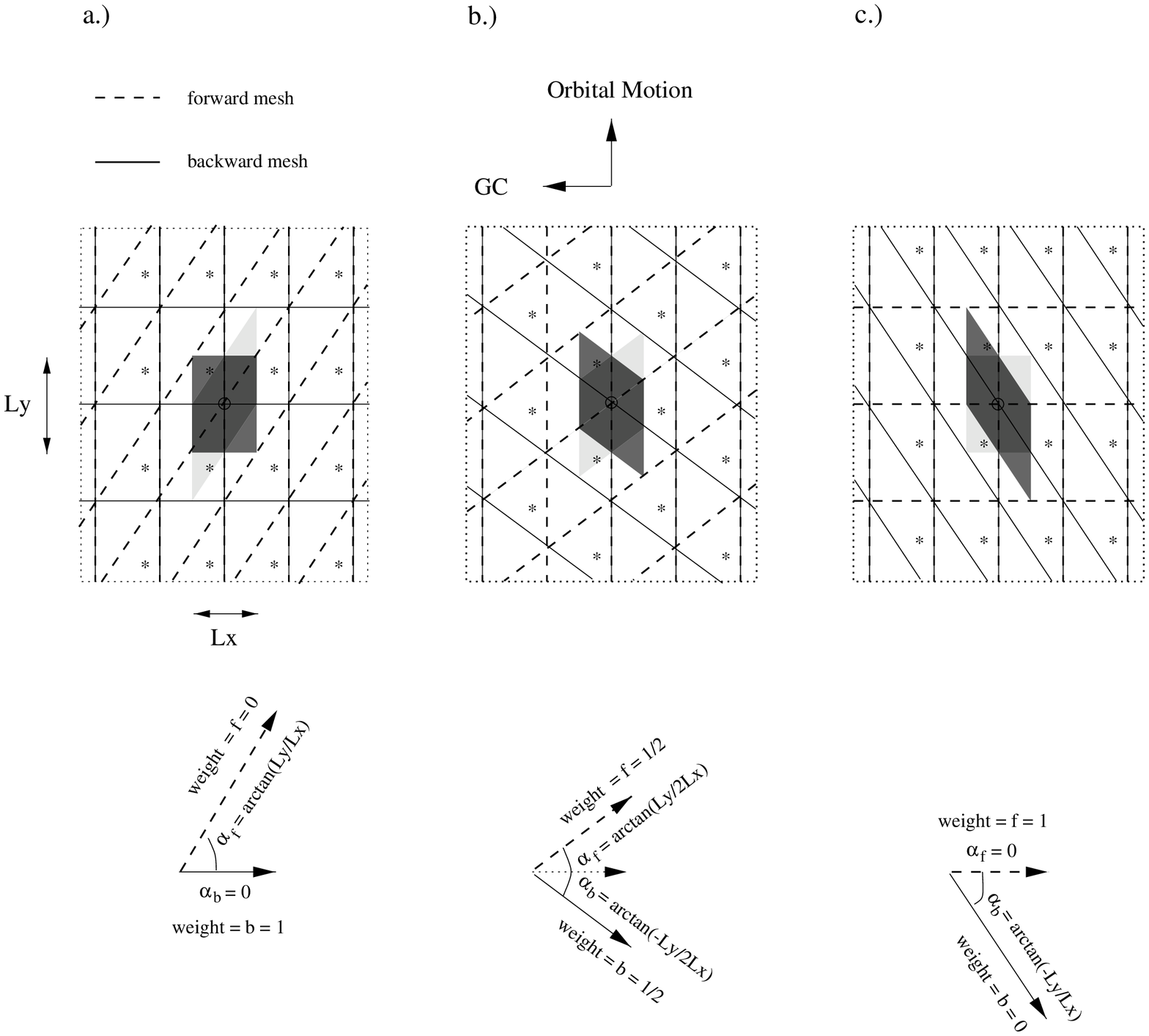,width=\hsize}}
\label{fig1}
{\bf Fig.~1a-c.} The dotted frame represents a section of the disk,
being infinite in two directions, seen from above.  The two meshes are
affine coordinate systems on which the mass distribution is periodic.
The dark and the light box represent the local computation box in
affine coordinates, i.e., in the forward and the backward mesh
respectively.  {\bf a:} The initial state of the two meshes
$(t=0)$. The inclinations of the meshes are $a_b=0$ and
$a_f=(L_y/L_x)$.  Thus the corresponding weighting factors are $b=1$
and $f=0$, respectively. Consequently, the forces in the Cartesian 
coordinates are for this situation, $F=F_b$.  Below the 
two meshes, the pitch
angles of the affine coordinate systems are indicated, $\alpha_f$
and  $\alpha_b$, respectively. 
The angle between the two meshes remains the same
at all times $(\alpha_f + \alpha_b = \arctan(L_y/L_x))$. {\bf b:} The
meshes and the weighting factors at $t=L_y/(2L_x \Omega_0)$. It is
valid, $a_b=-L_y/(2 L_x) = -a_f$ and thus $F=F_b/2+F_f/2$.  {\bf c:}
When the meshes reach these inclinations $(t=L_y/(L_x \Omega_0))$, 
they jump back to the positions shown in {\rm (a)} and the process 
starts again without introducing discontinuities in the dynamics.
\end{figure*} 

\stepcounter{figure}

The dark box in Fig.~1a represents the local box in a Cartesian
coordinate system (solid mesh). In the initial state a certain local
particle (star in the dark box) and its replicas (stars outside the
box) are periodic relative to the Cartesian coordinate system. But
then the particles are shifted by the shear and the periodicity
relative to a rectangular coordinate system is lost. However because
the shear is linear in $x$ there is for all times an affine coordinate
system $(x',y',z')$ on which the system is periodic. Thus we modify
our initially rectangular coordinate system with a time dependent
pitch angle.  The solid mesh in Fig.~1a-c represents an affine
coordinate system in which the periodicity of the system is
maintained. Its pitch angle is, $\alpha_b \le 0$, for all times. Thus
we call this coordinate system for convenience the {\it backward
mesh}. The inclination of the backward mesh $a_b$ is determine by the
shear,
\begin{equation}
\label{eqinc}
a_b=\tan\alpha_b=[(-\Omega_0 t)\mathop{\rm mod}(L_y/L_x)]\;,
\end{equation}
Fig.~1c shows the system at $t=L_y/(L_x \Omega_0)$. We can see that
the periodic arrangement of the particle images corresponds to those in
Fig.~1a. Consequently the system is again periodic on a rectangular
coordinate system and we can replace the backward mesh in Fig.~1c with
the one in Fig.~1a.  Thus the inclination of the affine coordinate
system jumps at $t=L_y/(L_x \Omega_0)$ from $a_b=L_y/L_x$ to $a_b=0$.
This accounts for the modulo function in Eq.~(\ref{eqinc}). Thus, if
$L_x/(L_y \Omega_0)$ is a multiple of the time-step, only a finite
number of affine coordinate systems is necessary. As a consequence the
corresponding kernels must be computed only once at the beginning of
the simulation and stored for subsequent use.

For the computation of the forces of the self-gravitating particles  
one coordinate system in which the matter distribution is periodic at
all times would in principle be enough (e.g., the backward mesh). 
However, in order to avoid discontinuities in the force field when the 
inclination of the coordinate system $a_b$ jumps back to zero, 
we compute the forces additionally in a second affine coordinate
system, in which the system is periodic as well. The dashed mesh
in Fig.~1a-c represents this second coordinate system. Because its
pitch angle is always, $a_f>0$, we call it the {\it forward mesh}. 
The inclination of the forward mesh $a_f$ can be deduced from those of
the backward mesh by:
\begin{equation}
a_f=\tan\alpha_f=a_b+(L_y/L_x)\;,
\end{equation}
The light box in Fig.~\ref{fig1} is the local computation box of the
forward mesh.  After the computation of the forces $F'$ in both
coordinate systems, we add them with weighting factors in order to
soften the effects of the abrupt transition at $t=L_y/(L_x \Omega_0)$
on the force field.  The forces computed in the forward and the
backward mesh are $F_b'$ and $F_f'$, respectively. Before adding the
forces with the corresponding weighting factors, we transform them to
Cartesian coordinates, $F_b'\rightarrow F_b, F_f'\rightarrow F_f$. The
single components of the forces are transformed as follows:
\begin{eqnarray}
F_{i,x}&=&F_{i,x}'\nonumber\\
F_{i,y}&=&F_{i,y}'+a F_{i,x}'\\
F_{i,z}&=&F_{i,z}'\;,\nonumber
\end{eqnarray}
where $i={b,f}$ for the backward, respectively for the forward mesh.
Then the forces are weighted and added, $F=b F_b + f F_f$. The
weighting factors $b$ and $f$ are normalized ($b+f=1$) and
proportional to the mesh inclination, $b=-a_b L_x/L_y$. Because
forces are additive such a weighted force summation is
permissible. The forces $F$ correspond now to those in
Eq.~(\ref{eq1}). That is, the inclined coordinate system are only used
to compute the forces of the self-gravitating particles with the
convolution method; then they are transformed to a Cartesian
coordinate system. The evolution of the system in the Cartesian
coordinate system is given by Eq.~(\ref{eq1}).

The weighting described above, not only softens the effect of the
abrupt change in time of the pitch angles, but also minimizes
asymmetry effects due to the mesh inclinations. Asymmetry effects
disappear for example completely when the inclination of one of the
meshes is zero or when both meshes have the same inclination. In the
first case (Fig.~1a) the weighting factor of the uninclined mesh is
one and thus the forces are computed exclusively in the rectangular
coordinate system.  In the second case (Fig.~1b) the asymmetry effects
in both inclined coordinate systems cancel each other out.
  
In an inclined coordinate system the gradient $\nabla$ depends on the
inclination. This must be taken into account by the calculation of
$F_b'$ and $F_f'$.  
The Euler-Lagrange equations yield then the forces
in an affine coordinate systems,
\begin{eqnarray}
\label{eq2}
F_{i,x}'&=&a_i \frac{\partial\Phi}{\partial y'}
-\frac{\partial\Phi}{\partial x'}\nonumber\\
F_{i,y}'&=&-(1+a_i^2)\frac{\partial\Phi}{\partial y'}+a_i
\frac{\partial\Phi}{\partial x'}\\
F_{i,z}'&=&-\frac{\partial\Phi}{\partial z'}\nonumber\;,
\end{eqnarray}
where $i=\{b,f\}$ for the backward, respectively for the forward mesh.

\subsection{Canonical Equations}
Pfenniger \& Friedli (\cite{Pfenniger93}) shown that the use of a
leap-frog finite difference approximation of Newton's equations in a
rotating reference frame with non-canonical variables lead to
instability (``complex instability'') in the sense of von Neumann.
This is not the case when canonical variables are used, then the
stability or instability character is conserved between the leap-frog
algorithm and the orbits. Therefore our model uses these equations.
The canonical equations of motion with the momenta $\{p_x, p_y, p_z\}$
are,
\begin{eqnarray}
\label{m1}
\dot{x}&=&	p_x+\Omega_0 y\nonumber\\
\dot{y}&=&	p_y-\Omega_0 x	\\
\dot{z}&=&	p_z  \hfill\nonumber, 
\end{eqnarray}
and 
\begin{equation}
\label{eq3}
\begin{array}{rrrrrrr}
\dot{p_x}&=&	(4\Omega_0 A_0 -\Omega_0^2)\, x	& + &F_x&+&\Omega_0 p_y\\
\dot{p_y}&=& -\Omega_0^2\, y 		& + &F_y&-&\Omega_0 p_x\\
\dot{p_z}&=& -\nu^2 \, z			& + & F_z &\;.\\
\end{array}
\end{equation}

These equations are invariant under the linear transformation
\begin{equation}
\label{eq4}
\begin{array}{rrrrrrr}
x&\rightarrow&x&+& k_x 		&&   		\\   
y&\rightarrow&y&-& 2A_0 \,t\, k_x	&+& k_y 	\\
p_x&\rightarrow&p_x&+&2 A_0\Omega_0 t k_x&-&\Omega_0 k_y  \\
p_y&\rightarrow&p_y&+&(\Omega_0-2A_0)k_x&&  \\
\;.&&& 
\end{array}
\end{equation}
where $k_x$ and $k_y$ are arbitrary numbers. Thus, whenever a particle
leaves the local box $L_x\times L_y \times L_z$ in the $x$ or
$y$-direction and its image enters somewhere on the opposite side (in
the affine meshes the image enters exactly at the opposite face), we
also have to transform the canonical momenta and their time
derivatives correspondingly to the rules given above.

\subsection{Kernel}
For the 2D simulations we use an isotropic interaction
potential. However, in order to resolve the flat disk vertically an
anisotropic kernel is necessary due to computational limits.  Thus
most of the 3D simulations are carried out with an anisotropic kernel
having the form of a parallelepiped.

\subsubsection{Isotropic Kernel}
In affine coordinates the softened isotropic interaction potential has
the form,
\begin{equation}
\Phi=\left\{
\begin{array}{l@{\quad:\quad}l}
\frac{1}{2\varepsilon}\left (3-
\frac{(1+a^2)x^2+y^2+2axy}{\varepsilon^2}\right)&r\le\varepsilon\\[4ex]
\frac{1}{r}&r>\varepsilon\;,\\
\end{array}
\right.
\end{equation}
where $a$ is the mesh inclination and $\varepsilon$ is the softening
length.  The advantage over a Plummer potential, used by TK and many
others, is that this potential become a correct $1/r$ gravitational
potential beyond the softening length. Thus there is no sum up of
small errors of the gravitational force due to the many distant
particles as in the case of a Plummer potential (Dehnen
\cite{Dehnen00}).

\subsubsection{Anisotropic Kernel}
\label{seckernel}
The simulation box, representing local dynamics of a disk galaxy on
the kpc scale, is rather flat $(L_z\ll L_x,L_y)$. Thus our 3D-model
needs an anisotropic force resolution and consequently an anisotropic
kernel. This will be explained more exactly in the following. To
calculate the forces of the self-gravitating particles we use a
particle-mesh method. This method consists of three steps. First, the
particle masses are assigned to the nodes of a mesh, which we call
{\it simulation mesh}. We do this in accordance with the cloud-in-cell 
(CIC) scheme (see, e.g., Hockney \& Eastwood 1981). The masses at the 
nodes of the simulation mesh can be considered as {\it new particles} 
representing the mass distribution of the original particles.  
Second, the forces for the new particles are calculated on the 
simulation mesh nodes via the convolution method 
(Hockney \& Eastwood \cite{Hockney81}):
\begin{equation}
\widetilde{\Phi}_{ijk} = \widetilde{K}_{ijk}\widetilde{\rho}_{ijk}\\
\end{equation}
\begin{equation}
\Phi = \widetilde{\widetilde{\Phi}}^{-1}_{ijk}\;,
\end{equation} 
where $\widetilde{}$ and $\widetilde{\widetilde{}}^{\;\:-1}$ are the
Fourier and the inverse Fourier transform, respectively. 
$K_{ijk}$ is the kernel and $\rho_{ijk}$ is the mass density at
a simulation mesh node $\vec{r}'_{ijk}$. If the mesh has 
$N_x\times N_y\times N_z$ nodes then 
$i=1,\ldots,N_x,;\: j=1,\ldots,N_y;\: k=1,\ldots,N_z$. 
The apostrophe $'$ indicates that the mesh is defined in inclined 
coordinates.  

The forces can now be calculated via Eqs.~(\ref{eq2}).
Finally the forces are interpolated at the original particle
positions.
 
\begin{figure}
\centerline{\psfig{file=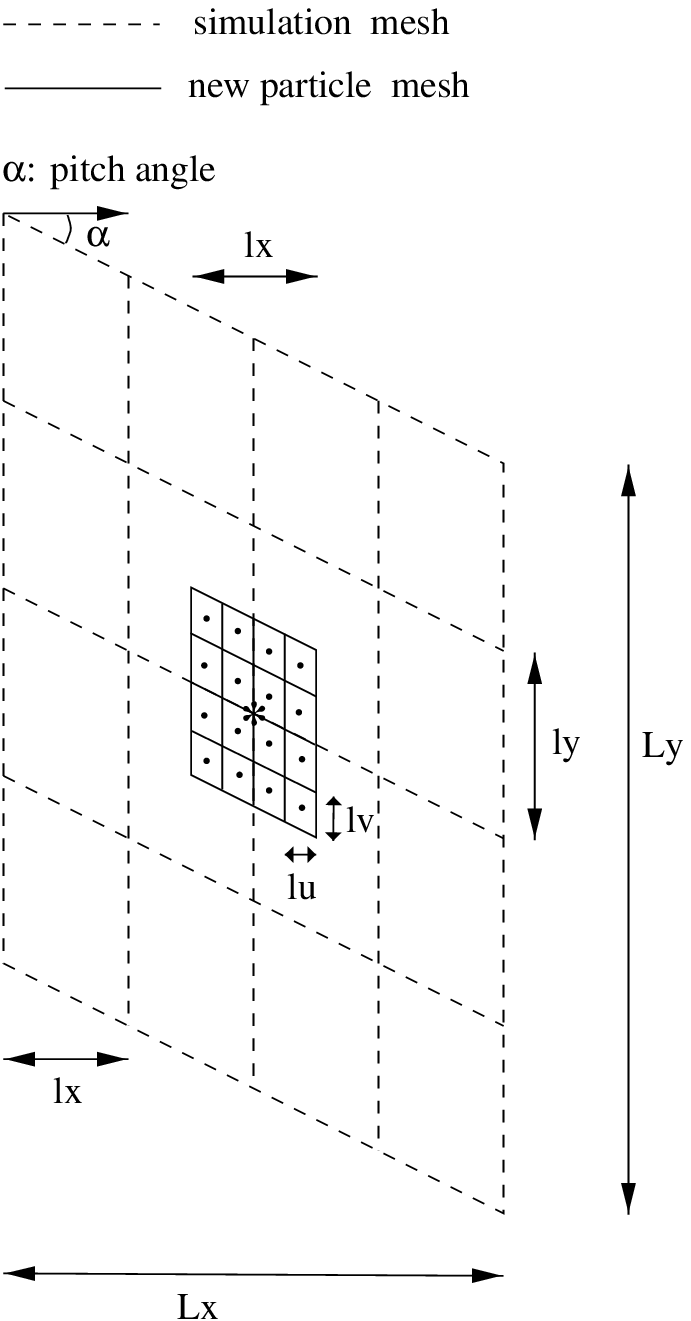,height=13 cm}}
\caption{\label{kernel} 2D representation of the simulation mesh and
  the new particle mesh depicting the discrete particle
  realization. The star indicates the origin.
  To calculate the kernel $K_{ijk}$ at the node $(i,j,k)$ of the
  simulation mesh, one has to sum up over all mass points. The mass
  points are represented by dots in the cell centers of the new
  particle mesh.}
\end{figure}

In order to avoid singularities and to approximate better physical
objects self-gravitating particle are considered to have an extent
and consequently a softened potential.  Pfenniger \& Friedli
(\cite{Pfenniger93}) optimized the softening by adapting the particle
extent as well as possible to the cell shape of the simulation
mesh.  In their polar-mesh simulations they approximated the cell
shapes with a uniform triaxial ellipsoid. Here we adopt the particle
extent to the cell shape as well. At a given time the cell shapes
are all identical, typically, because of the shear, a non-rectangular
parallelepiped. In the orthogonal case the corresponding analytical
form of the potential is known (McMillan \cite{McMillan58}). The
analytical expression of such a potential is however quite
cumbersome. Moreover we need also to describe the non-rectangular
case. Thus we use a discrete realization of the particle mass.  To
this end we distribute the mass of the new particles over a refined
discrete mesh having the same size as a cell of the simulation
mesh. We call the refined mesh, {\it new particle mesh}. Simulation
mesh and new particle mesh are shown in Fig.~\ref{kernel}. 
To calculate the kernel $K_{ijk}$ at position $r_{ijk}'$ one has 
to sum up over all mass points of the discrete particle realization,
\begin{equation}
\label{eqkernel}
K_{i j k}=\sum_{u=1}^{N_u}\sum_{v=1}^{N_v}\sum_{w=1}^{N_w}
          \frac{1}{\mid \vec{r}_{i j k}' - \vec{r}_{u v w}'\mid}\;,
\end{equation}
where $N_u \times N_v \times N_w$ is the number of mass points
representing a particle and 
$\vec{r}_{u v w}'$ are the cell-centers of the new
particle mesh (see Fig.~\ref{kernel}). The positions are given in
affine coordinates $\vec{r}'=(x',y',z')$. In order to calculate the
kernel the following coordinate transformation is thus carried out,
\begin{equation}
x' = x\;, \;\;\;\;\;\;  y' = y - ax\;, \;\;\;\;\;\; z' = z\;,
\end{equation} 
where $a$ is the the inclination of the affine coordinates.  The
inclination is fixed by the pitch angle, $a=\tan \alpha$.
Consequently the denominator in Eq.~(\ref{eqkernel}) has the form
\begin{eqnarray}
\lefteqn{\mid \vec{r}_{i j k}' - \vec{r}'_{u v w}\mid=}
\hspace{0.2cm}\nonumber\\ 
&&((1+a^2)(x_i'-x_u')^2+(y_j'-y_v')^2+(z_k'-z_w')^2+\\
&&\;\;2a(x_i'-x_u')(y_j'-y_v'))^{1/2}\nonumber\;.
\end{eqnarray} 

Since the kernel needs to be evaluated only once for every possible
inclination the cost of this procedure remains negligible.

Because the cell size of the simulation mesh determines the particle
extent, the softening is automatically fixed by the choice of the
simulation mesh.

It is important that the origin of the kernel represents the center of
the simulation box in order to avoid a non-zero temporal mean velocity
of the center of mass, which is introduced by an asymmetric
description of the centrifugal force.  Thus the positions
$r_{ijk}'=(x_i',y_j',z_k')$ are fixed as follows:
\begin{equation}
\begin{array}{rl@{\quad:\quad}l}
x_i'=& (i-\frac{N_x}{2}) l_x  &  i=1,\ldots,N_x\\[1ex]
y_j'=& (i-\frac{N_y}{2}) l_y  &  j=1,\ldots,N_y\\[1ex]
z_k'=& (i-\frac{N_z}{2}) l_z  &  k=1,\ldots,N_z\;,
\end{array}
\end{equation}
where $l_x\times l_y\times l_z$ is the size of a simulation mesh
cell.

The positions representing the discrete mass distribution of the new
particles $r_{uvw}=(x_u,y_v,z_w)$ are:
\begin{equation}
\begin{array}{rl@{\quad:\quad}l}
x_u'=& (u -\frac{N_u+1}{2}) l_u  &  u=1,\ldots,N_u\\[1ex]
y_v'=& (v -\frac{N_v+1}{2}) l_v  &  v=1,\ldots,N_v\\[1ex]
z_w'=& (w -\frac{N_w+1}{2}) l_w  &  w=1,\ldots,N_w\;,
\end{array}
\end{equation}
where $l_u\times l_v\times l_w$ is the cell size of the new particle
mesh.

The system is not periodic in the $z$-direction.  To suppress the
images introduced by the FFT we use the classical doubling-up
procedure (Hockney \& Eastwood \cite{Hockney81}), which by doubling
the size of the mesh over which the FFT must be performed exactly
cancels all the images. Thus only the lower half of the entire
mesh is relevant, i.e.,  only particles inside $-L_z/2\le z\le 0$
are active and particles leaving this zone are considered as escaped.

\subsection{Friction}
\label{friction}

\long\def\comment#1{} 

\comment{ To counteract the particle dynamical heating, TK proposed to
add an ad-hoc friction term playing the role of the weak dissipative
factors at work in the interstellar gas. 

Let us describe the properties of weakly dissipative systems: The
phase-space of our system has $6N={\cal N}$ dimensions, where $N$ is
the number of particles. The characteristic of dissipative systems is
to contract their phase-space volume.  The contraction continues until
the initial volume becomes zero and the system trajectory describe a
surface with dimension less than ${\cal N}$.  Loosely speaking, this
surface is called an attractor. There are only three types of
attractors for dissipative systems: fixed points, limit cycle and
strange attractors (Lichtenberg \& Lieberman \cite{Lichtenberg83}).
Strange attractors may have fractal dimensions\footnote{The fractal
dimension of an attractor in the phase-space is not the same as the
fractal dimension of a structure in the state space.} and describe
chaotic motions.

For our system consisting of $N$ coupled particles we expect
principally a chaotic motion and thus an evolution towards strange
attractors. The particular form of the strange attractor depends on
the initial state in phase-space, the dissipation strength and the
different parameters. For a given dissipation and parameter set the
phase space can be divided up in basins of attraction. All states
within such a basin approach the same attractor for
$t\rightarrow\infty$. Often there are a finite number of attractors.
Moreover the initial state of our system is statistically always the
same. Consequently all possible initial states fill a finite volume in
the ${\cal N}$-dimensional phase-space. Thus it may exist a limited
number of asymptotic behaviors with similar statistical properties.

A variation of a parameter in a dissipative system may change the
characteristic of the solution, i.e., when a critical parameter value
is passed a bifurcation occurs and the steady state flow is described
by an other attractor. The same holds for the dissipation strength
(Schmidt \cite{Schmidt87}). This means that there is only one
accessible attractor for a certain range of parameters and dissipation
strength. By varying the dissipation strength we can then explore the
different possible attractors. In order to access the different
attractors it is sufficient to alter the dissipation strength and
there is no need to know the exact form of the dissipation, as long as
the dissipation remains weak.
}

In the ISM the collisional rate must depend on the clumping state,
which must depend on the dissipation rate.  Consequently, we expect a
complex dependence of drag coefficients and mass density.  However, as
explained in Sect.~2, at the kpc scale the physics is dominated by the
conservative gravitational dynamics and its concrete behavior should
be weakly dependent on the particular dissipative factors, since
dissipation mainly acts to ensure the convergence of the system toward
the attractors determined by the conservative part of the system.
Thus, and following TK, as dissipation factors we adopt linear friction
terms, which should be weak in order to remain quasi-Hamiltonian
(Pfenniger \& Norman \cite{Pfenniger90}).

Yet the collisional properties of the interstellar medium can be
expected to differ along or transverse to the plane.  To minimize the
number of free parameters, we retain only two friction coefficients.
The linear friction terms $-C_x \dot{x}$ and $-C_z \dot{z}$ added to
the radial respectively to the vertical forces $(F_x, F_z)$ in
Eq.~(\ref{eq3}). There is no azimuthal friction in order to be
consistent with a global angular momentum conservation.

\subsection{Scaling, Units, Parameters}
In order to fix a scale, the origin of our local model is located at a
distance of ${\cal R}_0 = 8$ kpc from the galactic center and rotates
with an orbital frequency of $\Omega_0 = \theta_0/{\cal R}_0$, where
$\theta_0 = 210$ km/s.  We assume for the general case a flat rotation
curve.  Moreover, we assume that the active disk has a surface density
of $\Sigma_0=100$ M$_\odot$/pc$^2$.

As usual in local shear models of galactic disks the linear measure is
the critical wavelength, i.e., the longest unstable wavelength in a
zero-pressure disk,
\begin{equation}
\lambda_{\rm crit}=\frac{4\pi^2 G\Sigma_0}{\kappa_0^2}\;.
\end{equation}
The critical wavelength defines the scale for which the theory of
swing amplification predicts the strongest response (Toomre
\cite{Toomre81} , Julian \& Toomre \cite{Julian66}). 
For a flat rotation curve the epicyclic frequency
is, $\kappa=\sqrt{2}\Omega_0$ and consequently the critical wavelength
scales, with the parameter values indicated above, to $1\,\lambda_{\rm
crit}=12.32$ kpc. The disk scale height $z_0$ is then
$0.024\;\lambda_{\rm crit}$. However, the equations of motion are
scale free and the model can, with an appropriate choice of the
parameters, be rescaled at will.

The friction coefficients $C_x$ and $C_z$ of the friction terms $-C_x
\dot{x}$ and $-C_z \dot{z}$ are in this work indicated in units of
$1/\tau_{\rm osc}$, where $\tau_{\rm osc}$ is the period of the
unforced epicyclic motion. The cooling times of the radial and the
vertical damping are thus $t_{{\rm cool},x}=1/C_x\;\tau_{\rm osc}$ and
$t_{{\rm cool},z}=1/C_z\;\tau_{\rm osc}$. For all models presented
here, $\tau_{\rm osc} < t_{\rm cool}$ applies.

The time-step has to meet the following conditions:
\begin{eqnarray}
\Delta t &\le& 0.1\min\{l_i/\sigma_i\}\;,\;\; i=\{x,y,z\}\\
\Delta t &=& \frac{1}{k} \frac{L_x}{L_y \Omega_0}\;, 
\end{eqnarray} 
where $\sigma_i$ is the initial velocity dispersion ellipsoid, $l_i$
is the cell size and k is an integer.  According to Eq.~(\ref{eqinc})
the evolution of the inclination of the backward grid is periodic with
period $T=L_x/(L_y \Omega_0)$.  The second condition guarantees that
this period is a multiple of the time-step. Thus the number of
possible grid inclinations and consequently the number of kernels is
finite.  In order to satisfy the above conditions the time-step is
computed in two steps:
\begin{eqnarray}
k &=& \left\lceil \frac{10 L_x}{L_y \Omega_0 \min\{l_i/\sigma_i\}}
      \right\rceil\,\\
\Delta t &=& \frac{L_x}{L_y \Omega_0 k}\;,
\end{eqnarray}
where $\lceil \rceil$ means to round to the next higher integer.

\begin{table*}
\begin{center}
\begin{tabular}{|c|c|c|c|c|c|} \hline
Model & $L_x\times L_y$ & $n$ & Dynamical range & Potential/ 
& \#  Dimensions  \\ 

& $[\lambda^2_{\rm crit}]$ & $[1/\lambda^2_{\rm crit}]$  
& [dex] & Softening & \\ \hline \hline 

TK & $6.0 \times 8.0$ & 100-1200 & $0.6$ & Plummer & $2$\\ \hline 
\end{tabular} 
\caption{\label{tab1} Parameters characterizing the model of
  TK. Indicated are the size of the simulation zone, the number
  density of particles (surface density), the dynamical range, the
  gravitational potential of the particles and the number of dimensions.}
\end{center}
\end{table*}

\begin{table*}
\begin{center}
\begin{tabular}{|c|c|c|c|c|c|} \hline
Model & $C_x/10^{-3}\;\;[{\rm 1/\tau_{\rm osc}}]$  
& $N$ & $\varepsilon\;\;[\lambda_{\rm crit}]$  
& $\kappa/\Omega_0$ & $A_0/\Omega_0$ \\ \hline\hline

TK     & $3.5/n$ & $4800-57000$ & $0.20$ & $1.4$ & $0.5$ 
\\ \hline
\end{tabular} 
\caption{\label{tab2} Parameters of the TK model. The friction
coefficient $C_x$ is a function of the particle density $n$. $N$ is
the particle number, $\varepsilon$ is the softening length of the
Plummer potential. The epicycle frequency $\kappa$ and Oort's constant
$A_0$ are indicated in units of $\Omega_0$.}
\end{center}
\end{table*}

TK calculated the forces of the self-gravitating particles with direct
summation. Thus they had to introduce an upper cutoff in order to
limit the computational expenditure, meaning that beyond a certain
separation the particles lost their mutual gravitational interaction.
Their separation cutoff was equal to four times the softening
length. This limited the dynamical range of gravity to 0.6 dex. They
argued that a cutoff at larger separations did not affect the
resulting structures. 
\comment{
Consequently, the large scale correlations they
found can not be due to direct gravitational interaction.
} 
Thanks to the higher performance of the convolution method we can
extend the dynamical range without increasing the computation time.
This may be important in view of a self-similar matter organization in
self-gravitating systems.

\begin{table*}
\begin{center}
\begin{tabular}{|c|l|l|r|c|c|c|c|l|} \hline
Model 
& \multicolumn{1}{|c|}{$L_x\times L_y \times L_z$} 
& \multicolumn{1}{|c|}{$l_x\times l_y \times l_z$} 
& \multicolumn{1}{|c|}{$n$} 
&\multicolumn{2}{|c|}{Dynam. range [dex]} & Potential/ & \#  Dim. &
Var  \\ 

& \multicolumn{1}{|c|}{$[\lambda^3_{\rm crit}]$} 
& \multicolumn{1}{|c|}{$[\lambda^3_{\rm crit}]$}  
& \multicolumn{1}{|c|}{$[1/\lambda^2_{\rm crit}]$} 
& \hspace{0.2cm}plane\hspace{0.2cm} & vertically & Softening & & 
\\ \hline \hline
 
1 & $6.0 \times 6.0\;\;\lambda^2_{\rm crit}$ 
& $0.023 \times 0.023\;\;\lambda^2_{\rm crit}$ 
& 1820 & $1.5$ & - & Isotropic & $2$ & $C_x$\\ 

2 & $6.0 \times 6.0\;\;\lambda^2_{\rm crit}$ 
& $0.023 \times 0.023\;\;\lambda^2_{\rm crit}$ 
& 1820 & $1.5-2.5$ & - & Isotropic & $2$ & $\varepsilon$\\ 

3 & $6.0 \times 6.0 \times 0.8 $ & $0.188 \times 0.188 \times 0.013$ 
& 910 & $1.3$ & 0.5 & Isotropic & $3$ & $C_x$\\ 

4 & $6.0 \times 6.0 \times 0.8 $ & $0.094 \times 0.094 \times 0.013$ 
& 3640 & Var & Var & Isotropic & $3$ & $\varepsilon$\\ 

5 & $6.0 \times 6.0 \times 0.8 $ & $0.188 \times 0.188 \times 0.013$ 
& 910 & $1.5$ & 1.8 & Anisotropic & $3$ & $C_x,\nu=0.3$\\ 

6 & $6.0 \times 6.0 \times 0.8 $ & $0.188 \times 0.188 \times 0.013$ 
& 910 & $1.5$ & 1.8 & Anisotropic & $3$ & $C_x,\nu=3.0$\\ 

7 & $6.0 \times 6.0 \times 0.8 $ & $0.188 \times 0.188 \times 0.013$ 
& 910 & $1.5$ & 1.8 & Anisotropic & $3$ & $C_z$\\ 

8 & $6.0 \times 6.0 \times 0.8 $ & $0.188 \times 0.188 \times 0.013$ 
& 910 & $1.5$ & 1.8 & Anisotropic & $3$ & $\nu$\\ 

9 & $6.0 \times 6.0 \times 0.8 $ & $0.094 \times 0.094 \times 0.013$ 
& 3640 & $1.8$ & 1.8 & Anisotropic & $3$ & $C_x,\nu=0.3$\\ 

10 & $6.0 \times 6.0 \times 0.8 $ & $0.094 \times 0.094 \times 0.013$ 
& 3640 & $1.8$ & 1.8 & Anisotropic & $3$ & $C_x,\nu=3.0$\\ 

11 & $6.0 \times 6.0 \times 0.8 $ & $0.094 \times 0.094 \times 0.013$ 
& Var & $1.8$ & 1.8 & Anisotropic & $3$ & $N$\\ \hline

12 & $1.8 \times 1.8 \times 0.8 $ & $0.056 \times 0.056 \times 0.013$ 
& 10100 & $1.5$ & 1.8 & Anisotropic & $3$ & $C_x,\nu=0.3$\\ 

13 & $1.8 \times 1.8 \times 0.8 $ & $0.056 \times 0.056 \times 0.013$ 
& 10100 & $1.5$ & 1.8 & Anisotropic & $3$ & $C_x,\nu=3.0$\\

14 & $1.8 \times 1.8 \times 0.8 $ & $0.056 \times 0.056 \times 0.013$ 
& 10100 & $1.5$ & 1.8 & Anisotropic & $3$ & $C_z$\\ 

15 & $1.8 \times 1.8 \times 0.8 $ & $0.056 \times 0.056 \times 0.013$ 
& 10100 & $1.5$ & 1.8 & Anisotropic & $3$ & $\nu$\\

16 & $1.8 \times 1.8 \times 0.8 $ & $0.028 \times 0.028 \times 0.013$ 
& 40450 & $1.8$ & 1.8 & Anisotropic & $3$ & $C_x,\nu=0.3$\\

17 & $1.8 \times 1.8 \times 0.8 $ & $0.028 \times 0.028 \times 0.013$ 
& 40450 & $1.8$ & 1.8 & Anisotropic & $3$ & $C_x,\nu=3.0$\\

18 & $1.8 \times 1.8 \times 0.8 $ & $0.028 \times 0.028 \times 0.013$ 
& 40450 & $1.8$ & 1.8 & Anisotropic & $3$ & $C_z$\\

19 & $1.8 \times 1.8 \times 0.8 $ & $0.028 \times 0.028 \times 0.013$ 
& 40450 & $1.8$ & 1.8 & Anisotropic & $3$ & $A_0$\\ \hline

\end{tabular} 
\caption{\label{tab3} We use 18 models to explore the different
parameters. Besides the parameters presented in Table \ref{tab1} for
the TK model, we indicate here the mesh resolution $l_x\times l_y
\times l_z$ and the dynamical range vertical to the plane. Var
indicates the parameter, altered from run to run. The resulting
structure are then explored as a function of this parameter.  The
gravitational particle potential is either isotropic or it is deduced
from the discrete particle representation described in Sect.
\ref{seckernel}.}
\end{center}
\end{table*}

\begin{table*}
\begin{center}
\begin{tabular}{|c|c|c|r|c|c|c|c|r@{\hspace{0.1cm}=\hspace{0.1cm}}l|}
\hline 
Model & $C_x/10^{-3}$ & $C_z/10^{-3}$ & \multicolumn{1}{|c|}{$N$} 
& $\varepsilon$  & $\nu/\Omega_0$ & $\kappa/\Omega_0$ & $A_0/\Omega_0$ 
& \multicolumn{2}{|l|}{Var} \\

& $[1/\tau_{\rm osc}]$ & $[1/\tau_{\rm osc}]$ & & $[\lambda_{\rm crit}]$
&&&& \multicolumn{2}{|l|}{} \\\hline\hline

1 & Var & - & 65520 & 0.2 & - & 1.4 & 0.5 & $C_x$ &
$40-210\times10^{-3}$\\

2 & 100 & - & 65520 & Var  & - & 1.4 & 0.5 & $\varepsilon$ & 
$0.02-0.3$\\

3 & Var & 0.7 & 32760 & 0.3  & 0.3 & 1.4 & 0.5 & $C_x$ 
& $40-280\times10^{-3}$\\

4 & 100 & 0.7 & 131040 & Var  & 0.3 & 1.4 & 0.5 & $\varepsilon$ 
& $0.1-0.4$\\ 

5 & Var & 0.7 & 32760 & -  & 0.3 & 1.4 & 0.5 & $C_x$ & $40-280\times10^{-3}$\\

6 & Var & 0.7 & 32760 & -  & 3.0 & 1.4 & 0.5 & $C_x$ & $70-280\times10^{-3}$\\

7 & 140 & Var & 32760 & -  & 0.3 & 1.4 & 0.5 & $C_z$ & $0.04-40\times10^{-3}$\\

8 & 140 & 0.7 & 32760 & -  & Var & 1.4 & 0.5 & $\nu$ & $0.0-6.4$\\

9 & Var & 0.7 & 131040 & -  & 0.3 & 1.4 & 0.5 & $C_x$ & $40-210\times10^{-3}$\\

10 & Var & 0.7 & 131040 & -  & 3.0 & 1.4 & 0.5 & $C_x$ & 
$0-120\times10^{-3}$\\

11 & 70 & 0.7 & Var & -  & 0.3 & 1.4 & 0.5 & $N$ & $16000-128000$\\
\hline

12 & Var & 0.7 & 32720 & -  & 0.3 & 1.4 & 0.5 & $C_x$ & $10-70\times10^{-3}$\\

13 & Var & 0.7 & 32720 & -  & 3.0 & 1.4 & 0.5 & $C_x$ & $30-50\times10^{-3}$\\

14 & 50 & Var & 32720 & -  & 0.3 & 1.4 & 0.5 & $C_z$ & 
$0.04-40\times10^{-3}$\\

15 & 50 & 0.7 & 32720 & -  & Var & 1.4 & 0.5 & $\nu$ & $0.0-6.0$\\

16 & Var & 0.7 & 131050 & -  & 0.3 & 1.4 & 0.5 & $C_x$ & $20-40\times10^{-3}$\\

17 & Var & 0.7 & 131050 & -  & 3.0 & 1.4 & 0.5 & $C_x$ & $20-40\times10^{-3}$\\

18 & 30  & Var & 131050 & -  & 0.3 & 1.4 & 0.5 & $C_x$ &
$0.04-40\times10^{-3}$\\ 

19 & 30 & 0.7 & 131050 & -  & 0.3 & Var & Var & $\kappa$ & $1.4,\;1.7;\;\;
A_0=0.5,\;0.25$\\ \hline 
\end{tabular} 
\caption{\label{tab4} Parameters of model 1-19. Contrary to the TK
  model $C_x$ is a free parameter. Moreover we have, because of the
  extension to three dimensions, a vertical friction coefficient $C_z$ and a
  vertical frequency $\nu$. The parameter range for which we explore
  the models is indicated in the last column.}
\end{center}
\end{table*}

The parameters characterizing the model of TK are indicated in Table
\ref{tab1} and \ref{tab2}. They carried out numerical
shearing-sheet experiments
for different particle densities $n$. Their friction coefficient is a
function of the particle density $C_x = (3.5\times 10^{-3})/n\;{\rm
\tau^{-1}_{\rm osc}}$. In order to extend this study and to explore
the resulting structures in dependence of the different parameters, we
realize different versions of the shearing box model. These
model versions are characterized by different parameter
sets which fix the size of the simulation zone, the resolution,
the particle density etc..
For convenience we call these model versions in the following models.
That is, a model denotes in the following a version of
the shearing box model which is determined by a specific parameter set.
The parameters of the models are indicated in 
Table \ref{tab3} and \ref{tab4}.

To be able to do some statistics of structures produced on scales with
strongest swing amplification response we perform, like TK,
simulations with a quite large simulation zone. Model 1-11 have such a
large simulation box resp. simulation sheet in the 2D case, $L_x\times
L_y (\times L_z) = 6 \times 6 (\times0.8)\;\lambda^3_{\rm crit}$.
However since the dynamical range is limited due to computational
limits we perform also simulations for a smaller local box, in order
to resolve smaller scales. Therefore the simulation box of model 12-19
are reduced to $1.8 \times 1.8 \times 0.8\;\lambda^3_{\rm crit}$.

The time-step depends on the mesh resolution. The mesh resolution is
fixed by the number of particles and the size of the simulation zone.
The computation time of a simulation depends thus on the particle
density $n$. We increase $n$ with respect to previous models based on
direct force calculation up to a factor 30 and are furthermore able to
perform the simulations in 3D.  The code has been written in Matlab
for its ease of use, but clearly a compiled language program would
greatly improve its speed and memory usage.  A pseudo code is given in
Appendix A.

\subsection{Code Testing}

In order to check our code we carry out two-body simulations and
compare the results with analytical solutions of the Kepler
problem. We use a non-rotating inertial frame, thus
$\Omega_0=A_0=\nu=0$ in the equations of motion 
(Eq.~(\ref{m1}) and (\ref{eq3})). However we compute
the forces on the time dependent affine coordinate systems.  Thus
$\Omega_0 = \theta_0/{\cal R}_0$ in Eq.~(\ref{eqinc}), where the
inclination angles are determined.  That is, we calculate the forces
for a non-rotating isolated system with the help of ``shearing Fourier
meshes''. Furthermore we use the anisotropic kernel described in 
Sect.~\ref{seckernel}.

The vertical resolution of all simulations presented in this work is
$l_z = 0.013\;\lambda_{\rm crit}$, but the resolution in the plane
$l=l_x=l_y$ depends on the particle number.  Thus we test our code for
different resolutions $l$.  The particle extension is fixed by the
anisotropic kernel and is equal to the resolution.

In the initial state the velocities of the two particles are chosen,
in the way that they move on circular orbits. Accordingly to theory
the following holds at each time:
\parbox{8 cm}{
\begin{eqnarray*}
\Delta r & = & r-r_0  =  0\\ 
\ddot{r}_{\rm cm} & = & 0 \;,
\end{eqnarray*}
}\hfill
\parbox{8 mm}{\begin{eqnarray}\end{eqnarray}}
where $r$ is the relative particle distance at $t>0$ and
$r_0$ is the initial particle distance at $t=0$. 
$r_{\rm cm}$ is the center of mass.
The orbital period for a particle with mass $m$ is,
\begin{equation}
T = 2\pi \sqrt{\frac{r_0^3}{2m}}\;.
\end{equation}
Particle mass, distance and velocities chosen for these tests yield a
period of $T \approx 4$ galactic rotations.

\begin{figure}
\psfig{file=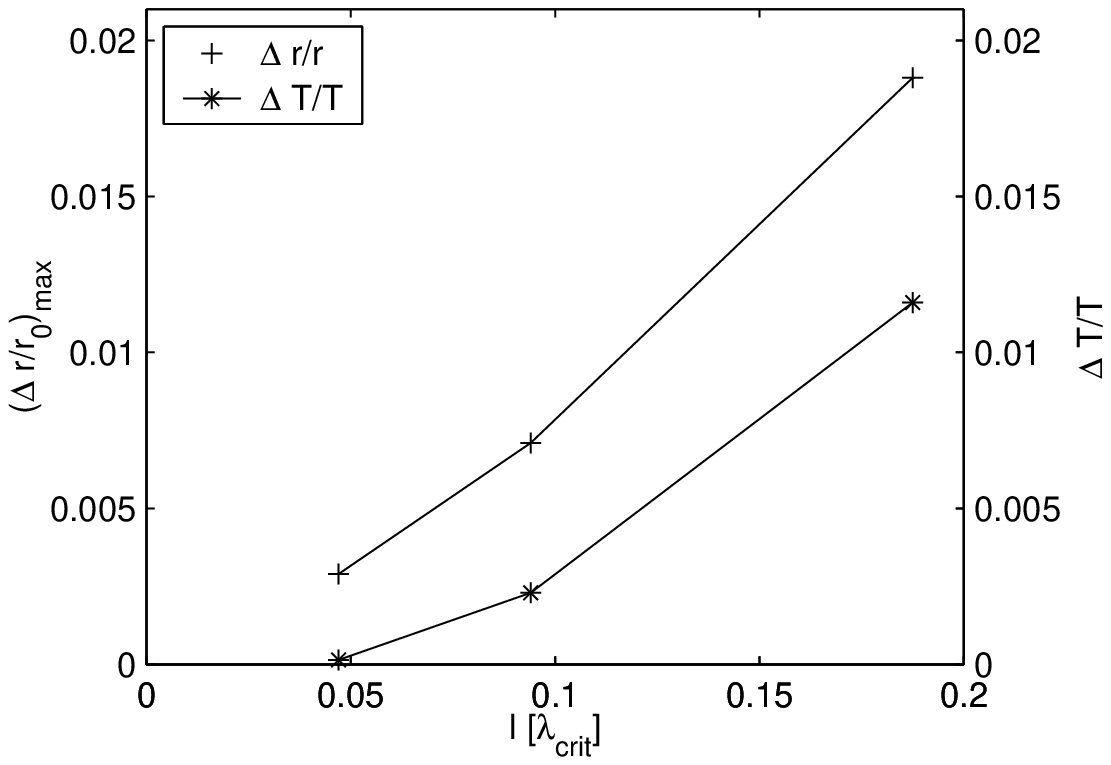,angle=90,width=\hsize}
\caption{\label{relfel} Relative errors, resulting from the simulation
  of two bodies on circular orbits, as a function of the resolution. 
  The simulations are performed for one period, 
  which corresponds to $\approx 4$ galactic rotations. 
  Crosses, left ordinate: The maximal relative error of the particle
  distance as a function of the resolution. The resolution is indicated in 
  units of $\lambda_{\rm crit}$. Stars, right ordinate: The relative error of 
  the orbital period.}
\end{figure}
 
\begin{figure}
\psfig{file=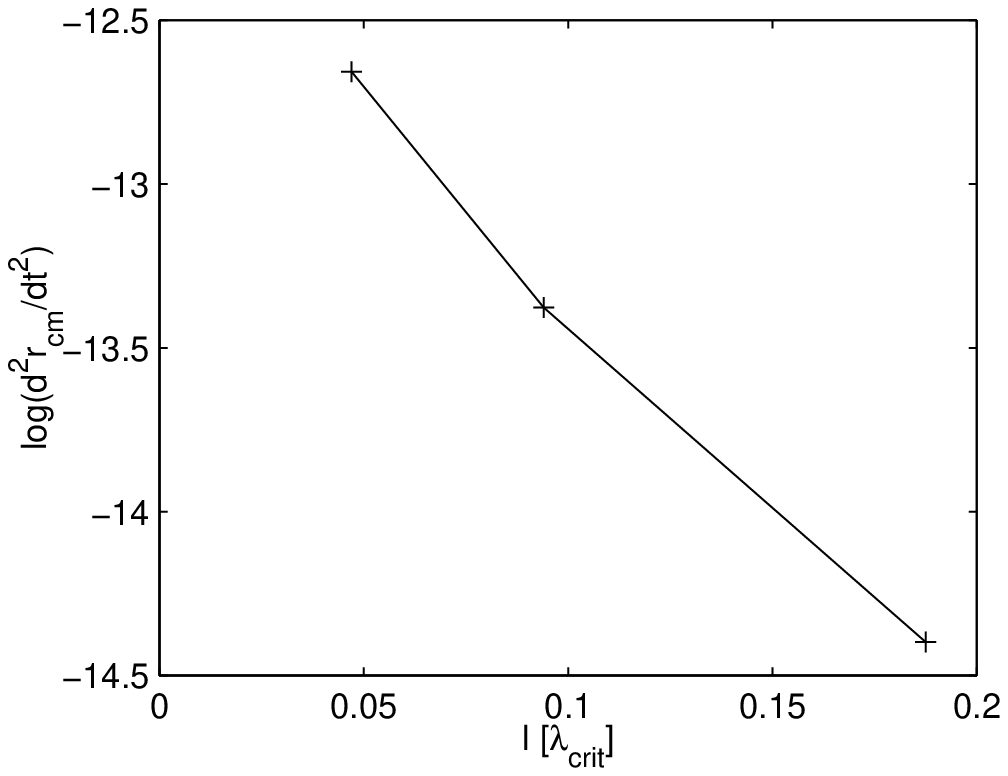,angle=90,width=7.7 cm}
\caption{\label{acccm} The maximal acceleration of the center of mass
  during the simulation of two bodies on a circular orbit as a
  function of the resolution. The
  simulations are performed for one orbital period.}
\end{figure}

\begin{table}
\begin{tabular}{|c|c|} \hline
$l$ $[\lambda_{\rm crit}]$ & $l/r_0$ \\ \hline 
$0.188$ & 0.094 \\
$0.094$ & 0.047 \\
$0.047$ & 0.024 \\ \hline
\end{tabular} 
\caption{\label{tab0} A small contribution to the deviation from the
  theoretical trajectories is due to the extension of our test
  particles. In this Table we indicate the ratio of particle
  extension and separation $l/r_0$ for the different resolutions $l$.}
\end{table}  

These theoretical results are compared with the experimental results,
i.e., with those resulting from our simulations. The code errors,
arising from this comparison are shown in Fig.~\ref{relfel} and
\ref{acccm} for different resolutions. The resolution is indicated in
units of $\lambda_{\rm crit}$.  During an orbital period $\Delta
r/r_0$ oscillates around zero. The maximal error of the particle
trajectory computed with the numerical model are then equal to the
amplitude of this oscillation. The amplitude $(\Delta r/r_0)_{\rm max}$ is
plotted in Fig.~\ref{relfel}. In this figure the relative error of the
orbital period $\Delta T / T$ is indicated as well.  Fig.~\ref{acccm}
reveals the acceleration of the center of mass $r_{\rm cm}$.  The here
presented errors must be considered as upper limits, because our
particles are not point-like but have an final extension, which is, as
mentioned above, equal the resolution $l$. In Table \ref{tab0} we
indicate the ratio $l/r_0$ for the different resolutions.

\subsection{Initial Conditions}
  
Because we are interested in the secular time behavior of the galaxy
disk, the simulations are performed for $t=10$ galactic rotations. In
the initial state at $t=0$ the particles are distributed uniformly in
the $x$-$y$-plane. In the $z$-direction the particle distribution
follows an isothermal law
\begin{equation}
\rho\propto\mathop{\rm sech}{}^2(z/z_0)\, , 
\end{equation}
where $\rho$ is the density and $z_0$ is the disk scale height.
The velocities at $t=0$ are determined by the shear
\begin{eqnarray}
\dot{x}&=&0\nonumber\\
\dot{y}&=&-2A_0 x\\
\dot{z}&=&0\nonumber
\end{eqnarray}
and the Schwarzschild velocity ellipsoid
\begin{eqnarray}
\sigma_x &=&\frac{3.36 G\Sigma_0 Q}{\kappa}\nonumber\\
\sigma_y &=&\frac{\sigma_x\kappa}{2 \Omega_0}\\
\sigma_z &=&\sqrt{\pi G \Sigma_0 z_0}\;, \nonumber 
\end{eqnarray}
where the Safronov-Toomre stability criterion is $Q \geq 1$ (Toomre
\cite{Toomre64}, Safronov \cite{Safronov60}).  This
velocity distribution is a permissible assumption, because we
represent the gas by dissipative particles.

\section{Structure Analysis}
In order to characterize the structures resulting from the shearing
box experiments, we determine the mass-size relation and the
velocity dispersion-size relation.

\subsection{Mass-Size Relation}
We choose randomly a set of particles with distances $r\le L/4$
from the center of the simulation box. The positions of the particles
in the set are restricted to $r\le L/4$ in order to avoid boundary
effects in the analysis of the mass-size relation. We will refer to
that at the end of this subsection. For each particle
in the set we count the number of neighboring particles $N(R)$ 
inside a certain radius (all particles in the simulation zone are
considered as possible neighbors).  
If we repeat this for other values of $R$ we can find the
structure dimension $D(R)$ via
\begin{equation}
D(R)=\frac{d {\rm ln}(N)}{d{\rm ln}(R)}(R)\;,
\end{equation}
where $R$ denotes the scale.
The mass-size relation is then
\begin{equation}
N(R)\propto M(R) \propto R^{D(R)}\;.
\end{equation}
If the structure dimension is independent of the scale, $D=D_f$, 
i.e., if $D$ is constant or oscillates around a mean value, then the 
mass-size relation is a power-law (Semelin \cite{Semelin00})
\begin{equation}
M \propto R^{D_f}\;.
\end{equation}
If furthermore $D_f$ is non-integer, the structure is fractal
(Mandelbrot \cite{Mandelbrot82}).  We will check if $D=D_f$ holds for
the structures resulting from the shearing box experiments. However
one has to take into account that these structures can never
correspond to an idealized mathematical set, generated by means of an
infinite number of levels.  Rather they are the result of a finite
simulation, modeling a finite physical system.  Thus the structures
resulting from our experiments can never be fractal beyond a lower and
upper cutoff.  An upper scale limit due to the numerical model is
given by the size of the simulation box in the $x$-$y$-plane.  On this
scale the system becomes periodic, meaning that it can not be fractal.
To avoid boundary effects as much as possible, we consider in our
analysis only particles inside the simulation box. Therefore the
fractal dimensions are only calculated for scales $R\le L/4$. Moreover
we determine the number of neighboring particles only for particles
with a distance $r\le L/4$ from the center of the simulation
box. Consequently even for particles at $r=L/4$ all neighboring
particles are inside the simulation box.  A lower scale limit is due
to the finite resolution of the simulation mesh. If the mesh cells
have the size $l_x\times l_y\times l_z$ and $l=l_x=l_y>l_z$ then we do
not expect to model correctly fractal structures below $2l$.

If however the structure dimension depends on the scale $D=D(R)$,
the structure dimension may simply be regarded as a statistical
measure describing the clumpyness on the corresponding scale.

\subsection{Velocity Dispersion-Size Relation}

There is observational evidence for Larson's law
\begin{equation}
\sigma \propto R^{\delta_L}
\end{equation}
on scales ${\cal O}(0.1)-{\cal O}(100)$ pc with $0.3\la\delta_L\la 0.5$
(e.g. Larson 1981, Scalo \cite{Scalo85}, Falgarone \& Perault
\cite{Falgarone87}, Myers \& Goodman \cite{Myers88}). This power-law
relation seems to extend beyond the 100 pc scale (Larson
\cite{Larson79}).

In order to check if our model can reproduce these or similar
velocity-size correlations
on the kpc scale, we determine the velocity dispersion-size 
relations for the resulting structures.
We use the same approach as for the determination of the fractal dimension,
but now we calculate  the velocity dispersion $\sigma$ of the particles 
inside a certain radius $R$. Then 
\begin{equation}
\delta(R)=\frac{d{\rm ln}\sigma}{d{\rm ln}R}(R)\;.
\end{equation}
What we said about the structure dimension applies also for $\delta$.
If $\delta$ is constant or oscillates around a mean value over a
certain scale range it may be regarded as the power-law index of
Larson's law on the kpc scale, $\delta=\delta_L$. 
If however $\delta$ depends on the scale, $\delta=\delta(R)$, 
it may be considered as a 
statistical measure determining the velocity correlation.

\section{Results}

\begin{figure*}
\psfig{file=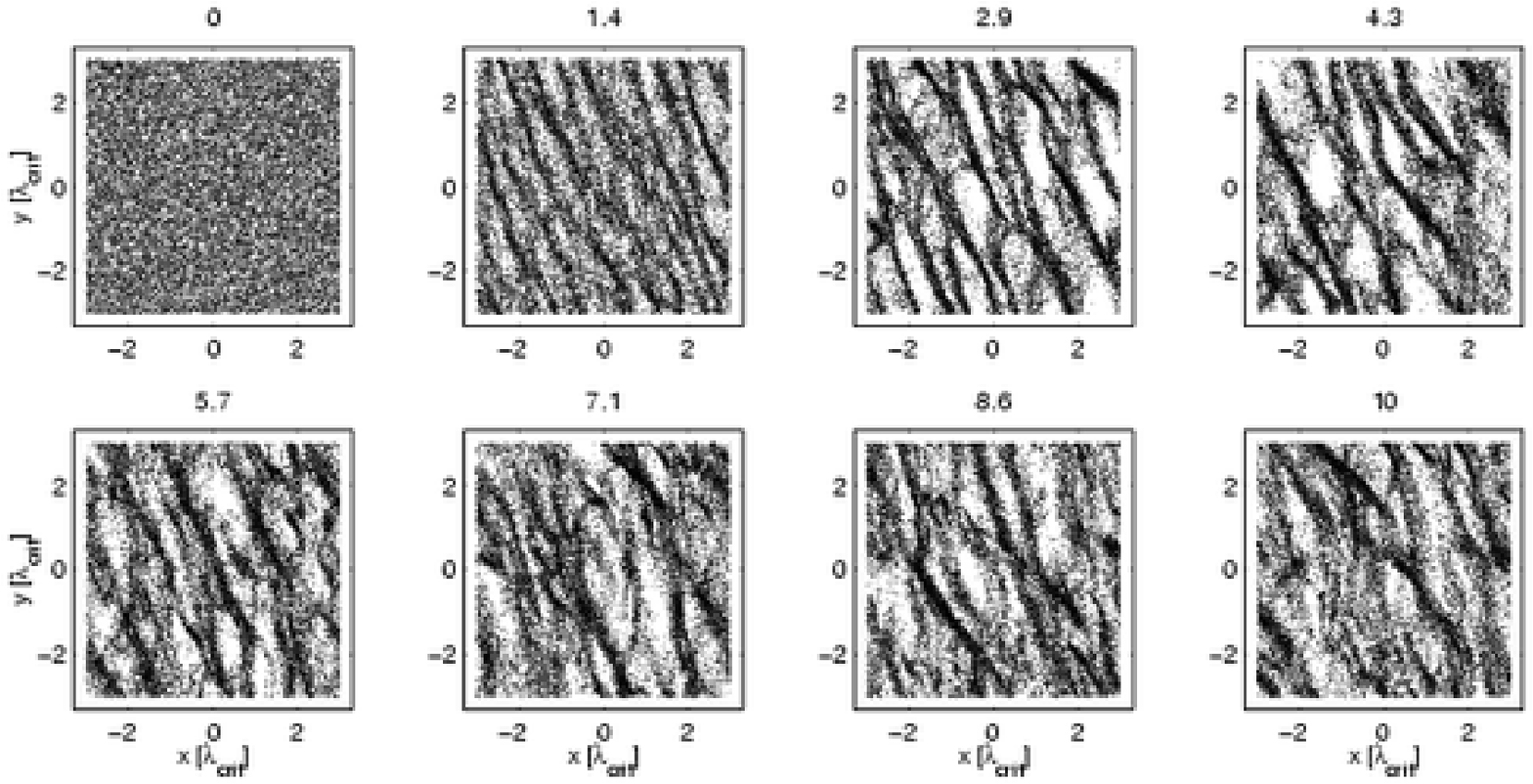,angle=90,width=\hsize}
\caption{\label{2dxy1} The evolution of the particle positions, resulting 
    from model 1 with a ``weak'' dissipation.The number of rotations of the
    shearing box around the galaxy center is indicated at the top of
    each panel. Shown is each second particle. Full resolution figures 
    available at http://obswww.unige.ch/Preprints/dyn\_art.html.}
\end{figure*}

\begin{figure*}
\psfig{file=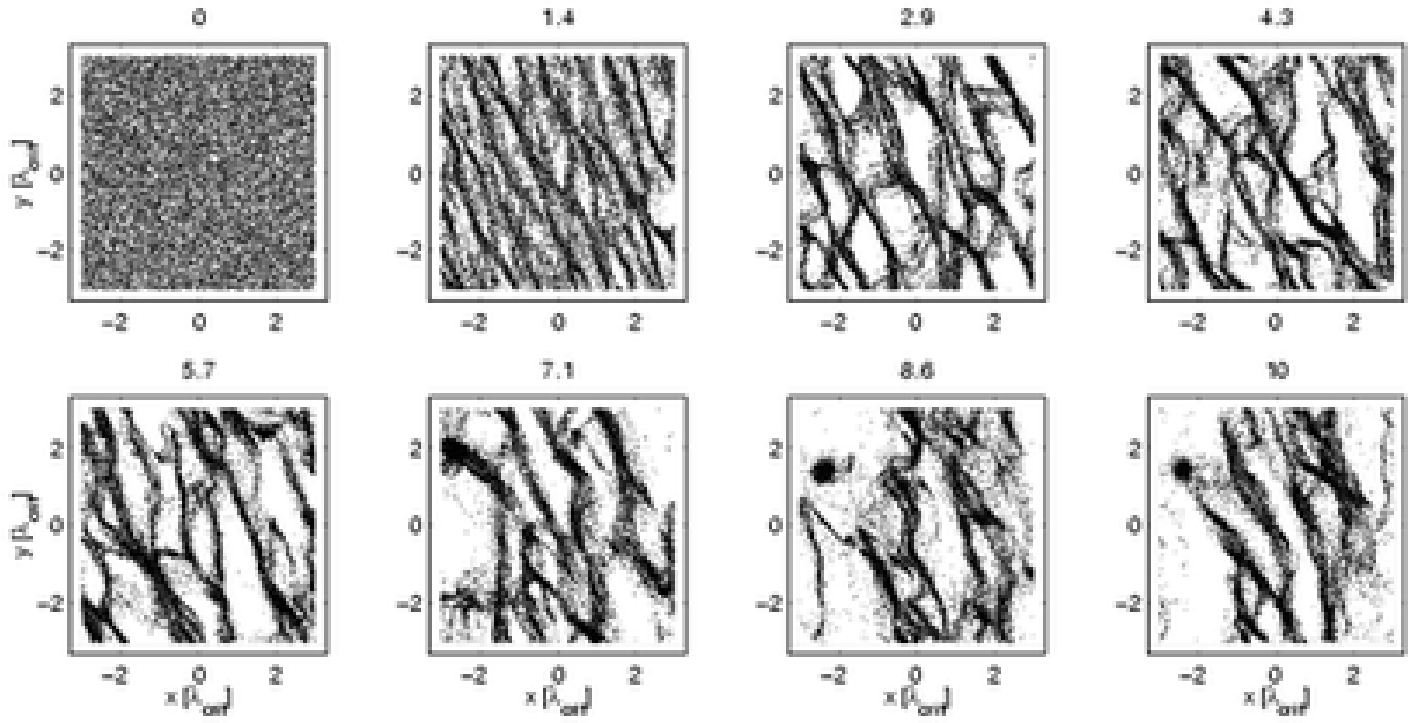,angle=90,width=\hsize}
\caption{\label{2dxy2} The evolution of the particle positions, resulting 
    from model 1  with a ``middle'' dissipation.}
\end{figure*}

\begin{figure*}
\psfig{file=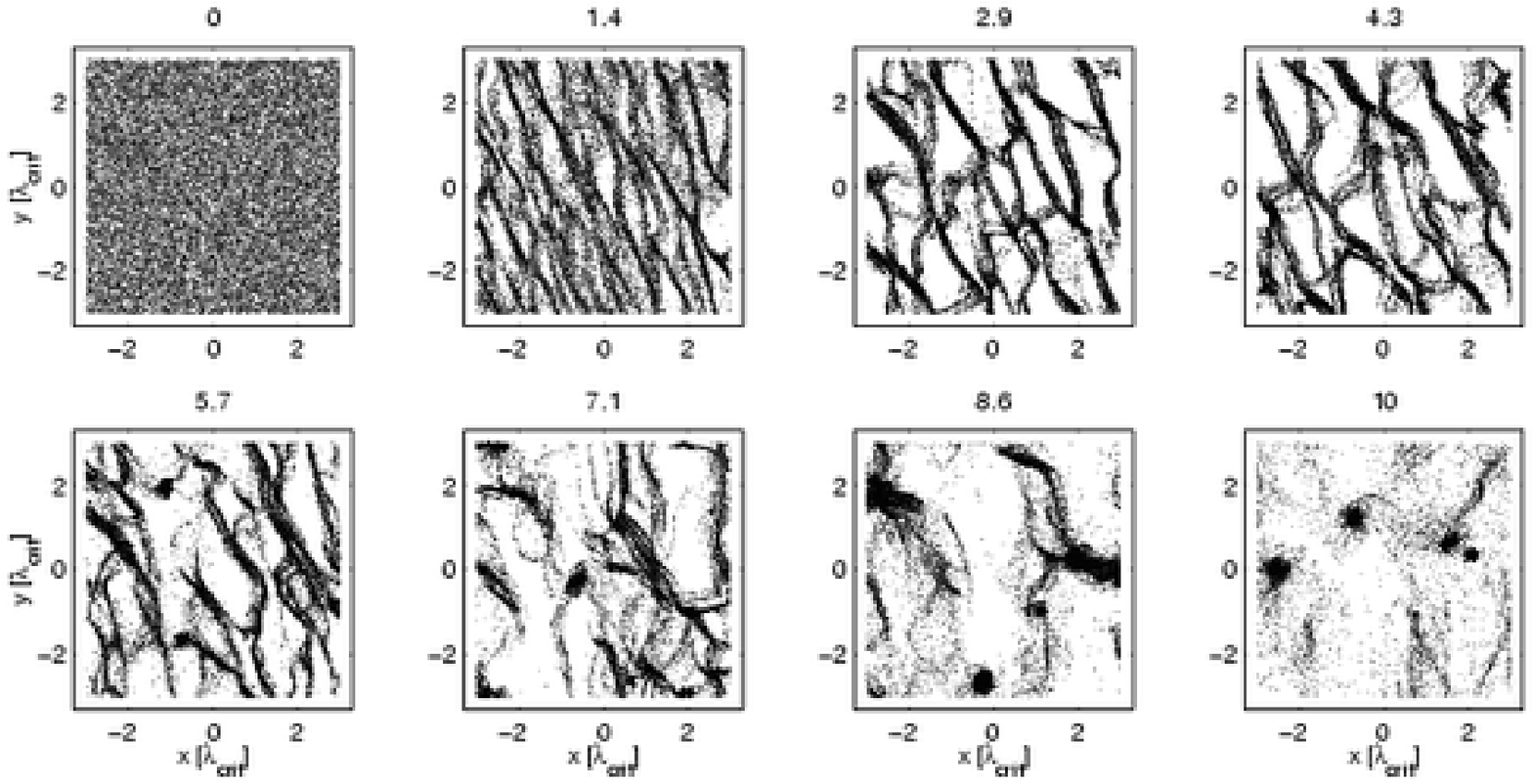,angle=90,width=\hsize}
\caption{\label{2dxy3} The evolution of the particle positions, resulting 
    from model 1  with a ``strong'' dissipation.}
\end{figure*} 

\subsection{2D Simulations}
\label{sc2d}   
To compare with earlier simulations, we start with 2D shearing sheet
simulations (model 1, 2 in Table~3, 4). The structures resulting from
these simulations depend mainly on the relative strength of the
competing gravitational and dissipation processes. 
Even without dissipation filamentary structures are already developed 
after $\approx 1/2$ galactic rotation. Yet, gravitational
instabilities lead to a conversion of bulk kinetic energy (shear-flow)
into random thermal motion. In this way the disk is heated up. 
Thus, if there is no or a too weak dissipation the 
initially arised filamentary structures are not
maintained and smear out. However, with an appropriate dissipation 
strength, filamentary structures can be maintained in a statistical
equilibrium. If the dissipation strength is increased beyond 
the ``equilibrium value'', the filaments
become denser and denser, and clumps in filaments may be formed. If
finally the dissipation dominates completely the heating process, hot
clumps, collecting almost all the matter of the simulation zone are
formed out of the filaments. Figs.~\ref{2dxy1}-\ref{2dxy3} show the
change of the structure morphology for three different radial friction
coefficients, i.e., dissipation strength. The friction coefficients
are, $C_x=70\times 10^{-3}\;\tau^{-1}_{\rm osc}$, $C_x=140\times
10^{-3}\;\tau^{-1}_{\rm osc}$ and $C_x=210\times 10^{-3}\;
\tau^{-1}_{\rm osc}$. To express the relative strength we call the
corresponding dissipations ``weak'', ``middle'' and ``strong''. All
three simulations reveal a fast fragmentation and structure
formation. After one rotation around the galaxy center the
characteristic striations appear already. The ``weak'' dissipation
leads to a statistical equilibrium of the structure. This statistical
equilibrium establishes after about 5 rotations and is maintained for
the rest of the simulation. It has a persistent pattern, formed by
transient structures. 
Contrary to the ``strong'' dissipation, where the structure
evolution is dominated by dissipation, i.e., the dissipated energy
can not be compensated by the heating mechanism.
where no statistical
equilibrium arises
 
\begin{figure*}
\psfig{file=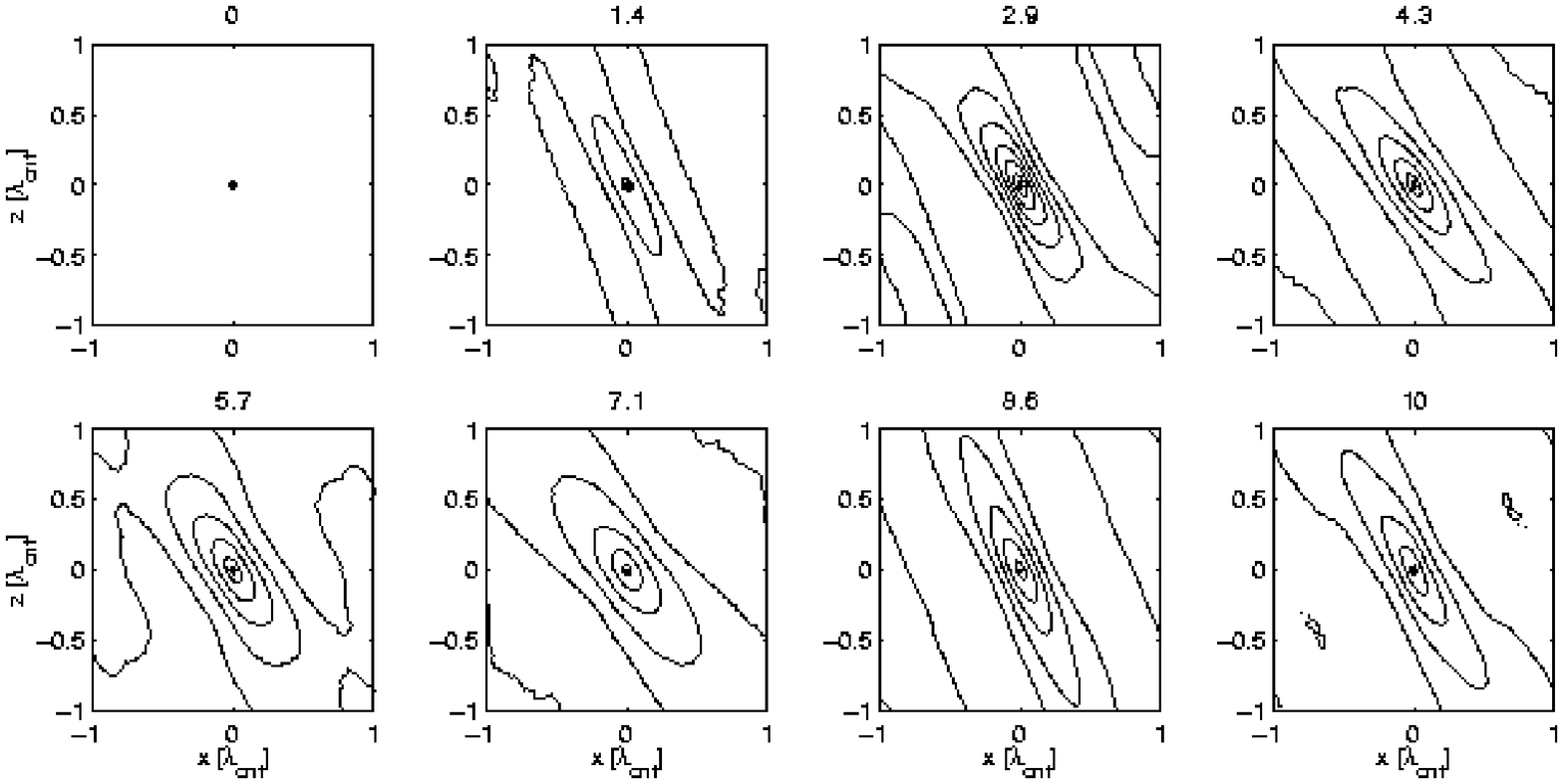,angle=90,width=\hsize}
\caption{\label{2dauto1} The autocorrelation function of the
  structures shown in Fig.~\ref{2dxy1}, resulting from the simulation 
  with the ``weak'' dissipation.}
\end{figure*} 

\begin{figure*}
\psfig{file=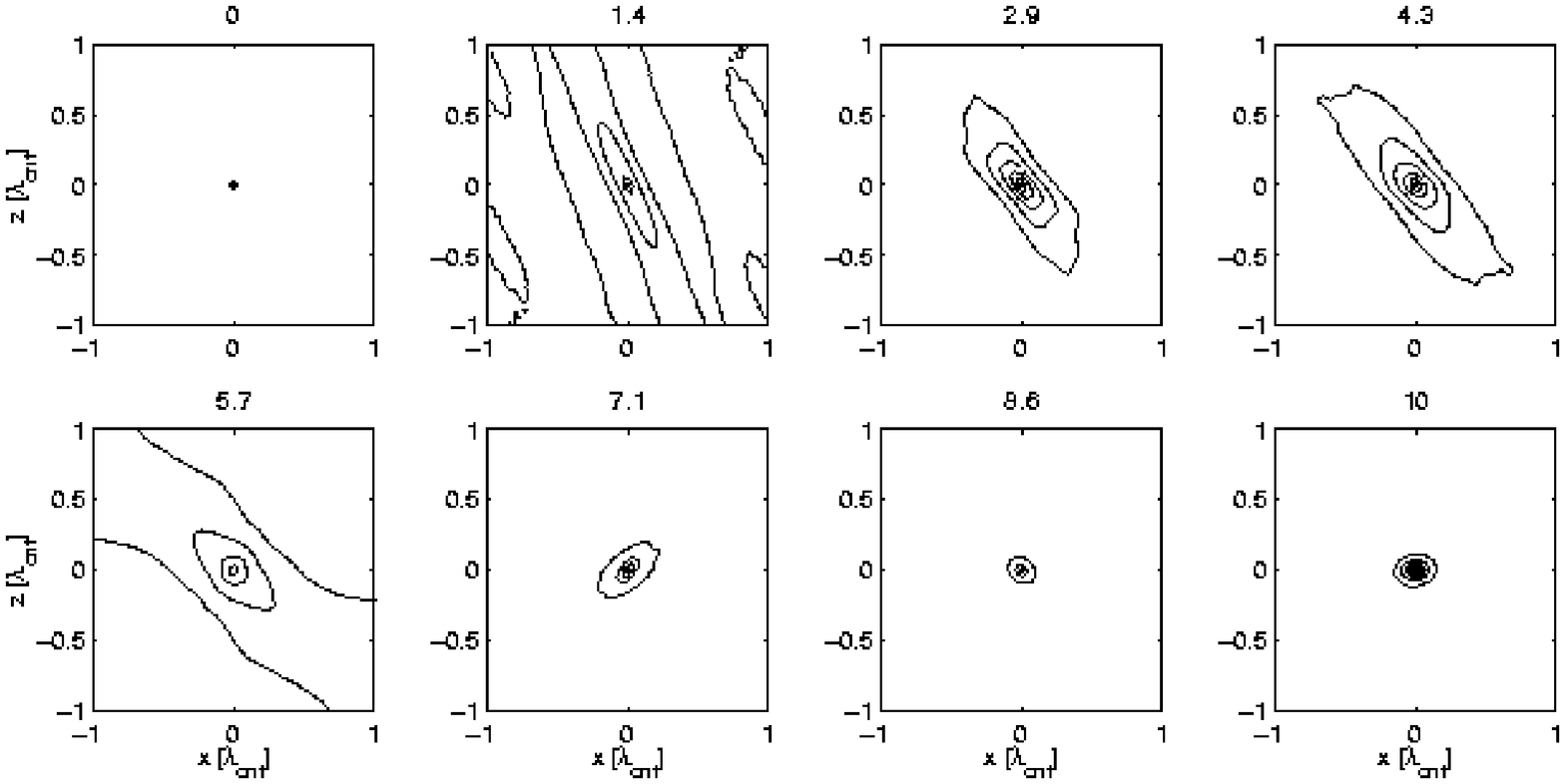,angle=90,width=\hsize}
\caption{\label{2dauto2} The autocorrelation function of the
  structures shown in Fig.~\ref{2dxy2}, resulting from the simulation 
  with the ``middle'' dissipation.}
\end{figure*}

\begin{figure*}
\psfig{file=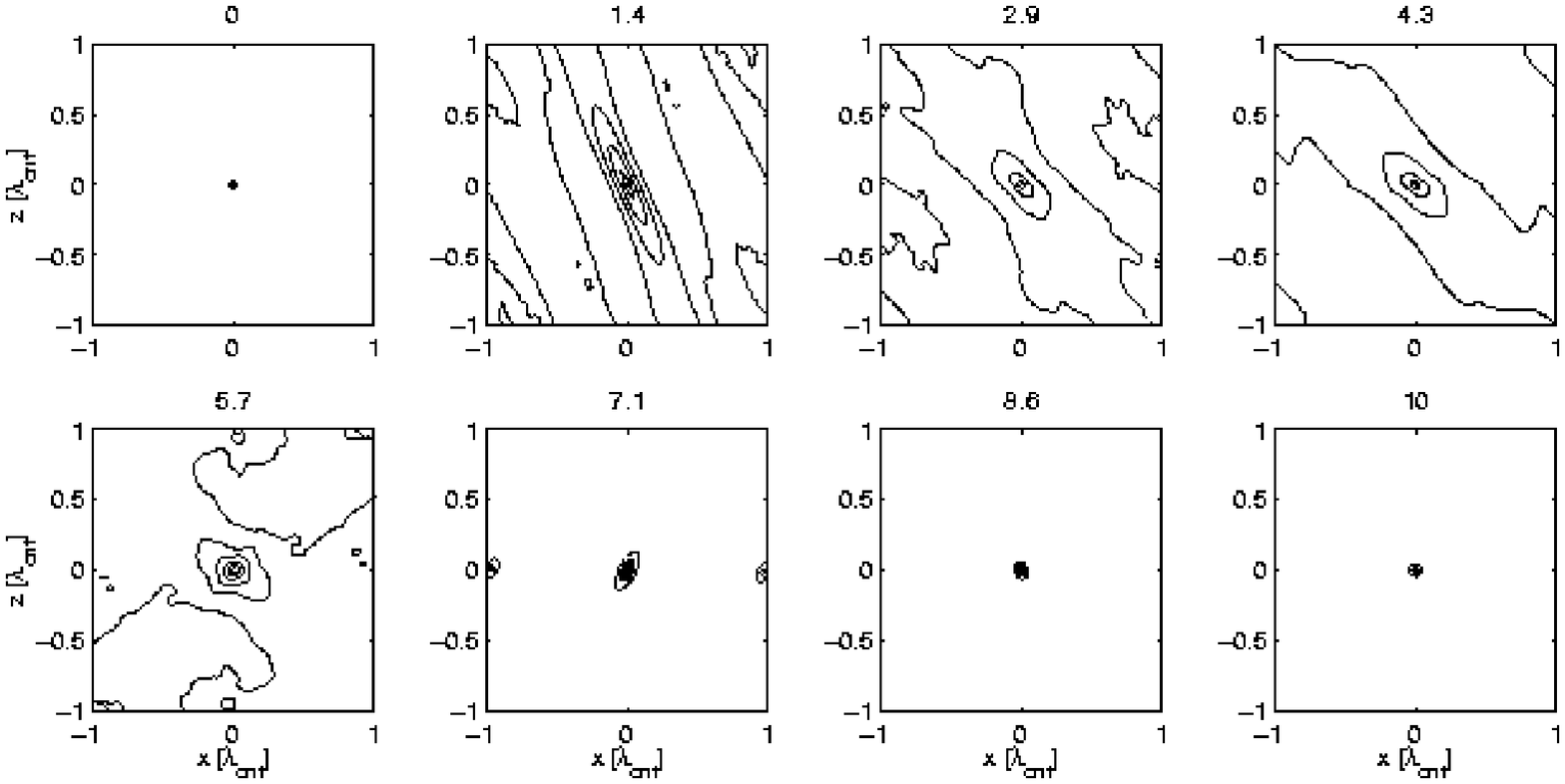,angle=90,width=\hsize}
\caption{\label{2dauto3} The autocorrelation function of the
  structures shown in Fig.~\ref{2dxy3}, resulting from the simulation 
  with the ``strong'' dissipation.}
\end{figure*} 

In order to characterize more precisely the resulting structures we
compute with Fourier transforms the 2D autocorrelation function of
the matter distributions shown in Figs.~\ref{2dxy1}-\ref{2dxy3}. 
The result is shown in Figs.~\ref{2dauto1}-\ref{2dauto2}. 
The figures reveal clearly the different morphology
resulting from the simulation with the ``weak'' and the ``middle''
dissipation. Whereas the autocorrelation function reveals striations 
with a characteristic inclination for the simulation with a ``weak''
dissipation, these striations disappear for the ``middle'' and the 
``strong'' dissipation. 

\begin{figure}
\psfig{file=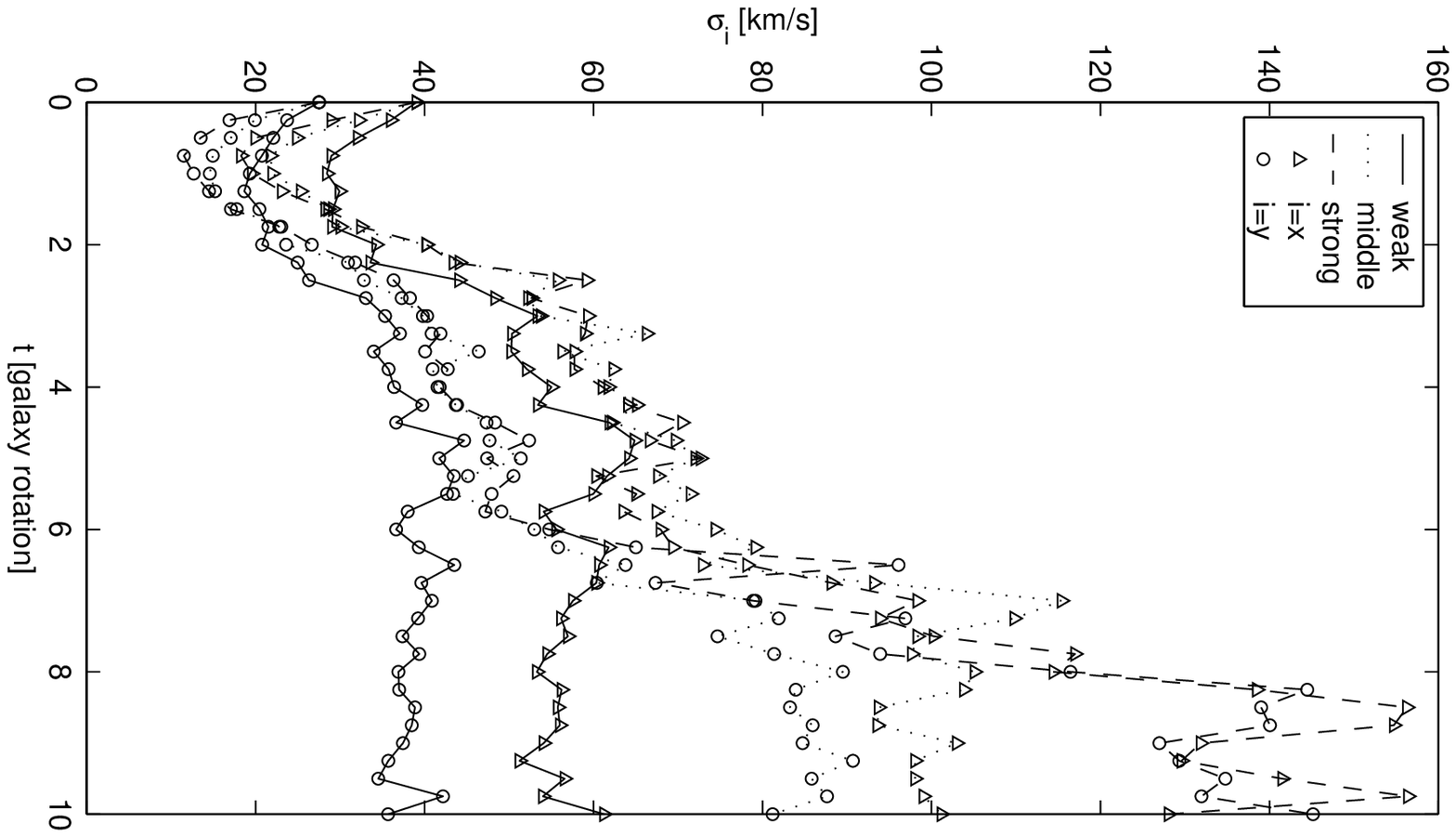,angle=90,width=\hsize}
\caption{\label{2ddisp1} The velocity dispersions
  $\sigma_x\;\;({\scriptstyle \triangle})$ 
  and $\sigma_y\;\;(\circ)$ resulting from
  the 2D simulations (model 1) with the ``weak'', the ``middle'' and
  the ``strong'' dissipation, respectively.}
\end{figure}

Because we deal with self-gravitating systems which have negative
specific heat for a certain energy range, energy dissipation does not
mean necessarily a system cooling. Indeed, the ``weak'', ``middle''
and the ``strong'' dissipation cool the corresponding system only
during the first rotation. Then the systems are heated up.  After some
rotations heating and energy dissipation are balanced out and the
velocity dispersions $\sigma_x$ and $\sigma_y$ reach a more or less
stable level (see Fig.~\ref{2ddisp1}).

It is interesting to note that $\sigma_x > \sigma_y$ always holds for
the ``weak'' and the ``middle'' dissipation strength. The same holds
also for the ``strong'' cooling during the first six rotations, then
this ordering is obviously destroyed by the formation of hot clumps.

\begin{figure}
\psfig{file=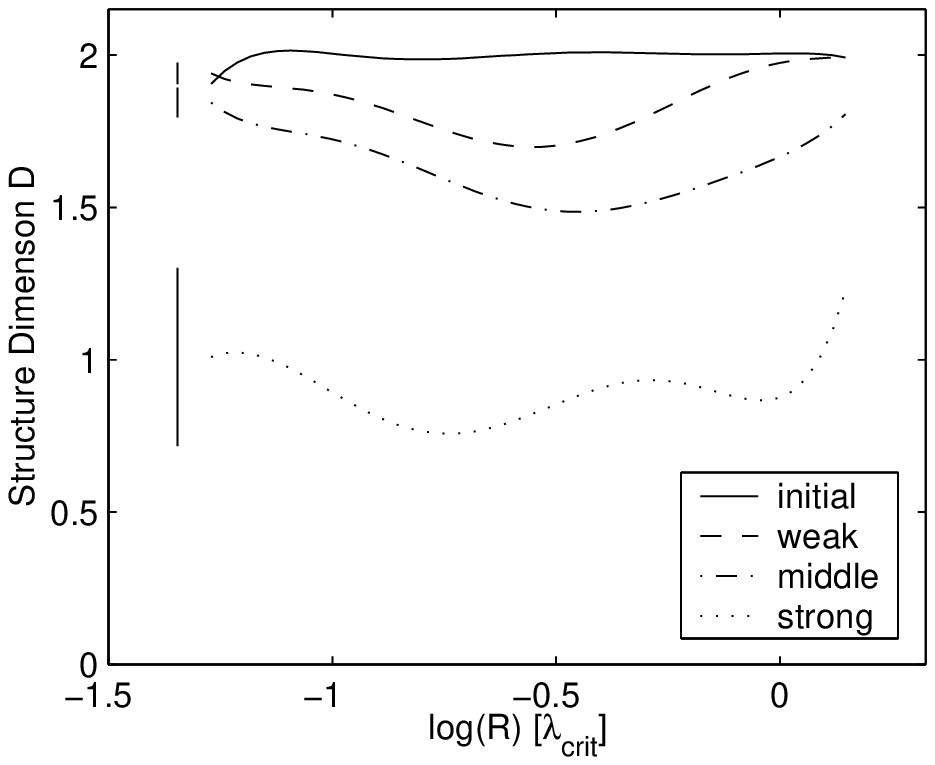,angle=90,width=\hsize}
\caption{\label{2dfrac} The structure dimension $D$ as a function of
  the scale $R$. The corresponding structures result from the 2D
  simulations (model 1) with the ``weak'', the ``middle'' and 
  the ``strong'' dissipation, respectively. The error bars at the
  left of each curve indicate the mean $1 \sigma$ errors.}
\end{figure}

\begin{figure}
\psfig{file=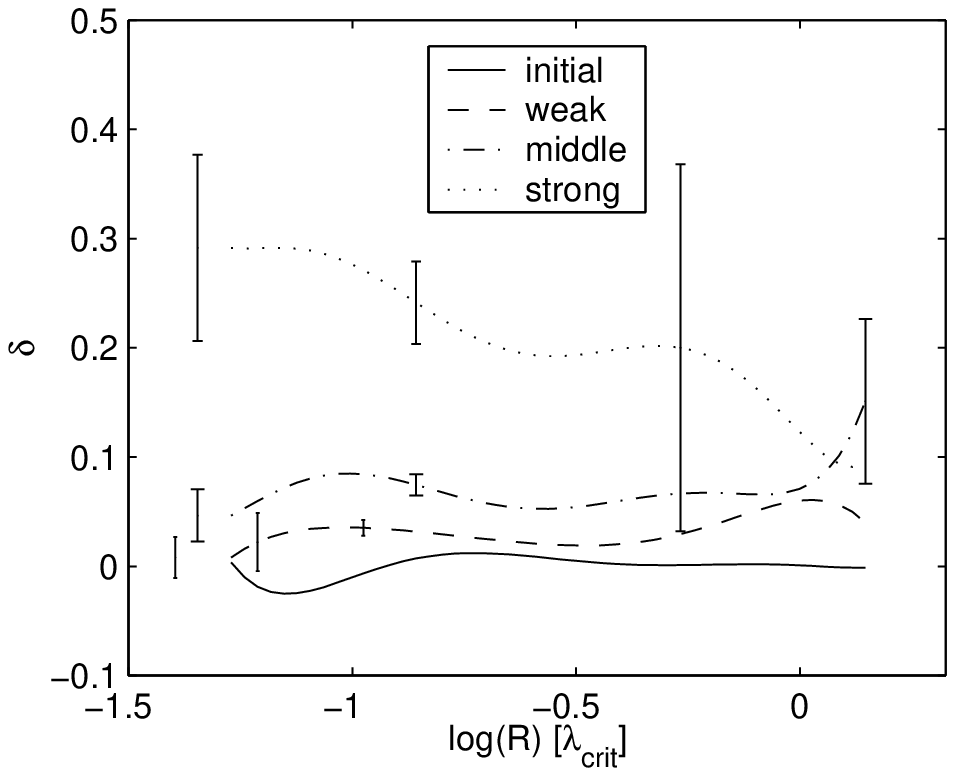,angle=90,width=\hsize}
\caption{\label{2dlarson} The index $\delta$ as a function of the
  scale $R$. The corresponding structures result from the 2D
  simulations (model 1) with the ``weak'', the ``middle'' and
  the ``strong'' dissipation, respectively. At the left of each curve
  the $1 \sigma$ mean error bars are indicated. Furthermore, 
  the location and the size of the maximal and the minimal error bars
  are indicated.}
\end{figure}

In order to characterize the structure of the simulation terminal
phase we determine the structure dimension $D$ and the index $\delta$.
The longer term evolution of the structures may be superimposed by
fluctuations\footnote{To obtain an idea about these fluctuations see
the evolution of the Schwarzschild velocity ellipsoid in
Fig.~\ref{2ddisp1} and Fig.~\ref{3ddisp1}.} on time-scales of the
order of $\sim 1/2\; \tau_{\rm rot}$, where $\tau_{\rm rot}$ is the time for a
rotation around the galactic center.  In order to eliminate these
fluctuations we indicate in this paper mean values of the structure
dimension $D$ and the index $\delta$, determined during the last two
rotations. In figures showing $D$ or $\delta$ we indicate in addition
$1 \sigma$ error bars.
    
\begin{figure*}
\psfig{file=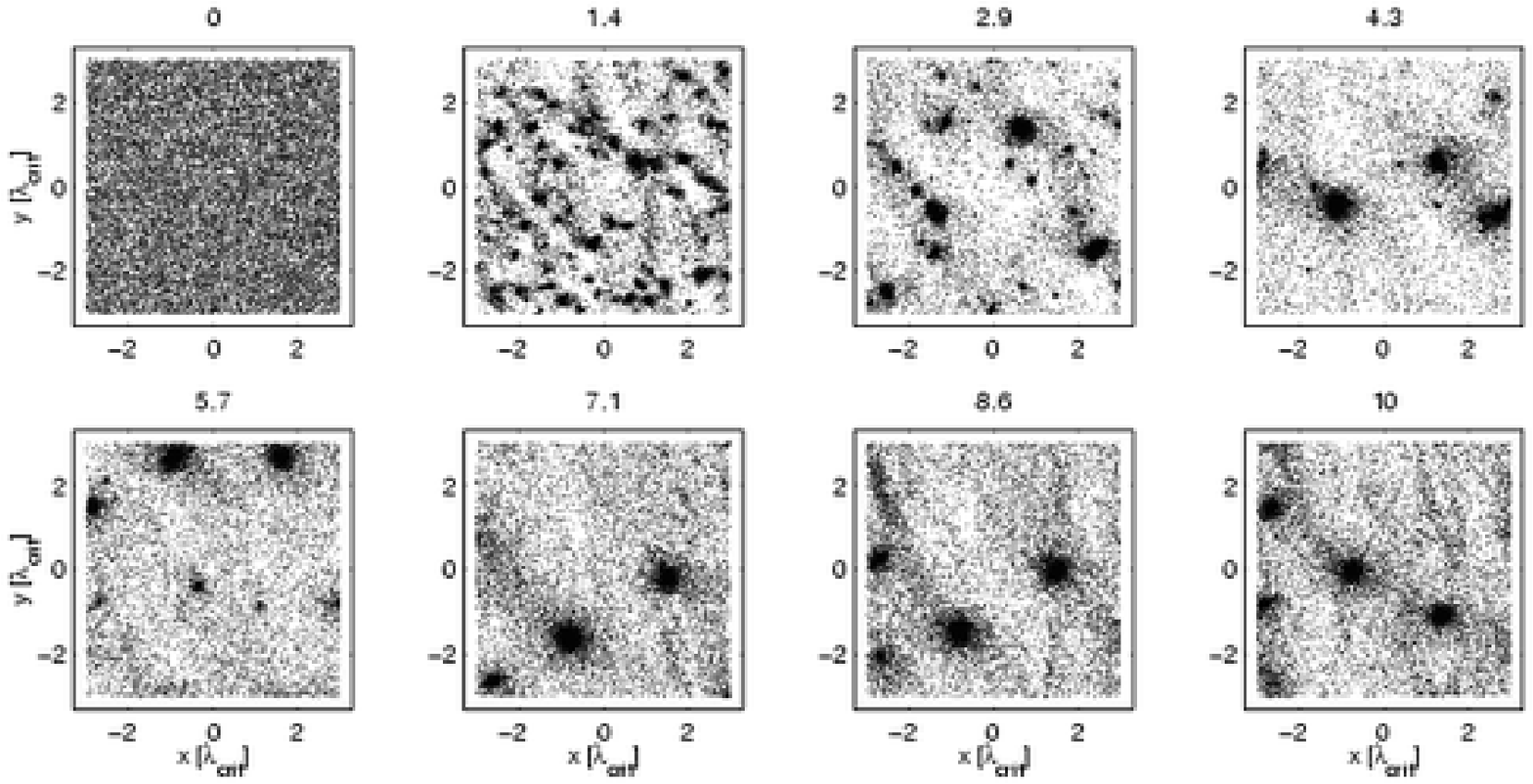,angle=90,width=\hsize}
\caption{\label{2dxy4} The evolution of the particle positions, resulting 
    from a simulation performed with model 2. The softening length is, 
    $\epsilon=0.02\;\lambda_{\rm crit}$. 
    Shown is each second particle.}
\end{figure*} 

\begin{figure}
\psfig{file=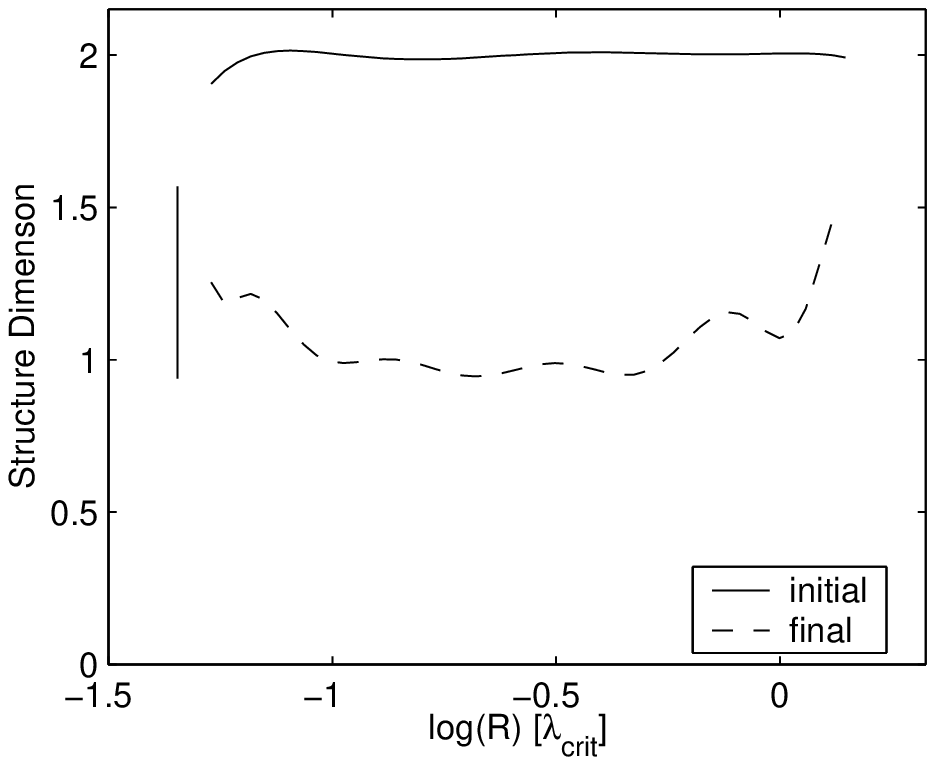,angle=90,width=\hsize}
\caption{\label{2dfrac2} The structure dimension $D$ as a function of
  the scale $R$. The corresponding structures result from a 2D
  simulation performed with model 2. The softening length is, 
  $\varepsilon=0.02\;\lambda_{\rm crit}$. The
  dynamical range of the simulation is 2.5 dex.}
\end{figure}

The dimension $D$ and the index $\delta$ resulting form the structures
shown in Fig.~\ref{2dxy1}-\ref{2dxy3} are plotted in
Fig.~\ref{2dfrac} and Fig.~\ref{2dlarson}, respectively, 
as a function of the scale $R$. The vertical lines at the left
of the curves are the $1 \sigma$ error bars. Contrary, to
the structure dimension $D$ the error bars of the index $\delta$
can vary up to a factor 7. Thus we indicate in Fig.~\ref{2dlarson}
in addition the position and the size of the largest and the
smallest error bars.

The stronger the dissipation is, the more filamentary resp. clumpy
are the resulting structures and consequently the lower is the structure
dimension. A comparison of the structure dimensions $D$ with the
indices of the velocity dispersion-size relation $\delta$ shows that 
$\langle \delta(R) \rangle$ increases with decreasing
$\langle D(R) \rangle$, where $l<R<L$ (resolution: $l=l_x=l_y$,
box size in the plane: $L=L_x=L_y$).  

For the strong dissipation the mass-size relation can be approximated
by a power-law for a scale range of roughly 1 dex, but 
the error bars are relatively large and the scale
range is too small to call the corresponding structure scale-free
or fractal.

As long as the structures are not completely
dominated by hot clumps (strong dissipation) 
the velocity dispersion-size relation may be
approximated by a power-law.
However, also here the mean error bars are quite large with respect to the
value of $\delta$, especially for the weak dissipation, where
the resulting $\delta$-value would also be compatible with
an uncorrelated velocity dispersion-size relation.

For the middle dissipation there is a little more evidence for
a power-law relation over roughly 1 dex. However, also if there
is a power-law relation the value of $\delta$ would be far away
from the index of Larson's law measured in  
molecular clouds $(0.3<\delta_L<0.5)$.

The softening length in model 1 is quite large and corresponds to
those of the TK model. In model 2 (Table 3 and 4) we reduce the
softening length $\varepsilon$, but we pay attention that
$\varepsilon>l_x=l_y$ is always valid, i.e.,  that the softening length
is always larger than the cell size of the simulation mesh. As
expected the general tendency is that a smaller softening yields a
stronger clustering and thus a smaller structure dimension.  The
structures resulting from simulations with a small softening length
$(\varepsilon<0.1)$ become relatively fast very clumpy (after 2-3
rotations).  In contrast to simulations with a strong dissipation,
where also clumps are formed, the number of clumps remains from a
certain moment on nearly constant, i.e., it does not decrease due to
mergers.  This is shown in Fig.~\ref{2dxy4}.  The structure dimension
for this simulation is shown in Fig.~\ref{2dfrac2}. 

\subsection{3D Simulations}

\begin{figure}
\psfig{file=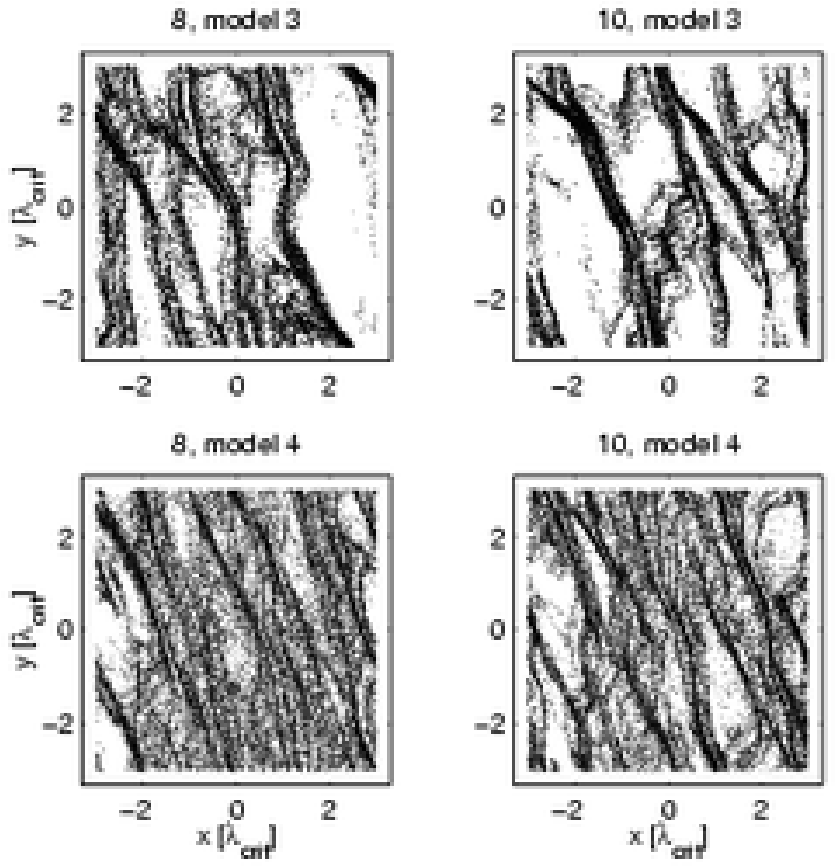,angle=90,width=\hsize}
\caption{\label{3dharmxy} The particle positions after 8 and 10
    rotations, resulting 
    from model 3 and model 4. For model 4 we only show each forth
    particle. Upper panels (model 3): 
    $C_x=140\times 10^{-3}\;\tau^{-1}_{\rm osc},\;\varepsilon=0.3
    \;\lambda_{\rm crit},\;N=32760$.
    Lower panels (model 4): 
    $C_x=110\times 10^{-3}\;\tau^{-1}_{\rm osc},\;\varepsilon=0.3
    \;\lambda_{\rm crit},\;N=131040$.}
\end{figure} 

\begin{figure}
\psfig{file=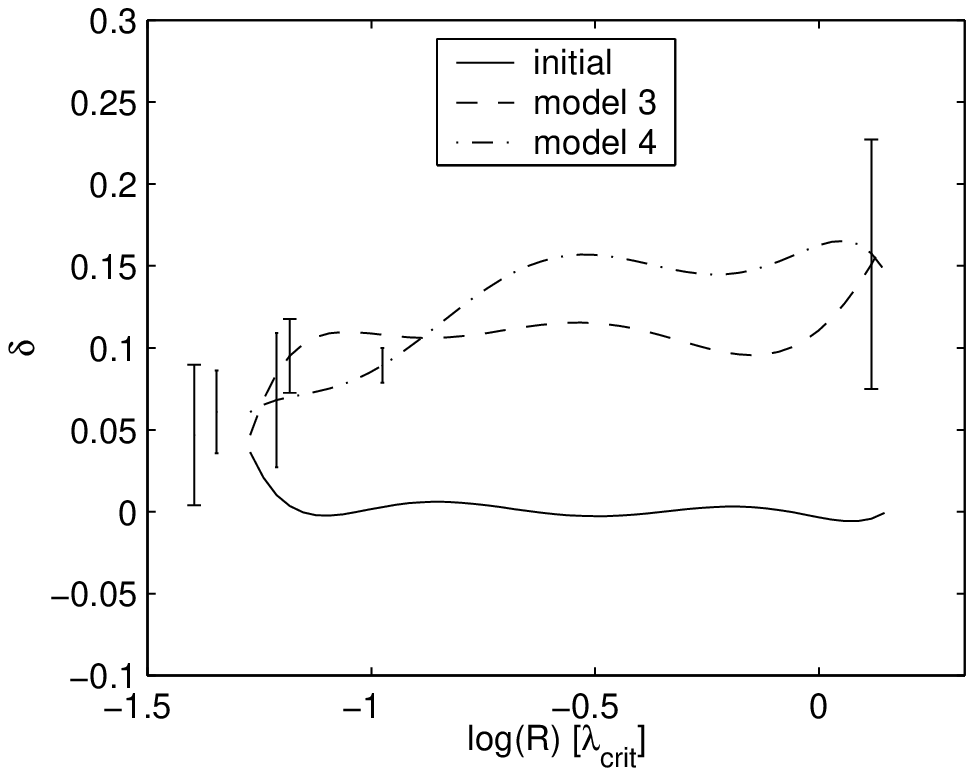,angle=90,width=\hsize}
\caption{\label{larson3dharm} The index $\delta$ as a function of the
  scale $R$. The corresponding structures result from simulations
  performed with model 3 and
  4. The index resulting from model 3 shows a power-law relation with
  $\delta_L\approx 0.1$.}
\end{figure}

\subsubsection{Isotropic Kernel}

We extend the models to 3 dimensions and carry on using an isotropic
particle potential (model 3 and 4).

In models using a particle-mesh method the number of particles is
determined by the number of mesh-cells and vice versa.  Due to
computational limits and in order to do a reasonably sized parameter
study on the available machines we have to limit the number of
particles to $N\approx 130000$. Thus it is not possible to resolve the
system vertically with a softening length $\varepsilon \approx
l_x=l_y$. The models using an isotropic potential reproduce thus only
2D dynamics in a 3D space. As expected we find therefore the same
parameter dependence as in model 1 and 2, respectively. 
Compared with previous
models we carry out here also simulations for a slightly larger
softening length. These simulations reveal for a limited scale range
approximately a velocity dispersion-size power-law relation. 
Fig.~\ref{3dharmxy} shows the
structures resulting from 2 simulations with $\varepsilon =
0.3\;\lambda_{\rm crit}$. The indices $\delta$ resulting from these 2
simulations are shown in Fig.~\ref{larson3dharm}.

\subsubsection{Anisotropic Kernel}

We use now the anisotropic  kernel described in Sect.~\ref{seckernel}.
Thus the softening length $\varepsilon$ is here no longer a free
parameter. The softening of the gravitational potential is 
given by the resolution of the simulation mesh. With the anisotropic 
kernel we can resolve the system vertically, so that there is a
vertical dynamical-range of 1.8 dex for all models with anisotropic
kernel. Since the third dimension is now resolved we explore the structure
in dependence on the friction coefficient $C_z$  (model 7) and the
vertical frequency $\nu$ (model 8). 

\begin{figure}
\psfig{file=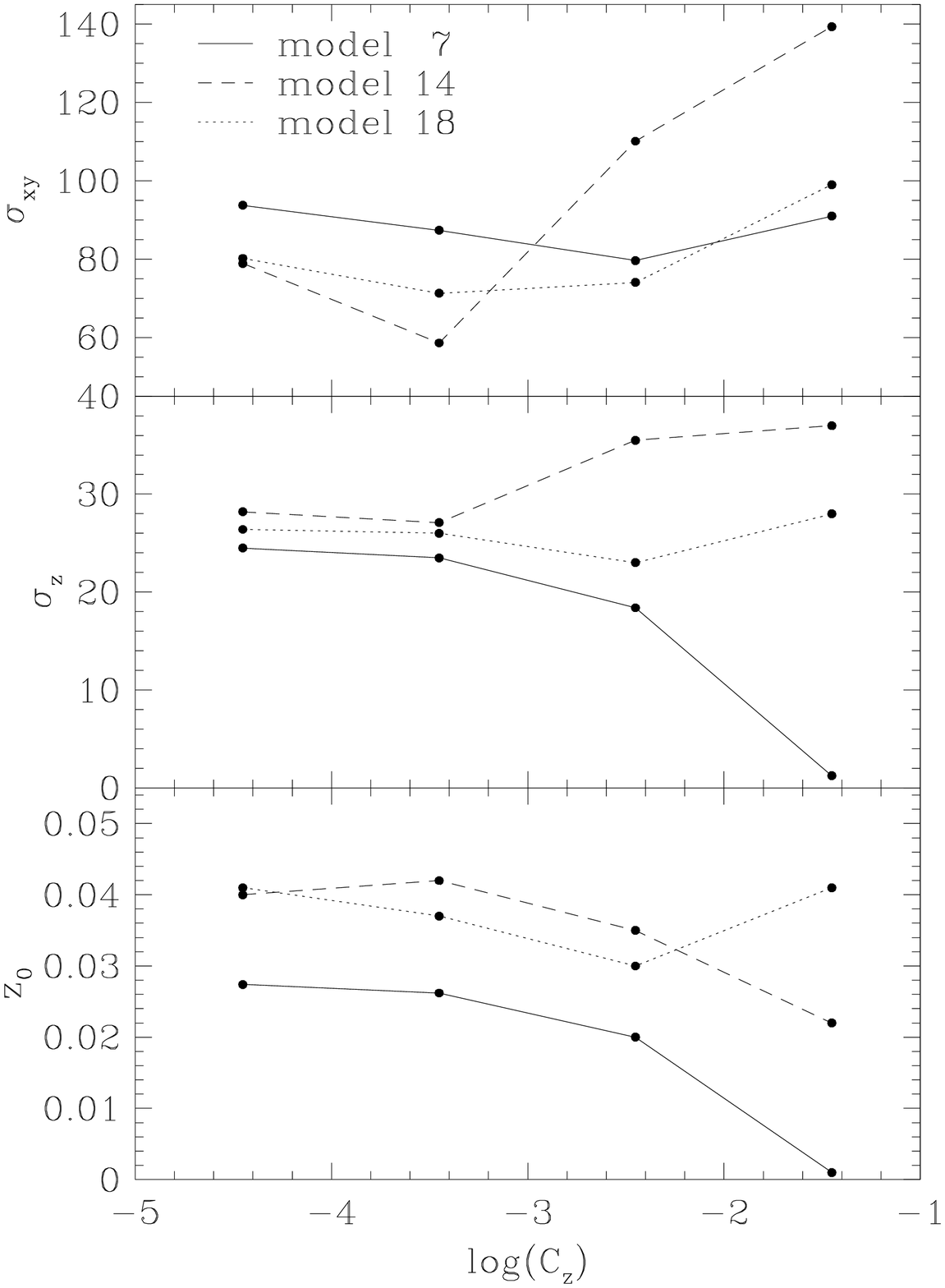,width=\hsize}
\caption{\label{hz1} The velocity dispersions in the plane
  $\sigma_{xy}$ and vertically to it, $\sigma_z$,  as a function
  of the vertical friction coefficient $C_z$ for model 7, 14 and 18
  (upper panel, middle panel). The lower panel shows the disk scale
  hight $z_0$. The parameter values are mean 
  values ascertained during the last 
  two rotations. As soon as the dissipation leads to the formation of
  clumps the disk is heated up. The structure of model 14, eg., becomes
  clumpy for $C_z=50\times 10^{-3}\;\tau_{\rm osc}$.}
\end{figure}

\paragraph{Vertical Friction Coefficient $C_z$.}
We start with simulations, where $C_x=C_z$. However, these simulations
show that with such a dissipation it is not possible to maintain
simultaneously strong density fluctuations and the disk thickness. 
That is, either the density fluctuations are maintained and 
the disk scale height tends towards
zero, or the disk scale height is maintained nearly constant and the
density fluctuations smear out. Therefore we choose for all further simulations
$C_x > C_z$, i.e., $t_{{\rm cool},x}<t_{{\rm cool},y}$. These results 
justify a posteriori the choice of two friction coefficients. The velocity
dispersions in the plane $\sigma_{xy}=(\sigma_x^2+\sigma_y^2)^{1/2}$ 
are mainly controlled via $C_x$, whereas $\sigma_z$ and with it the disc scale
height $z_0$ is principally driven by $C_z$. However, as soon as
structures are formed with sizes comparable or smaller than the disk
thickness, the dynamics in the plane and vertically to it are no
longer independent.  Thus the effect of $C_z$ on the structure depends
also on $C_x$.  The general effect of $C_z$ on the self-gravitating
disk is the following: As long as the structure remains filamentary an
increase of $C_z$, diminishes $z_0$ and $\sigma_z$. The solid curve in
Fig.~\ref{hz1}, resulting from a simulation performed with model 7,
represents such a behavior. The effect of $C_z$ is also studied in
model 14 and model 18. There the structure changes from filamentary to
clumpy due to an increase of $C_z$. As a consequence these systems may
be heated up by further dissipation (see Fig.~\ref{hz1}).

\paragraph{Vertical Frequency $\nu$.}

\begin{figure}
\psfig{file=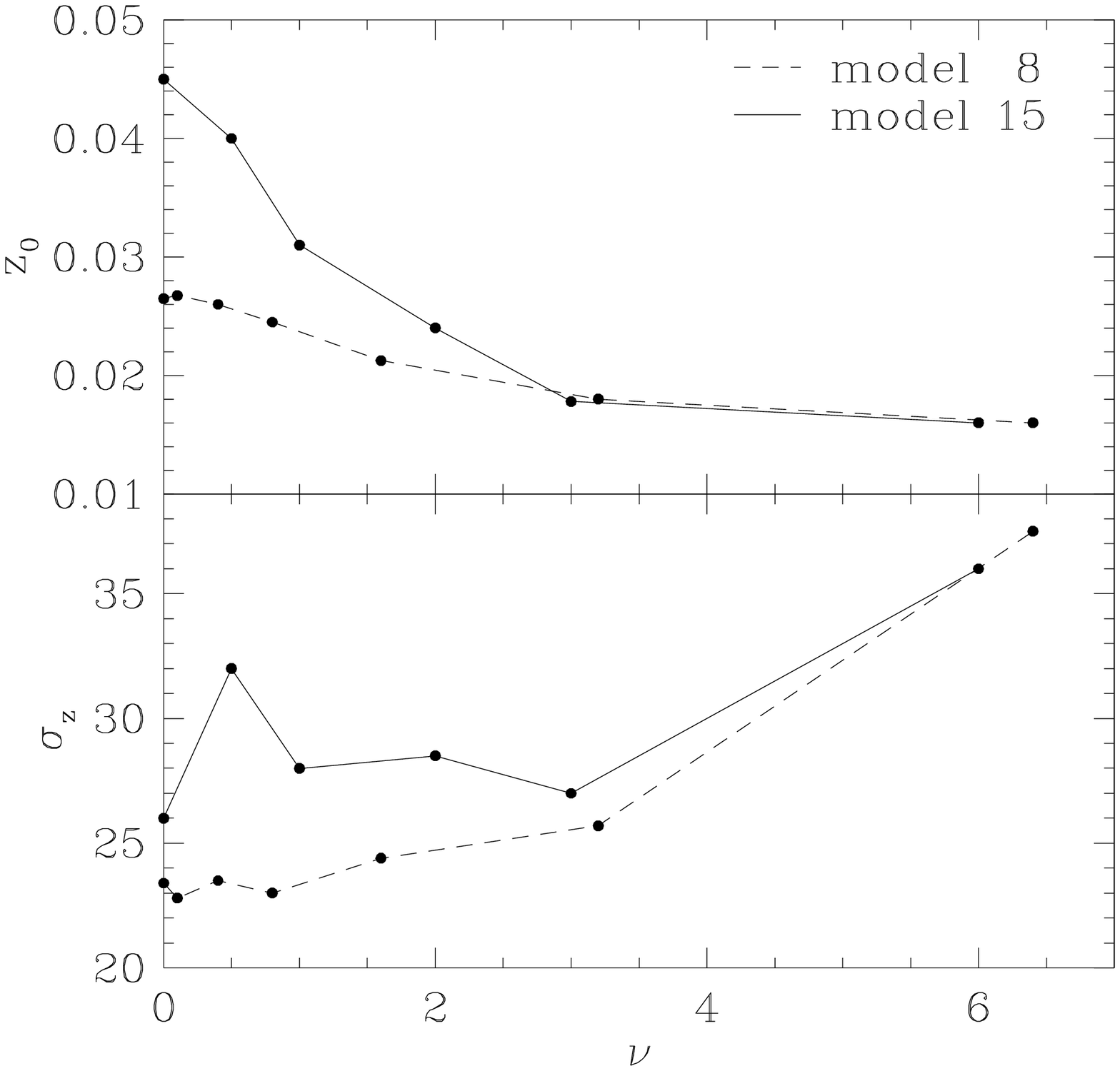,width=\hsize}
\caption{\label{nu1} The disk scale height $z_0$ and the vertical
  velocity dispersion $\sigma_z$ as a function of the vertical frequency 
  $\nu$ deduced from model 8 and model 15, respectively. 
  $z_0$ and $\sigma_z$ are mean values calculated during the
  last two galaxy rotations.}
\end{figure}

In model 8 and 15 we study the effects of the vertical frequency $\nu$ on
structure and dynamics. The vertical frequency determines the strength
of the backward force due to a displacement from the galaxy plane.
The backward force stems from the external galactic potential.
Thus an increase of $\nu$ binds the particle stronger to the disk and 
diminishes $z_0$ (see Fig.~\ref{nu1}). 

In all 3D models the mean particle-velocity vertical to the 
plane $\langle v_z \rangle$ is not exactly zero, but
oscillates with an amplitude of 
$\approx 0.1\;{\rm kms^{-1}}$ and a frequency equal to the 
vertical frequency $\nu$. 

\begin{figure}
\psfig{file=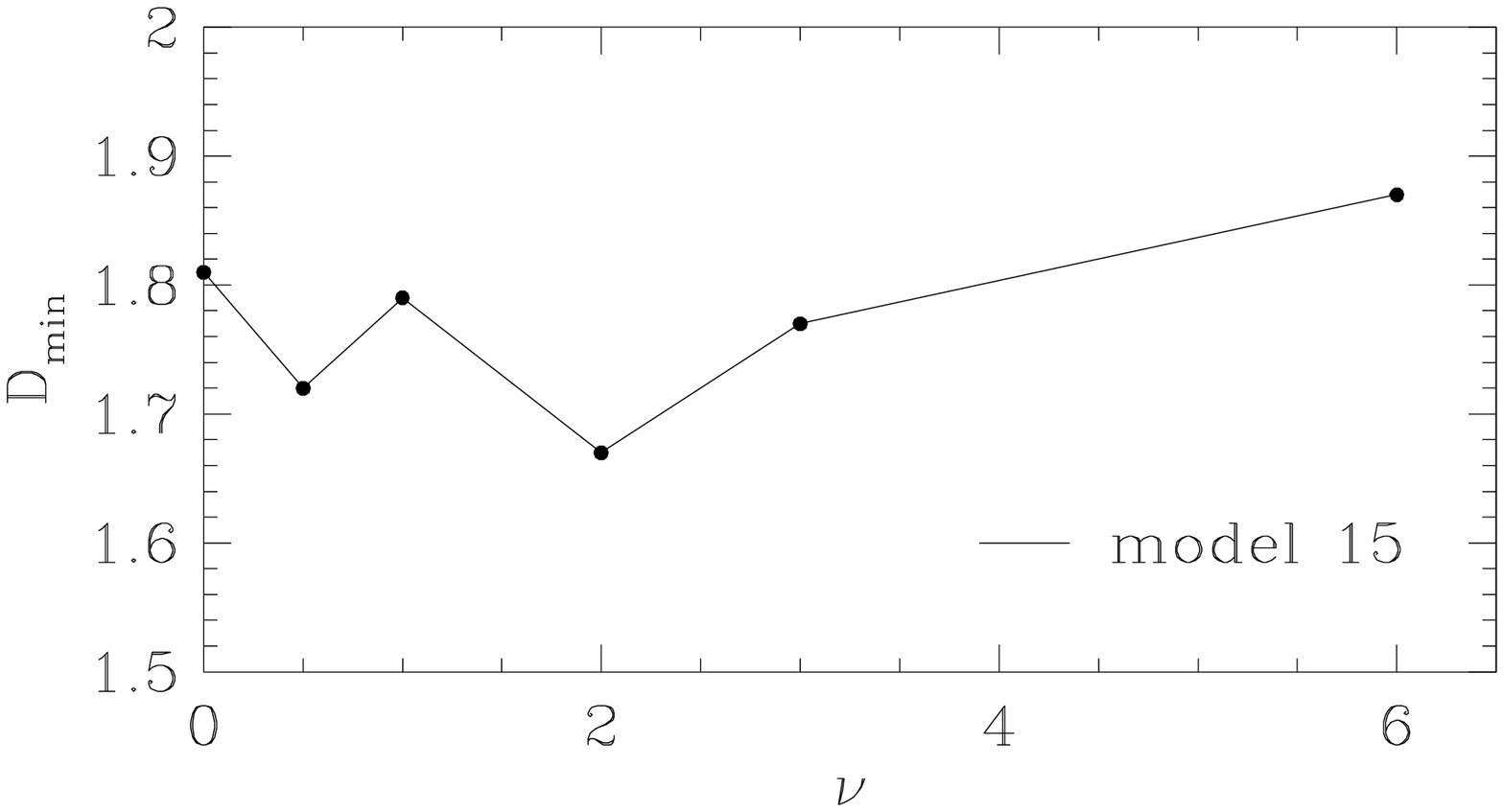,width=\hsize}
\caption{\label{nu2} The minimal structure dimension $D_{\rm min}$ as a
  function of $\nu$. The structure dimension is a mean value
  determined during the last two galaxy rotations. The corresponding
  structures result from simulations performed with model 15.}
\end{figure}

Concerning the effect of $\nu$ on the structure, there are frequencies 
producing a clumpy structure,
whereas some higher or lower frequencies produce a more filamentary
structure. To show this we determined the minimal structure dimension,
\begin{equation}
\label{dmineq}
D_{\rm min}=\{\min[D(R)]: l<R<L\}\;,
\end{equation}
where $L=L_x=L_y$ (box size in the plane) and $l=l_x=l_y$ (resolution).
The minimal structure dimension determines how strong the structures
differ from a homogeneous matter distribution, i.e., the lower
$D_{\rm min}$ the more filamentary resp. clumpy the structure.
We calculate $D_{\rm min}$ for the structures resulting from simulations
with different $\nu$. The result is shown in Fig.~\ref{nu2}.
The two simulations with $D_{\rm min}\approx 1.7$
have a more clumpy structure than the ones resulting from the
other simulations, which are more filamentary.  
  
\paragraph{Particle Number $N$.}
With model 11 we determine the structure dimension and the disk scale
height as a function of the particle number $N$. We only alter the
particle number. The mesh resolution and the friction coefficients
$C_x$ and $C_y$ are kept constant. The 2D simulations of TK shown
that a higher particle number leads to a stronger dissipation. Thus we
expect a decrease in the disk scale height and a smaller structure
dimension for higher $N$. We find a clear decrease of the disk scale
height for an increasing particle number, but there is no clear
trend of the structure dimension. This is because the disk is heated
up in the plane due to the decreasing disk thickness.

We find that the effect on the structure due to a change of the
particle number can be compensated with an appropriate choice of the 
dissipation strength, i.e., if we use a weaker dissipation for an increased
particle number, the statistical properties of the resulting
structures remain unchanged.

\begin{figure*}
\psfig{file=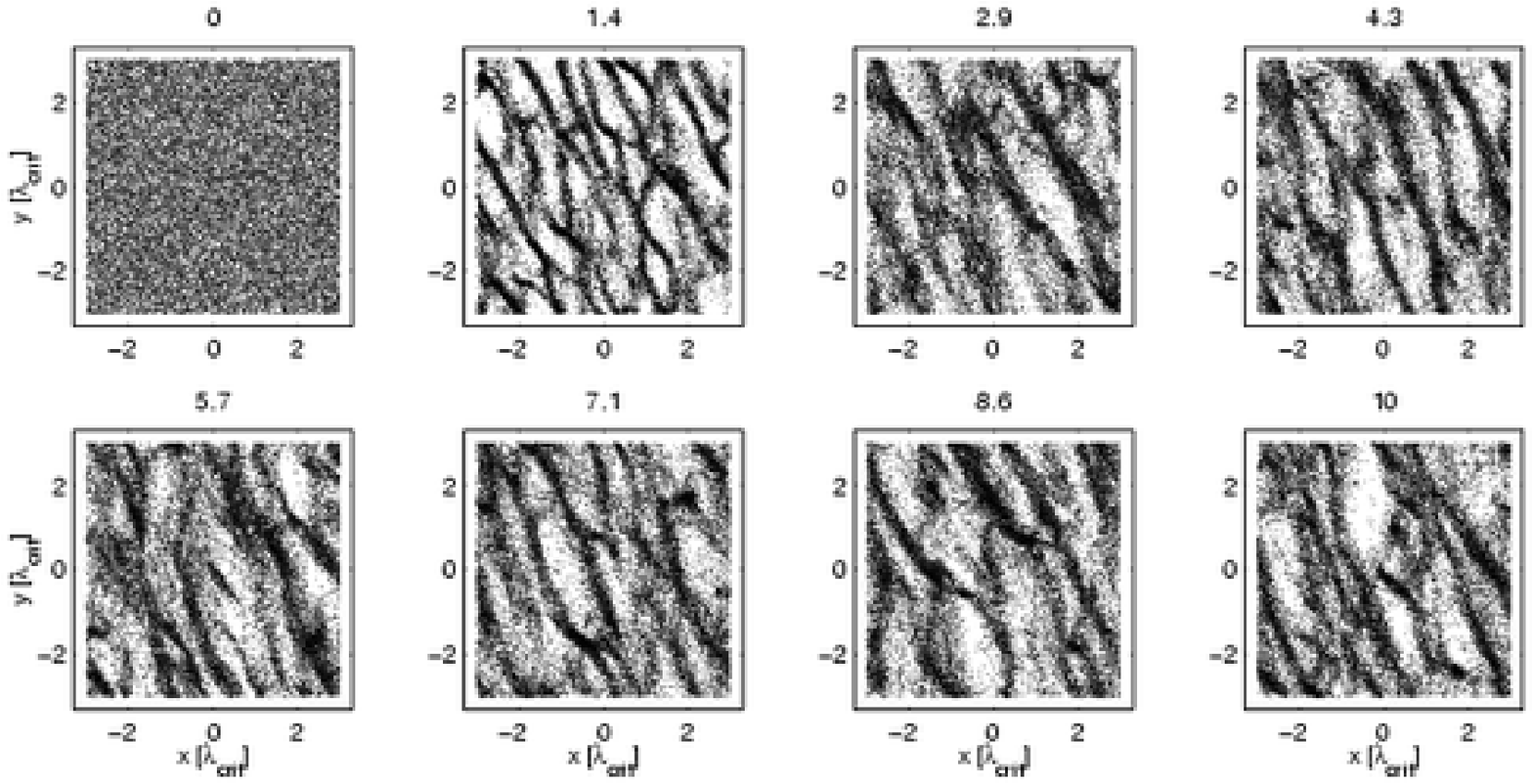,angle=90,width=\hsize}
\caption{\label{3dgrossxy1} The evolution of the particle positions
  seen from above the galaxy plane. The structures result from a
  simulation of model 10. The friction coefficient is 
  $C_x=64\times 10^{-3}\;\tau^{-1}_{\rm osc}$ (``weak'' dissipation).
  The number of rotations of the shearing box around the 
  galaxy center is indicated at the top of each panel. Shown is 
  each fourth particle.}
\end{figure*} 

\begin{figure*}
\psfig{file=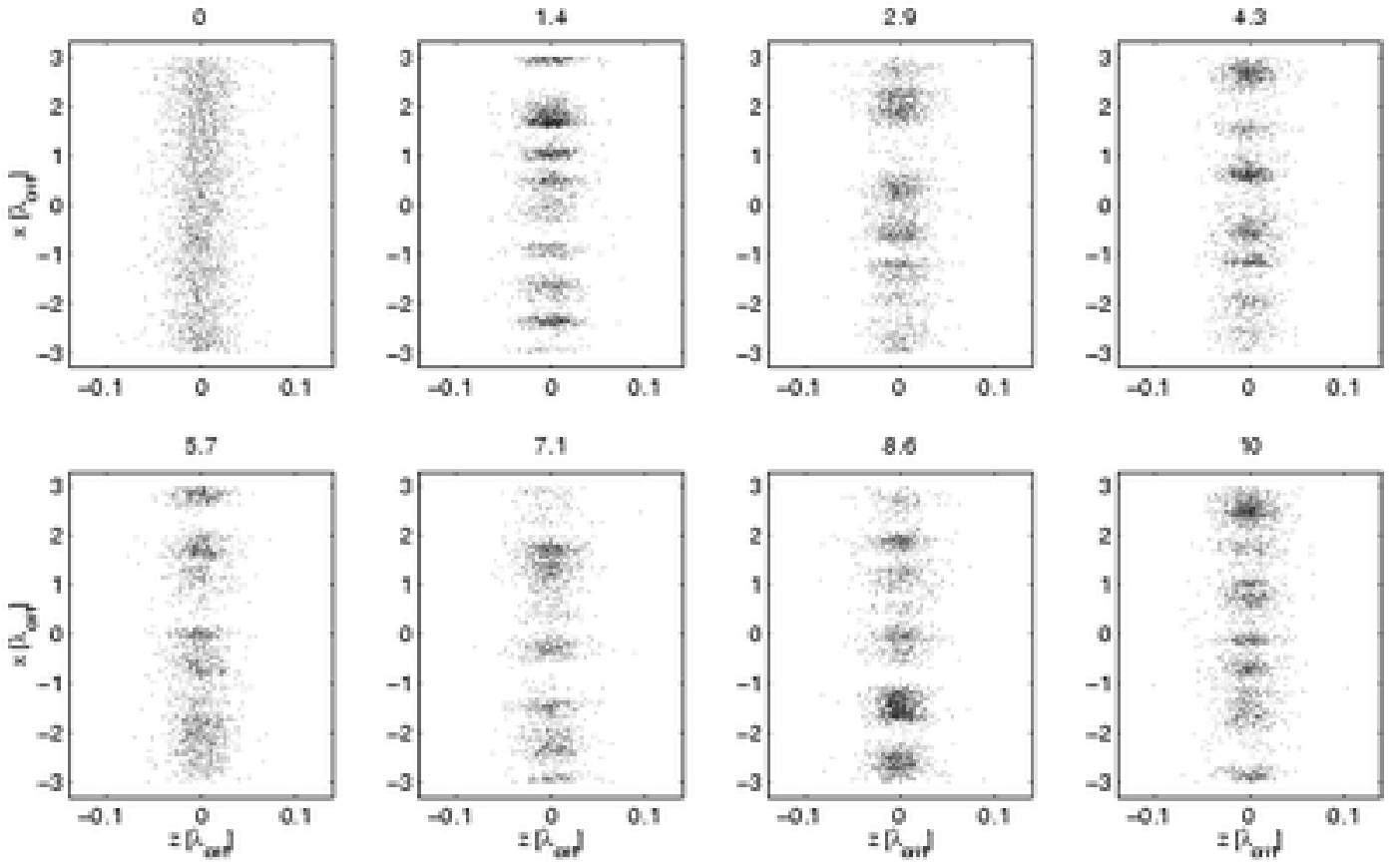,angle=90,width=\hsize}
\caption{\label{3dgrossxz1} The particle positions inside the slice,
  $-0.05 \lambda_{\rm crit}<y<0.05 \lambda_{\rm crit}$, seen along the 
  direction of orbital motion (model 10, ``weak'' dissipation).} 
\end{figure*} 

\begin{figure*}
\psfig{file=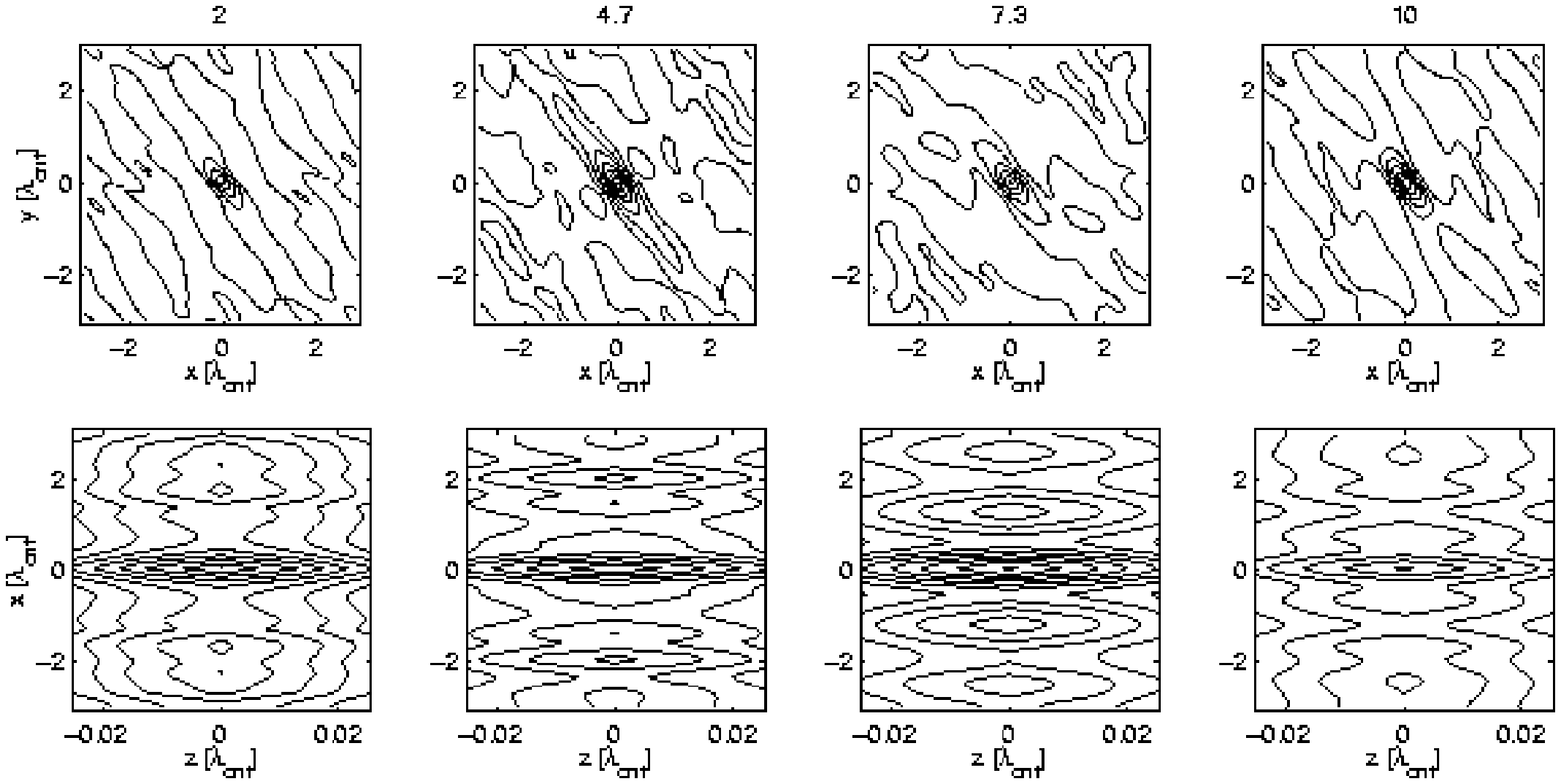,angle=90,width=\hsize}
\caption{\label{3dauto} The autocorrelation function of the particle
  distribution, presented in Fig.~\ref{3dgrossxy1} and
  \ref{3dgrossxz1} (model 10, ``weak'' dissipation). Upper panels:
  Autocorrelation function in the x-y-plane. Lower panels: 
  Autocorrelation function in the z-x-plane.}
\end{figure*} 

\begin{figure}
\psfig{file=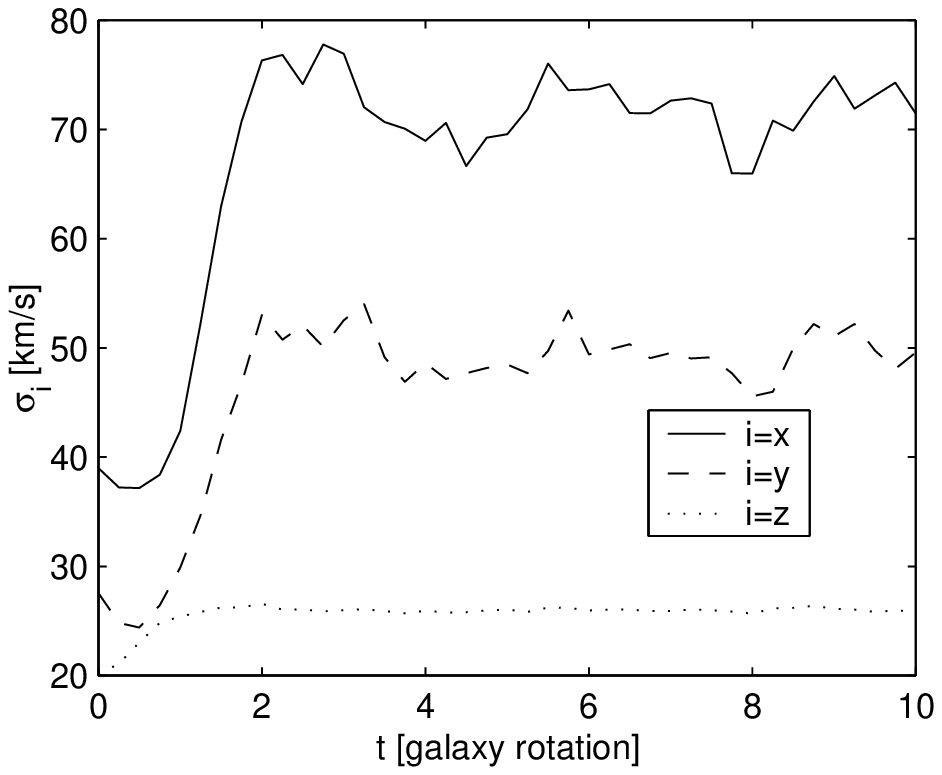,angle=90,width=\hsize}
\caption{\label{3ddisp1} The velocity dispersion components
  $\sigma_x$, $\sigma_y$ and $\sigma_z$ as a function of time $t$
  (indicated in galaxy rotation units). The dispersions result from a
  simulation performed with model 10 and ``weak'' dissipation.}
\end{figure}

\begin{figure}
\psfig{file=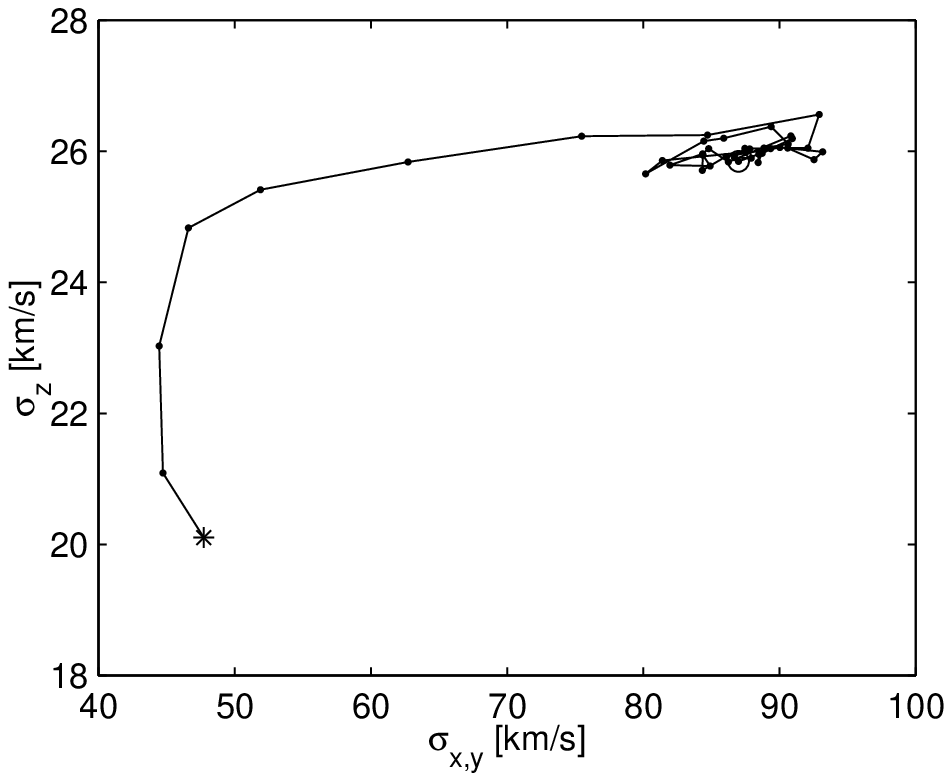,angle=90,width=\hsize}
\caption{\label{3ddisp2} The vertical dispersion $\sigma_z$ as a
  function of the dispersion in the plane $\sigma_{xy}$ (model 10,
  ``weak'' dissipation). This plot shows clearly how the system
  attains a stable state. The star and the circle denote the starting
  and the ending point respectively.}
\end{figure}

\begin{figure}
\psfig{file=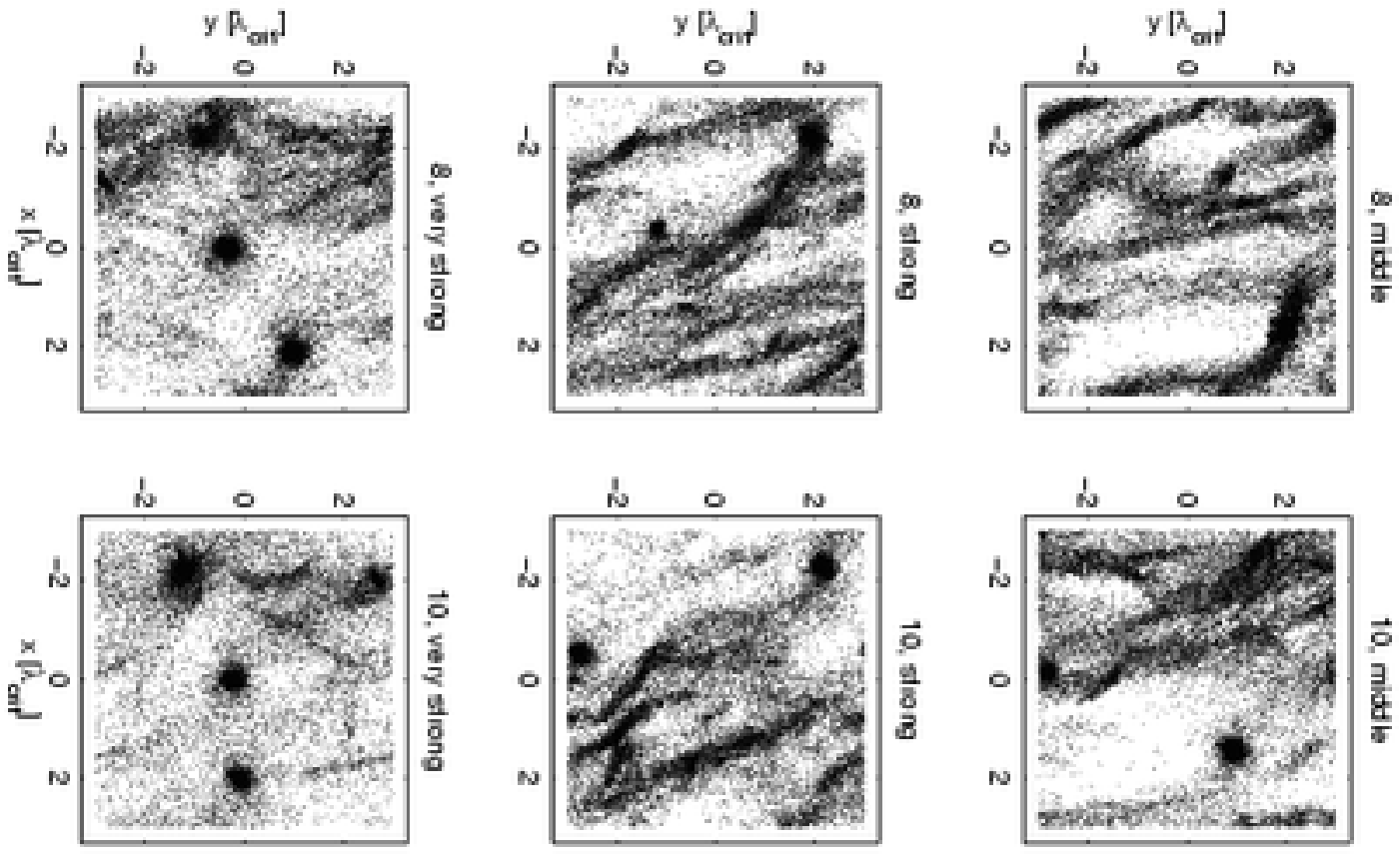,angle=90,width=\hsize}
\caption{\label{3dgrossxy2} The particle positions after 8 and 10
    galaxy rotations. The structures result from simulations performed 
    with model 10. Upper panels: ``middle'' dissipation. Middle
    panels: ``strong'' dissipation. Lower panels: ``very strong''
    dissipation. Shown is each forth particle.}
\end{figure}

\begin{figure}
\psfig{file=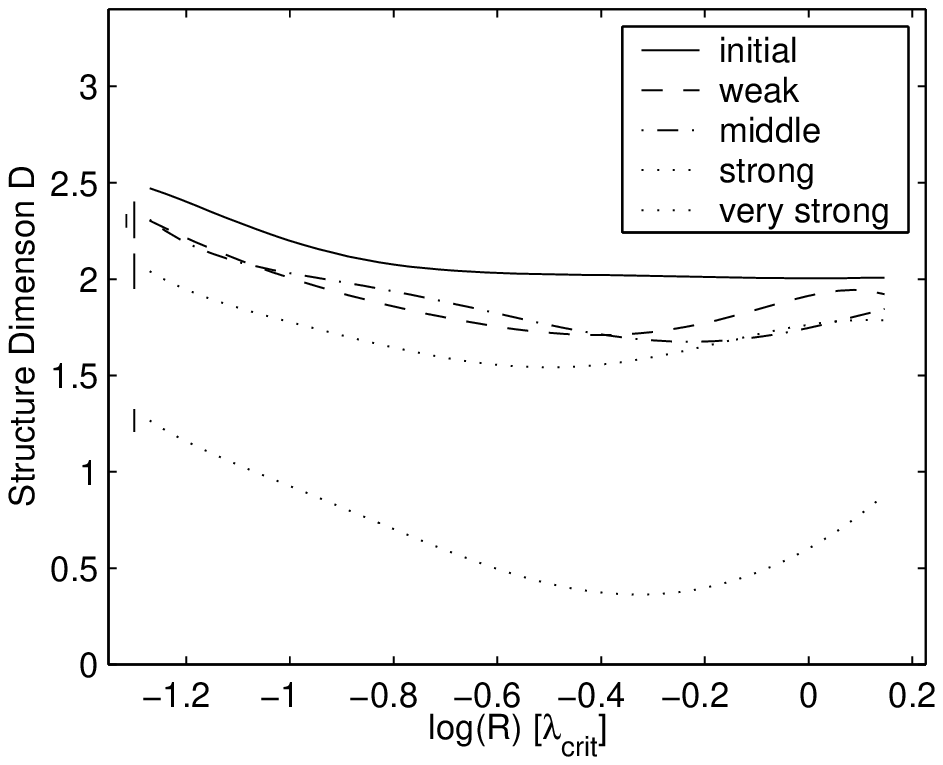,angle=90,width=\hsize}
\caption{\label{3dfrac1} The structure dimension $D$ as a function of
  the scale $R$. The corresponding structures result from 3D
  simulations of model 10. The structure dimension is indicated for 
  the simulation with the ``weak'' (see Fig.~\ref{3dgrossxy1}), 
  the ``middle'', ``strong'', and the ``very strong'' dissipation 
  (see Fig.~\ref{3dgrossxy2}).}
\end{figure} 

\paragraph{Radial Friction Coefficient $C_x$ / Large Simulation Zone.}
In Sect. \ref{sc2d} we explored for the 2D models how the structure
formation depends on the radial friction coefficient $C_x$.  Here we
study the connection between structure and radial friction in 3D.
Considered are structures resulting from models with large simulation
zone and anisotropic kernel (model 5, 6, 9 and 10).

Concerning the structure in the plane we find qualitatively the same
dependence as in the 2D simulations. That is, the initially formed 
structures smear out if we don't dissipate energy. 
However we still find correlations in the matter distribution 
after ten galaxy rotations.
For an increasing radial dissipation the structures become 
denser and denser until finally hot clumps are formed out
of filaments.  Fig.~\ref{3dgrossxy1} and \ref{3dgrossxz1} show, as an
example, the structures resulting from a simulation performed with
model 10.  The structures reach in this simulation very fast (after 2
rotations) a statistical equilibrium. The autocorrelation functions
revealing the underlying characteristics of these structures are
shown in Fig.~\ref{3dauto}. Fig.~\ref{3ddisp1} and \ref{3ddisp2} show
the evolution of the velocity dispersions. These Figures confirm that a
statistical equilibrium is attained after $\approx 2$ rotations around
the galaxy center.

The filaments resulting from these simulations $(C_x=64\times
10^{-3}\;\tau_{\rm osc})$ are very dense and they are first signs of
clump formation.  A slight increase of the dissipation would thus turn
the structure from filamentary to clumpy.  Indeed, the structures in
Fig.~\ref{3dgrossxy2} resulting from simulations performed with the
same model but with slightly higher radial dissipations $C_x=
66,\;68,\;70\times 10^{-3}\;\tau_{\rm osc} $, are already clumpy. For
convenience we call the relative strength of the radial dissipations
used in model 10 ``weak'', ``middle'', ``strong'', and ``very
strong''.  In Fig.~\ref{3dfrac1} the structure dimension $D$ is shown
for the four different dissipation strengths. The structure dimensions
are not constant over the corresponding dynamical range and are thus
not fractal. However, the structure dimension resulting from the
simulation with the ``strong'' dissipation has a dimension $1.5<D <1.8$
over the whole dynamical range in the plane and remains smaller than 2 
also on scales where the disk thickness becomes important.
Furthermore, the structure dimension has 
at $0.06\;\lambda_{\rm crit}$ the same dimension as
at $1.6\;\lambda_{\rm crit}$ with a minimum in between. This is
qualitatively a different behavior than those of the
initial state. The slight increase of the structure
dimension at the left and at the right may be induced by the small
scale (resolution) and large scale cutoff 
(periodicity), respectively. This behavior is quite general for
our simulations and
it may be that a larger dynamical range produce a more constant
structure dimensions, $D(R)\approx$ const..

\begin{figure}
\psfig{file=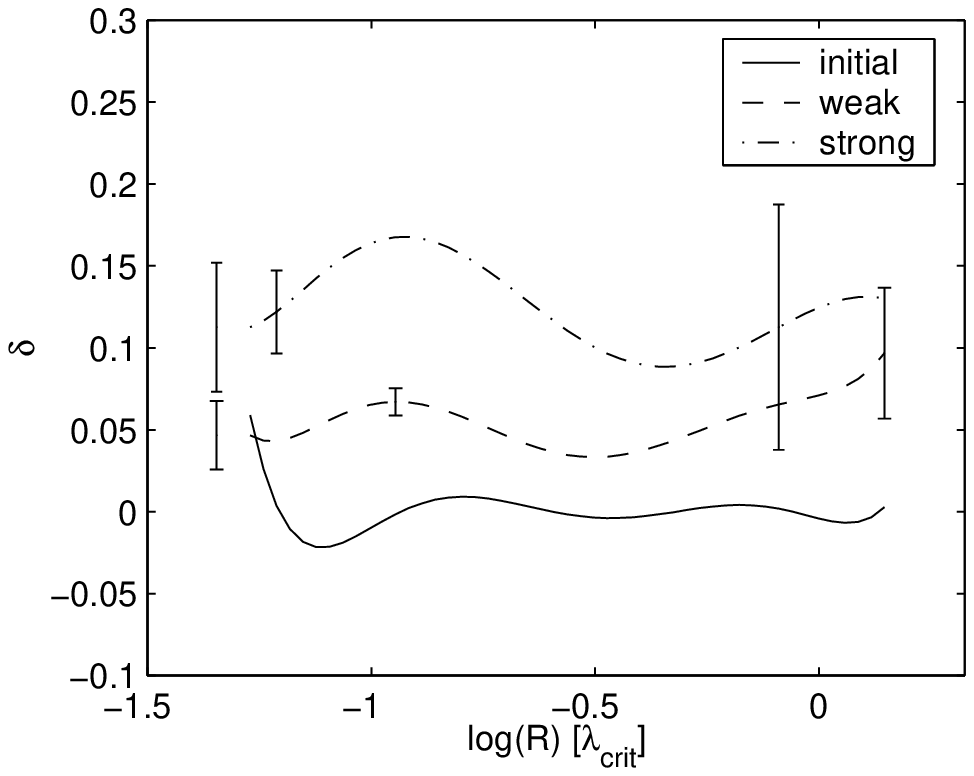,angle=90,width=\hsize}
\caption{\label{larson3ddis} The index $\delta$ as a function of the
  scale $R$. The corresponding structures result from 2 simulations
  performed with model 6. One simulation is performed with a ``weak''
  dissipation and produce a filamentary structure whereas the other
  simulation is performed with a ``strong'' dissipation, producing a
  clumpy structure.}
\end{figure}

We do not find a velocity dispersion-size relation similar 
to a power-law for the 
simulations performed with model 10. However we do find some hints for
such a relation in the structures resulting from model 6. 
Fig.~\ref{larson3ddis} shows the
velocity dispersion-size relation for two different simulations of model
6. The simulation with the weak dissipation produces 
filamentary structures whereas the simulation with the strong
dissipation produces filaments and clumps, thus $\delta$ varies
stronger and the error bars are larger for these structures. 
However, both simulations produce a velocity dispersion-size relation 
which may be approximated to first order by a power-law. Of course,
the error bars are too large and the scale range is too small
to call the corresponding velocity dispersion-size relations scale free. 
The resolution in model 6 is larger than those in model 10. Why this
softer gravitation seems to reproduce better a power-law velocity 
dispersion-size relation is at the moment unclear.

\begin{figure}
\psfig{file=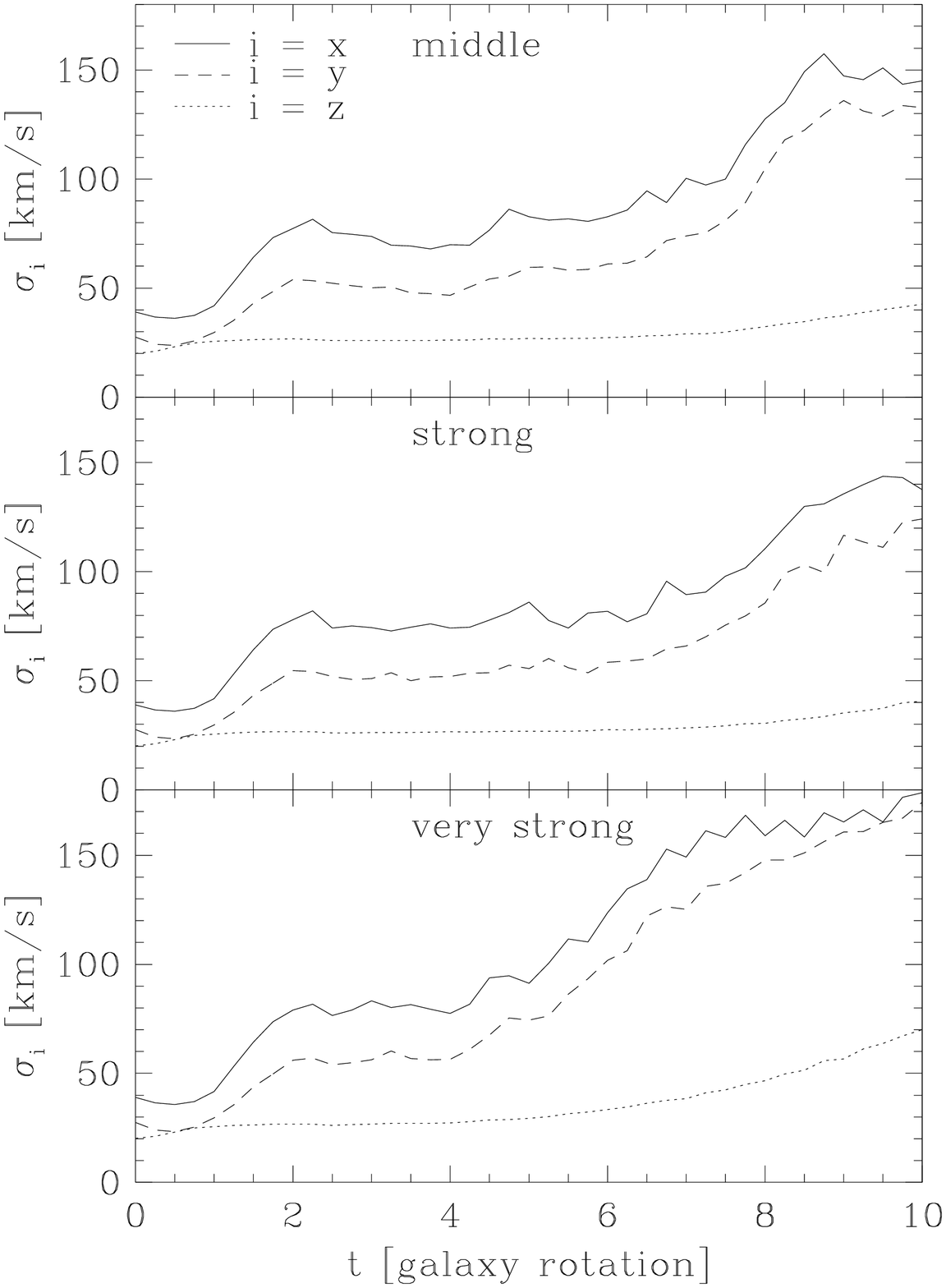,width=\hsize}
\caption{\label{3ddisp3} The velocity dispersion components
  $\sigma_x$, $\sigma_y$ and $\sigma_z$ as a function of time. The
  dispersions result from the simulation with the ``middle'', ``strong''
  and ``very strong'' dissipation strength, respectively (model 10). 
  The matter distribution of the corresponding simulations are shown in 
  Fig. \ref{3dgrossxy2}.} 
\end{figure}

Analog to the 2D case we find also in 3D simulations a systematic
ordering of the velocity dispersion components. The ordering,
$\sigma_x > \sigma_y > \sigma_z$, holds for quite a large range of
dissipation strength. This is shown by simulations performed with
model 10. The velocity dispersion components resulting from the 
simulation with the ``weak'' dissipation strength is shown 
in Fig. \ref{3ddisp1} and those with the ``middle'', ``strong'' 
and ``very strong'' dissipation strengths are shown in Fig. \ref{3ddisp3}.
Even if we don't dissipate energy this ordering is still observed 
after 10 galaxy rotations. Only strongest clump formation can destroy
the systematic ordering. 
From the definition of $x$ and $y$ in Sect. \ref{principle}
it follows $\sigma_x = \sigma_R$ and $\sigma_y = \sigma_\phi$.  To sum
up, one can therefore say, that as long as our model produces
structures resembling the lumpy matter distribution in spirals, it
also produces an ordering of the velocity dispersion components
$(\sigma_R > \sigma_\phi > \sigma_z)$ corresponding to those of the
solar neighborhood, or of $N$-body simulations of spiral disks
(Pfenniger \& Friedli 1991).

\begin{figure*}
\psfig{file=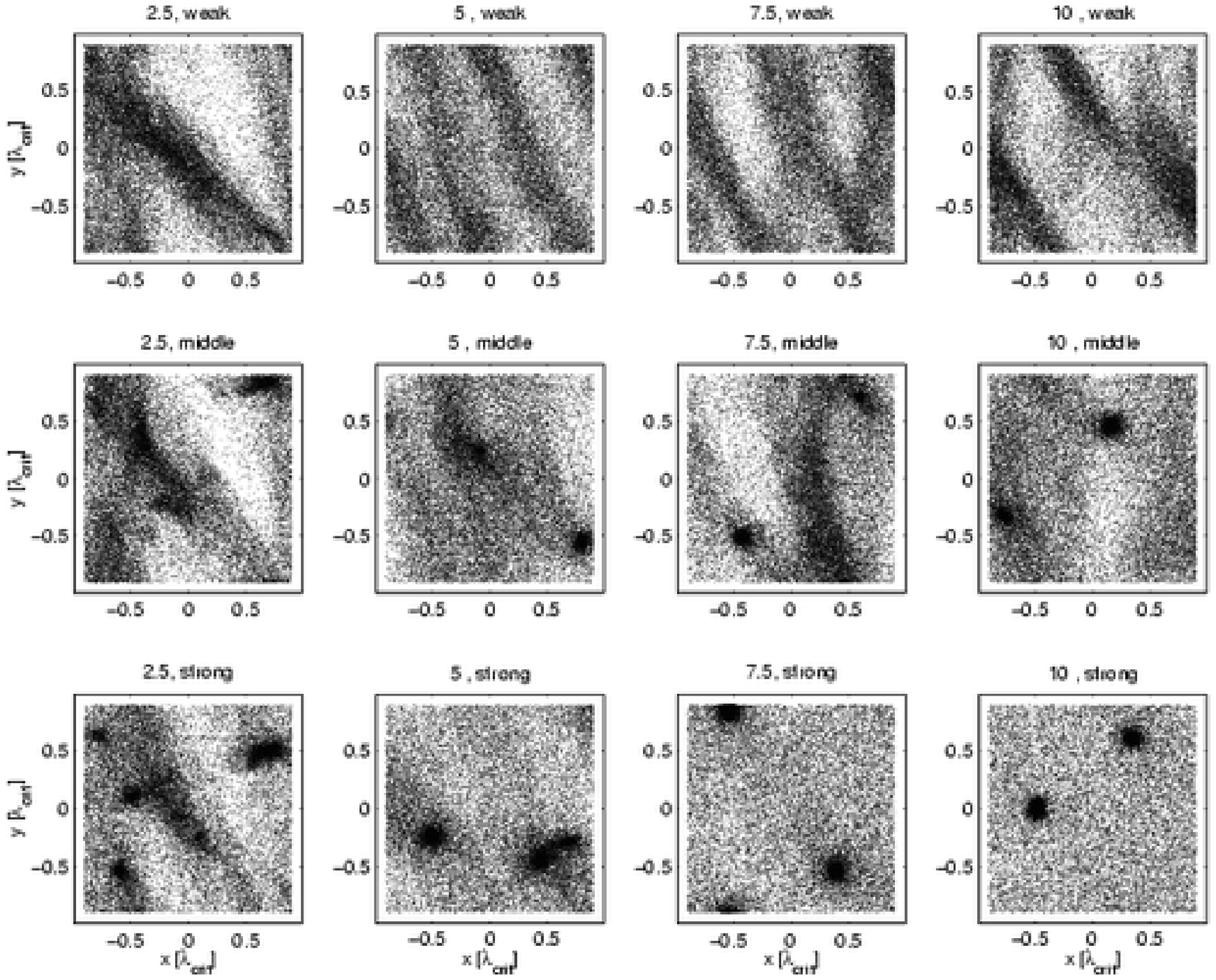,angle=90,width=\hsize}
\caption{\label{3dsmallxy} The evolution of the particle positions
  seen from above the galaxy plane. Shown is each forth particle.
  The structures result from 
  simulations performed with model 16. We show 3 simulations with
  different dissipation strength. Upper panels: ``weak''
  dissipation. Middle panels: ``middle'' dissipation. Lower panels:
  ``weak'' dissipation. The number of rotations of the shearing box
  around the galaxy center is indicated at the top of each panel.} 
\end{figure*}

\paragraph{Radial Friction Coefficient $C_x$ / Small Simulation Zone.}
In order to increase the resolution in the plane and to approach the
vertical resolution we decrease the size of the simulation box. The
structures resulting from these simulations differ from those in the
large simulation box. We still find filaments for a ``weak''
dissipation and clumps for a ``strong'' dissipation (see
Fig.~\ref{3dsmallxy}).  However, filaments and clumps, respectively,
appear less numerous. From this point of view the structures in the
small simulation box are not a fractal continuation of those in the
large simulation box. However, this does not exclude the possibility
that a simulation with a dynamical range incorporating the scale of
the large and the small simulation box would produce scaling laws over
the whole dynamical range.

\begin{figure*}
\psfig{file=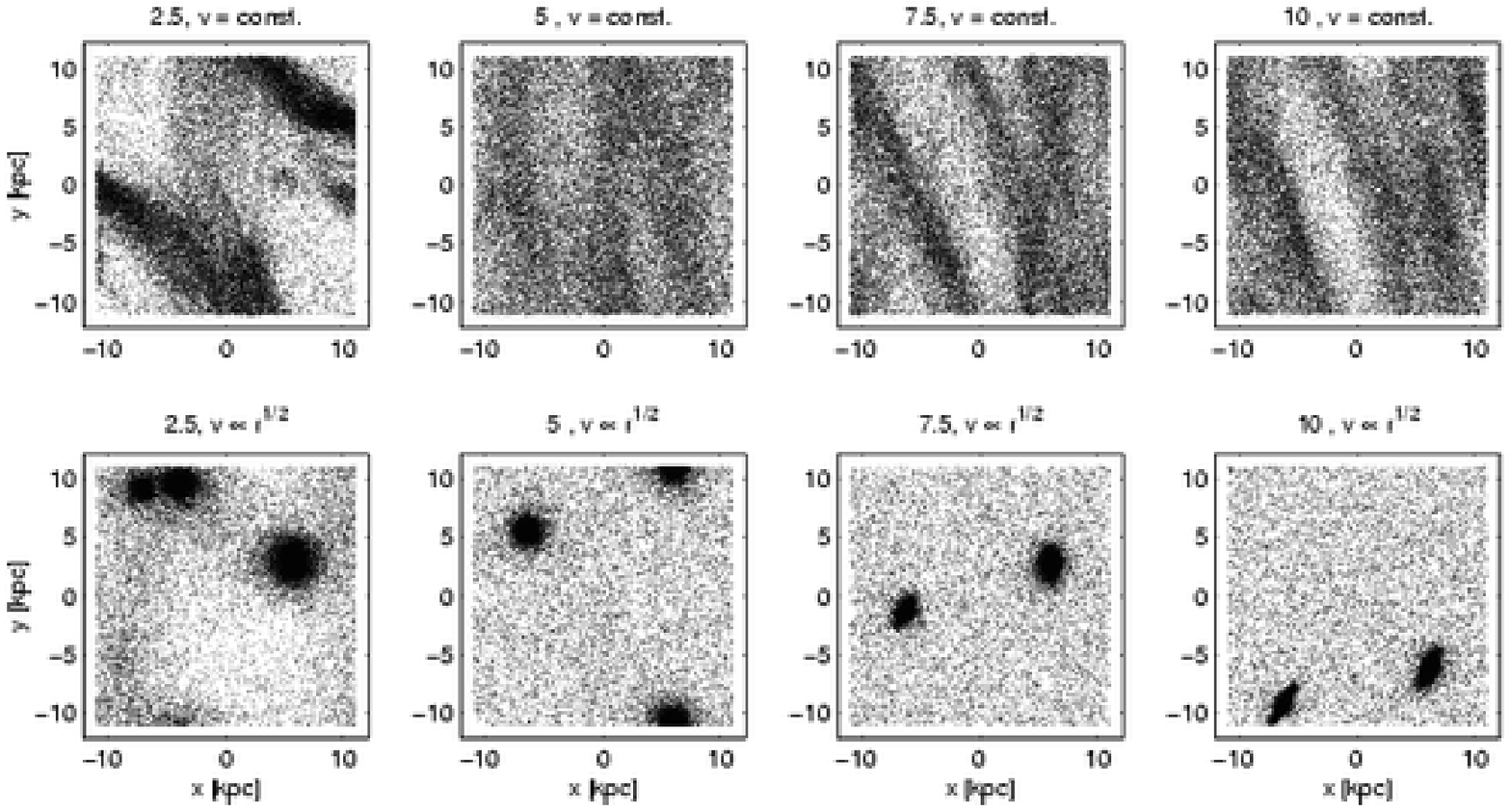,angle=90,width=\hsize}
\caption{\label{flach} Evolution of the matter distribution, seen from 
  above the galaxy plane. The structures result from simulations with
  different rotation curves performed with model 19. Upper panels: flat 
  rotation curve $(v = const.)$. Lower panels: increasing rotation
  curve $(v\propto \sqrt{r})$. The number of rotations around the
  galaxy center is indicated at the top of each panel. Shown is each
  forth particle.} 
\end{figure*}

\paragraph{Shear.}
The inhomogeneous structures appearing in the shearing-box simulations
can only be maintained when the dissipated energy is compensated by an
energy injection. The source of this energy is the shear motion. If
the shear is reduced, the dissipated energy can no longer be balanced
by the energy injection and the system collapses. A rotation curve,
increasing with the square root of the radius $(v\propto \sqrt{r})$
reduces the shear-flow, $\dot{y}=-2A_0 x$, by a factor two. With model 19
we perform simulations for different rotation curves. That is, for the
usual flat curve with $A_0=0.5\:\Omega_0,\; \kappa=\sqrt{2}\:\Omega_0$
and for a curve increasing with the square root of the radius,
$A_0=0.25\:\Omega_0,\; \kappa=\sqrt{3}\:\Omega_0$.  Such a choice of
the rotation curve doesn't reflect a realistic case but serves to
explore the influence of the shear on the structure formation.
However, a change of the rotation curve alters also the epicyclic
frequency and consequently the Schwarzschild velocity ellipsoid as
well as the critical wavelength $\rm \lambda_{crit}$. Because we only
want to examine the effect of the shear, simulations with different
rotation curves are carried out with the same initial velocity
dispersion.  Moreover, the distances are indicated in units of kpc.
The effect of a decreased shear is revealed in Fig.~\ref{flach}. The
same friction coefficients as in the simulations with the flat
rotation curve lead for the slightly increasing $(v\propto \sqrt{r})$
curve to a collapsed system.

\begin{figure}
\psfig{file=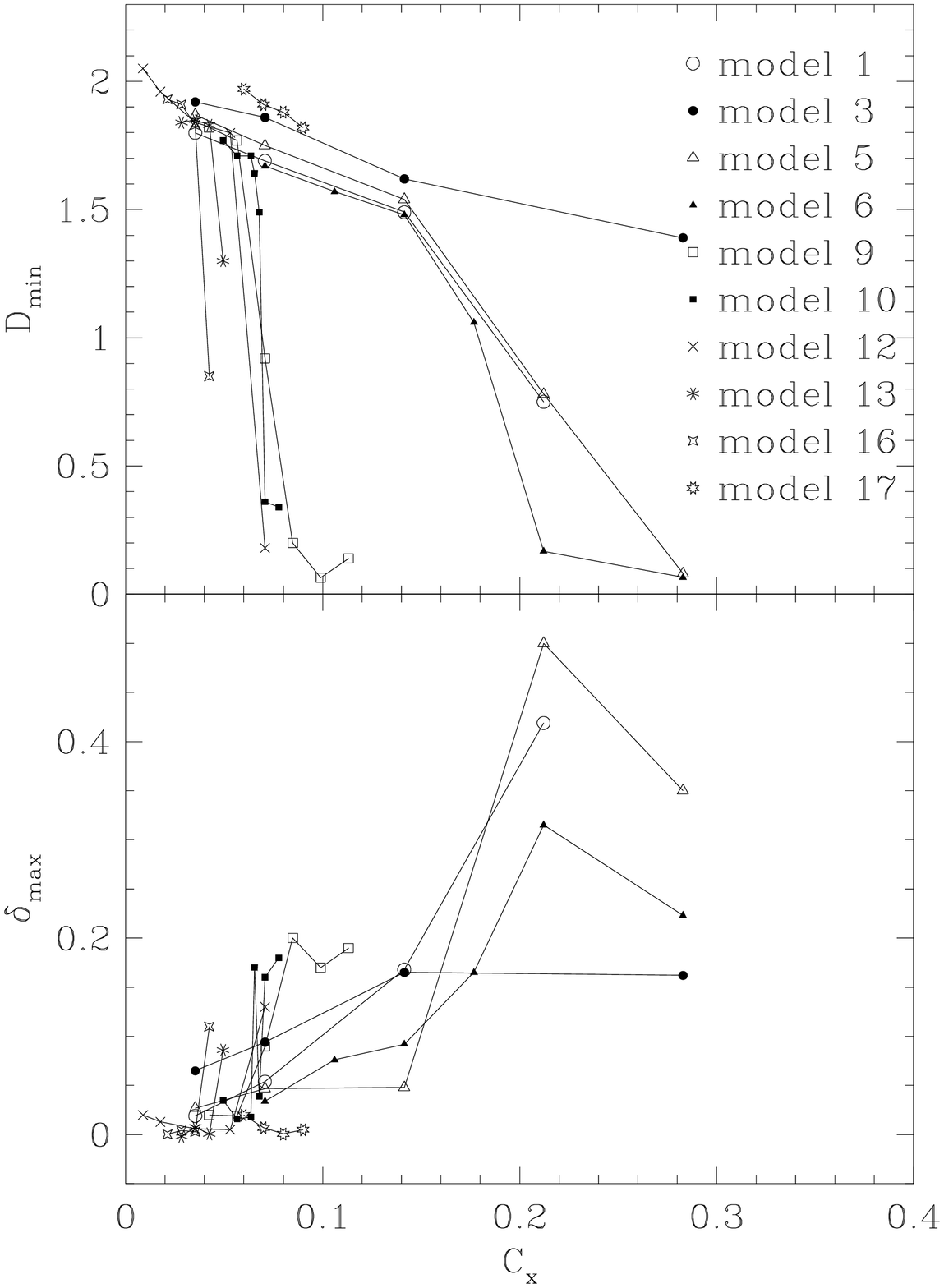,width=9.6cm}
\caption{\label{dmin} Upper panel: The minimal structure dimension
  $D_{\rm min}$ as a function of the radial friction coefficient $C_x$.
  Lower panel: The maximal index $\delta_{\rm max}$ of the velocity dispersion-size
  relation as a function of the radial dissipation strength. $D_{\rm min}$ 
  and $\delta_{\rm max}$ are mean values calculated during the last two
  galaxy rotations.}
\end{figure}

\subsubsection{Minimal Structure Dimension and Maximal Index $\delta$}

In this paper we study mainly the structure dimension in dependence of 
the radial friction coefficient $C_x$, because this parameter
determines principally the degree of structure inhomogeneity. 
Indeed, the parameters $C_z$ and $\nu$ serve mainly to prevent 
the disk from vertical dissolution and a change of the differential
rotation or the particle number can be balanced by an appropriate 
choice of the dissipation strength. 

The structure characteristics  
in dependence of $C_x$ are studied for different simulation boxes 
and resolutions. In order to compare the structures and the 
dynamics as a function of $C_x$
resulting from the different models we calculate the minimal structure
dimension and the maximal index $\delta$ of the velocity dispersion-size
relation.  The minimal structure dimension is defined in
Eq.~(\ref{dmineq}).  Correspondingly, the maximal index is,
\begin{equation}
\delta_{\rm max}=\{\max[\delta(R)]: l<R<L\}\;.
\end{equation}
The results are summarized in Fig.~\ref{dmin}.  The general trend is
that the structures become denser and more clumpy for higher radial
dissipation, leading to a smaller structure dimension and to higher
velocity differences on the different scales of the system, i.e., to
higher $\delta_{\rm max}$.

\section{Discussion}

With respect to the models with the large simulation zone (model 1-11), 
the models with the small simulation zone (model 12-19) produce less
fragmentation. That is, only a few weak filaments appear in
simulations of these models as long as the dissipation is weak 
and an increase of the dissipation
strength can not produce a more fragmented or filamentary structure,
because clumps appear very quickly. These clumps become rapidly
unphysically hot and collect almost all the matter of the simulation
zone. The clumps may thus hinder the formation of a more fragmented
structure in these models.

The formation of non-transient collapsed clumps, which have the
tendency to attract more and more matter were also found in other
numerical studies using dissipative or sticky particles in order to
model the ISM (Huber \cite{Huber01b}, Semelin \& Combes
\cite{Semelin00}). The appearance of these clumps may thus be a
generic problem of gravitational clustering simulations with
dissipative particles.

In the following we discuss some aspects related with the formation of
these non-transient and in our opinion unphysical hot clumps, which
may hinder the evolution towards a structure, being hierarchical over
a more extended scale-range. We discuss numerical, dynamical, and
physical aspects of the problem.
 
\subsection{Numerical Problems}
The size of dissipative particles, given by the softening length, is
much larger than the smallest clumps in the interstellar medium. Thus
the nearly homogeneous mass distribution (smear out) over the
softening length is not justified and it is not a priori clear that
the inhomogeneous structures below the resolution scale do not affect
the structures on larger scales.

Observations of the ISM reveal filamentary structures down to the
smallest resolvable scales, thus it is unclear down to what scales the
highly inhomogeneous structures continues. Moreover, a large scale
range has probably to be taken into account in order to reproduce the
observed structures. Thus it is an open question whether in the
following years a resolution will be reached such that basic clumps
can indeed be represented by dissipative particles.  Taking into
account the ubiquitous trend of gravitationally unstable media to
produce sheets and filaments not only in the ISM but also in
cosmological simulations, it might be that the particle model of the
basic mass unit, as a rigid body conserving mass, is not adequate.

\subsection{Dynamical aspects}
One could argue that when the ``mean free path'' of cloud clumps is
larger than their size and a particle description seems to be
justified, physical cloud clumps collide and dissolve because they
contain internal degrees of freedom due to the smaller subclumps
moving inside them. Thus dissipative particle simulations should
incorporate collisions with mass exchange (Pfenniger
\cite{Pfenniger94}).  This is particularly relevant if the clouds
collide with supersonic speed, 
that is, with velocities larger than their internal velocities.

Inherent dynamical properties of gravitational unstable media may
cause further problems. Let us discuss two of them:\vspace{1ex}

\noindent 1.) Typically clumps are subject to an anisotropic
gravitational contraction that alter its morphological
characteristics, i.e., clumps may become pancakes which become
filaments. This was shown by Zeldovich (\cite{Zeld70}) and
numerically confirmed by Kuhlman et al. (\cite{Kuhlman96}, hereafter
KU). In their numerical experiments KU replaced particles in dense
regions by clouds, made up of $2^3$ resp. $2^5$ particles.  
They found that only a small fraction collapses along all three
directions and forms dense clumps. A further
subdivision of the particles would probably lead to the same result.
If the transformation from clumps to pancakes and to filaments on
small scales is important for the appearance of the global structure,
it may be problematic to model gravitational unstable media with
particles, because a higher particle number would not solve the
inherent problem of anisotropic clustering.
\vspace{1ex}      

\noindent 2.) A further problem when simulating gravitational
clustering is related to the exponential propagation of two-body
relaxation in hierarchical $N$-body systems.  If the mass distribution
of the considered system is self-similar over a range of scales, at
each scale the effective bodies can be viewed as the corresponding
clumps.  The hierarchical structure acts as a strong two-body
relaxation amplifier since at each scale the effective number of
bodies is strongly reduced.  If this effective number is ${\cal
O}(10)$ the relaxation time at each level is of the order of the
crossing time, so two-body relaxation is a major driver of evolution
throughout the scales.
\vspace{1ex}

\noindent 3.)  Let us describe in more detail the related problem of
error propagation in a hierarchical system.  First we show that in a
gravitational unstable medium developing a hierarchical structure,
matter on smallest scales evolves faster.

A hierarchical mass distribution satisfies the scaling relation
between two levels $L$ and $l$:
\begin{equation}
\frac{M_L}{M_{l}}=\left( \frac{R_L}{R_l} \right)^D ,
\end{equation}
where $D$ is the mass scaling exponent, which would be the fractal
dimension in a self-similar hierarchical system. $D$ is restricted to
stay in the interval $[0-3]$ since mass is positive and space filling 
can not exceed the third power of scale.
Then the density scales as
\begin{equation}
\frac{\rho_L}{\rho_l}=\left(\frac{R_L}{R_l}\right)^{D-3}.
\end{equation} 
and consequently the crossing time $\tau_{\rm dyn} \propto (G
\rho)^{-1/2}$ scales as
\begin{equation}
\frac{\tau_{{\rm dyn},L}}{\tau_{{\rm dyn},l}}=
\left(\frac{R_L}{R_l}\right)^{(3-D)/2}.
\end{equation}
So in a hierarchical model the dynamical time (or crossing, or
free-fall time) always decreases at smaller scales since $0<D<3$
(Pfenniger \& Combes \cite{PfenComb94}).  Thus the low scale 
structures evolve faster.

Self-gravitating $N$-body systems are chaotic and neighboring
trajectories diverge exponentially (Miller \cite{Miller64}). 
This means also that any error propagates exponentially:
\begin{equation}
\Delta x \propto \Delta x_0 e^{\lambda t}\;,
\end{equation}
where $\Delta x_0$ is a small initial error at time $t=0$, and $\Delta
x$ is the error at time $t$.  Now, the degree of chaos and thus the
error evolution is determined through the maximum Liapunov exponent
$\lambda$, which for small $N$-body systems is approximately inversely
proportional to the dynamical time, $\lambda \propto 1/\tau_{\rm dyn}$
(Miller \cite{Miller94}).

With this estimate, let us see how errors are amplified through the
two adjacent levels.  If $\Delta x_l$ is in fact the initial error at
the lower level $l$, we get after one crossing time at level $L$,
$\tau_{{\rm dyn},L}$:
\begin{equation}
\frac{\Delta x_L}{\Delta x_l} = 
\exp\left( \frac{\tau_{{\rm dyn},L}}{\tau_{{\rm dyn},l}} \right) = 
\exp\left( \left(\frac{R_L}{R_l}\right)^{(3-D)/2} \right)  .
\end{equation}
The error amplification becomes rapidly huge as soon as $D < 3$.  Say,
if $\frac{R_L}{R_l}=2$ and $D=1.5$, then $\frac{\Delta x_L}{\Delta
x_l} = 5.38$.  Across $n$ levels the error ratio at the highest level
goes as $5.38^n$ and becomes much larger than the scale ratio, $2^n$.
(It is easy to show that for $D<2.264$ the error amplification is
always larger that the scale ratio, while for $\frac{R_L}{R_l}>2.72$
and $D>2.265$ one can find hierarchical systems for which this problem
does not occur).

Thus we find an inherent dynamical problem, which can not be solved by
using a higher resolution, because the increase of resolution
exponentially increases the errors through the scales.  Possibly such
hierarchical systems might be dominated by numerical errors in
simulations, and by small scale physics in real systems.

\subsection{Additional physics}
Small scale physics, supporting the dissolving process of dense
clumps, is not taken into account in our simple model.  However,
stellar winds, supernovae, jets and outflows may be important in the
overall mass transport across the scales.  These processes may ensure
a cyclic matter flow by giving matter back to larger scales, which
would condense back via gas cooling.  Such a matter-flow may be
crucial for sustaining the transient nature of hierarchical clumps for
extended time.

\section{Conclusions}

The structure resulting from the local simulations of self-gravitating
disks can be homogeneous, filamentary or clumpy depending on the
relative strengths of the competing gravitational and dissipation
processes. As long as the structure is mainly filamentary
self-gravitation and dissipation ensure a statistical equilibrium,
where repeated transient patterns are formed.  If the dissipative
processes begin to dominate the evolution, the structures turn from
filamentary to clumpy. During the subsequent evolution the clumps
become hotter and more massive.

In general, clumpy structures do not evolve towards a statistical
equilibrium.  However model 2 does show that it is also possible to
establish a persistent pattern of clumps.  These clumpy structures may
show signs of a power-low mass-size relation.  Some of our 2D as well
as 3D simulations suggest also a power-law velocity dispersion-size relation.

However the scale range of the simulations appears too small to draw
final conclusions about a fractal structure and an extension of
Larson's law beyond the size of molecular clouds.  We can suggest a
few reasons causing the discrepancy:\vspace{1ex}

\noindent 1.) The numerical resolution should extend over several 
more decades of scale before the fragmentation stops to be dominated 
by finite scale range boundary effects.

\noindent 2.) A fundamental law is associated with the rigid point particle
representation of mass which, by forcing a particular lowest scale
boundary condition, would prevent the bottom-up building of scaling
relations to match the observed ones.  
Also the propagation of errors may be super-exponential in a
hierarchical organized medium
because the error evolution at largest scales aren't determined
by the dynamical time of these scales, but by
the much smaller dynamical time of the smallest scales.   
Systems with dimension lower than about
2.2 are particularly concerned.  This point is relevant for particle
simulations of gravitationally unstable systems, such as cosmological
and thin disk ($D<2$) simulations.
 
\noindent 3.) A key physical ingredient, such as mass cycling through 
the scales due to star formation, stellar activity, and gas cooling 
could be essential to sustain a fractal state of the ISM.\vspace{1ex}

Finally, it is interesting to note that the anisotropy of the
velocity-dispersion ellipsoid, resulting from our simulations, has
systematically the same ordering $(\sigma_R>\sigma_\phi>\sigma_z)$ and
relative amplitudes as observed in the Galaxy and in $N$-body
simulations of spirals.  Since the models are deliberately a
simplified representation of reality, we learn from this that this
ordering may be due to a very general property of galactic disks, to
be substantially self-gravitating in $z$, and to rotate differentially
with a similar shear rate set by a constant rotation curve.
 
\begin{appendix}

\section{Pseudo code of the shearing box program}

\begin{tabbing}

\hspace{0.3cm}\=\hspace{0.3cm}\=\hspace{0.3cm}\=\hspace{0.3cm}\= \kill

\noindent
$\bullet$ Initialization of the particle positions and velocities.
\\[0.2cm]
$\bullet$ Calculation of the canonical momenta.
\\[0.2cm]
$\bullet$\hspace{1ex}\parbox[t]{8.45cm}{
          The inclination of the meshes is \\
          $a_b=[(-\Omega_0 t)\rm{mod}\;(L_y/L_x)]$
          and \\
          $a_f=a_b+(L_y/L_x)$, respectively.\\
          Thus there is a finite number of
          possible inclinations, $n=T/\Delta t$, where 
          $T=L_y/(L_x\Omega_0)$ and $\Delta t$ is the time-step.}
\\[0.3cm]

{\bf do i=1,n}
\\[0.2cm]

\> $\bullet$\hspace{1ex}\parbox[t]{8.15cm}{
       The derivations of the Kernel $\vec{\nabla}K(a_f)$ for the different 
       possible inclinations $a_f$ of the forward mesh are
       calculated. $K$ in an $n_x \times n_y \times n_z$-matrix,
       where $n_x n_y n_z$ is the number of cells in the local
       simulation box.}
\\[0.2cm]
\>$\bullet$\hspace{1ex}\parbox[t]{8.15cm}{
       The points: $(n_x/2, n_y/2, 1:n_z)$ of $\vec{\nabla}K(a_b)$ are 
       calculated for the backward mesh. The other Kernel points can
       be deduced from $\vec{\nabla}K(a_f)$ of the forward mesh, making 
       use of their symmetry.}
\\[0.3cm]

{\bf enddo}
\\[0.3cm]

{\bf do i=1,nt} (time propagation loop)
\\[0.3cm]

\>{\bf do $g=1,2$} (Do for forward and backward mesh, respectively)
\\[0.3cm]

\>\>{\bf if} (g=1)
\\[0.3cm]

\>\>\>$\bullet$\hspace{1ex}\parbox[t]{7.55cm}{
     $a=a_f=[(-\Omega_0 t)\mathop{\rm mod}(L_y/L_x)]+L_y/L_x$. The
     inclination $a$ corresponds to the forward mesh.}
     \\[0.2cm]

\>\>\>$\bullet$\hspace{1ex}\parbox[t]{7.55cm}{
     The particle positions of the Cartesian coordinates are saved
     $(y_c=y).$}
     \\[0.2cm]

\>\>\>$\bullet$\hspace{1ex}\parbox[t]{7.55cm}{
     Transformation of the particle position: Cartesian
     coordinates $\rightarrow$ Forward mesh $(y=y-ax)$.}
     \\[0.2cm]

\>\>{\bf else}
     \\[0.3cm] 

\>\>\>$\bullet$\hspace{1ex}\parbox[t]{7.55cm}{
     $a=a_r=a_f-L_y/L_x$. The inclination $a$ corresponds to the 
     backward mesh.}
     \\[0.2cm]

\>\>\>$\bullet$\hspace{1ex}\parbox[t]{7.55cm}{
     $y=y_c$. The particle positions are again those in Cartesian
     coordinates.}
     \\[0.2cm]

\>\>\>$\bullet$\hspace{1ex}\parbox[t]{7.55cm}{
     $y=y-ax$.Transformation of the particle position: Cartesian
     coordinates $\rightarrow$ Backward mesh.}
     \\[0.2cm]

\>\>\>$\bullet$\hspace{1ex}\parbox[t]{7.55cm}{
     The missing points in the $\vec{\nabla}K(a_b)$-matrix are
     deduced from $\vec{\nabla}K(a_f)$, making use of their symmetry.}
     \\[0.3cm]

\>\>{\bf endif}
\\[0.3cm]

\>\>$\bullet$\hspace{1ex}\parbox[t]{7.9cm}{
     Transformation of $\vec{\nabla}K(a)$ into Fourier space.
     $(\vec{\nabla}K(a)\rightarrow\widetilde{\vec{\nabla} K(a)})$} 
     \\[0.2cm]

\>\>$\bullet$\hspace{1ex}\parbox[t]{7.9cm}{  
    $y=y\;\mathop{\rm mod} L_y$. All the particles should lie inside the
    local box of the forward and the backward mesh, respectively.}
     \\[0.2cm]

\>\>$\bullet$\hspace{1ex}\parbox[t]{7.9cm}{  
    Mass to mesh assignment and transformation of the density $\rho$ into
    Fourier space $(\rho \rightarrow \tilde{\rho})$.}
     \\[0.2cm]

\>\>$\bullet$\hspace{1ex}\parbox[t]{7.9cm}{
    Calculation of the potential differential
    $(\vec{\nabla}\Phi=
    \widetilde{\widetilde{\vec{\nabla}K(a)}\widetilde{\rho}}^{-1})$,
    where 
    $\;\widetilde{}\; ^{-1}$ denotes the inverse Fourier transform.}
     \\[0.2cm]

\>\>$\bullet$\hspace{1ex}\parbox[t]{7.9cm}{   
    Calculation of the forces acting on the nodes of the affined meshes.}
     \\[0.2cm]

 \>\>\>$F_x=a \frac{\partial\Phi}{\partial y}-
        \frac{\partial\Phi}{\partial x}$
        \\[0.15cm]

\>\>\>$F_y=-(1+a^2)\frac{\partial\Phi}{\partial y}
       +a\frac{\partial\Phi}{\partial x}$
       \\[0.1cm]

\>\>\>$F_z=\frac{\partial\Phi}{\partial z}$
       \\[0.1cm]
  
\>\>$\bullet$\hspace{1ex}\parbox[t]{7.9cm}{
    Force interpolation in accordance with the CIC-method.}
     \\[0.2cm]

\>\>$\bullet$\hspace{1ex}\parbox[t]{7.9cm}{
    Transformation of the forces. Affine mesh $\rightarrow$
    Cartesian coordinate system.}
     \\[0.3cm]

\>\parbox[t]{\hsize}{
  {\bf enddo} $(g=1,2)$}
  \\[0.3cm]

 \>$\bullet$\hspace{1ex}\parbox[t]{8.15cm}{
  The particle positions are again those in Cartesian
  coordinates $(y=y_c)$}
     \\[0.2cm]

\>$\bullet$\hspace{1ex}\parbox[t]{8.15cm}{
  Force weighting}
     \\[0.2cm]

\>$\bullet$\hspace{1ex}\parbox[t]{8.15cm}{
  Calculation of the new canonical momenta and particle
  positions by means of the implicit canonical finite difference
  approximation (Pfenniger \& Friedli \cite{Pfenniger93}).}
  \\[0.2cm]

\>{\bf if}\hspace{1ex}\parbox[t]{8.1cm}{
    ($z$-coordinate of a particle lies outside the
   local box.)}
   \\[0.3cm]

\>\>$\bullet$\hspace{1ex}The particle is considered as escaped.
  \\[0.3cm]

\>{\bf elseif}\hspace{1ex}\parbox[t]{7.5cm}{
   ($x$- and/or $y$-coordinate lies outside the local box.)}
  \\[0.3cm]

\>\>$\bullet$\hspace{1ex}\parbox[t]{7.9cm}{
    Reentrance at the opposite side with appropriate canononical
    momenta (see Eqs.~(\ref{eq4})).}
    \\[0.3cm]

\>\parbox[t]{8.1cm}{
  {\bf endif}}
  \\[0.3cm]

\>\parbox[t]{8.1cm}{
  {\bf if} (storage condition)}
  \\[0.3cm]

\>\>$\bullet$\hspace{1ex}\parbox[t]{7.9cm}{
    Calculation of the particle velocities.}
  \\[0.2cm]

\>\>$\bullet$\hspace{1ex}\parbox[t]{7.9cm}{
    Storing of the particle positions and velocities to the disk.}
  \\[0.3cm]

\>{\bf endif}
  \\[0.3cm]{\bf enddo} (time propagation loop)
 
\end{tabbing}

\begin{acknowledgements}

We thank the referee, Ralf Klessen, for the careful reading
of the manuscript.
This work has been supported by the Swiss National Science Foundation.

\end{acknowledgements}

\end{appendix}


\begin{thebibliography}{}
\bibitem[1972]{Aronson72} Aronson E.B., Hansen C.J., 1972, ApJ 177,
  145
\bibitem[2000]{Beuther00} Beuther H., Kramer C., Deiss B., Stutzki J., 
                          2000, A\&A, accepted for publication
\bibitem[1994]{Binney94} Binney J., Tremaine S., 1994, Galactic
  Dynamics, Princeton University Press, Princeton, New Jersy
\bibitem[1977]{Brahic77} Brahic A., 1977, A\&A 54, 895
    Principles of Stellar Dynamics, 
    Dover Publications, New York
\bibitem[1982]{Casoli82} Casoli F., Combes F., 1982, A\&A 110, 287
\bibitem[1998]{Combes98} Combes F., 1999, CeMDA 72, 91
\bibitem[2000]{Dame00} Dame T.M., Hartmann D., Thaddeus P., 2000, to
  be published in ApJ, astro-ph/0009217
\bibitem[2000]{Dehnen00} Dehnen W., 2000, accepted for publication in MNRAS, 
  astro-ph/0011568
\bibitem[2000]{Elmegreen00} Elmegreen B.G., Sungeun K, 
                            Staveley-Smith L., scheduled for ApJ 548,
                            Feb 10, 2001, astro-ph/0010578
\bibitem[1996]{Elmegreen96} Elmegreen B.G., Falgarone E., 1996, ApJ
  471, 816
\bibitem[1987]{Falgarone87} Falgarone E., Perault M., 1987, in
  Physical Processes in Interstellar Clouds, eds. Morfill TG.E.,
  Scholer M., Dordrecht: Reidel, 59
\bibitem[1992]{Falga92} Falgarone E., Puget J-L., Perault M., 1992, A\&A,
  257, 715
\bibitem[1965]{Goldreich65} Goldreich P., Lynden-Bell D., 1965,
  MNRAS, 130, 125
\bibitem[1999]{Griv99} Griv E., Rosenstein B., Gedalin M., Eichler D., 
  1999, A\&A 347, 821
\bibitem[1999]{Heit99} Heithausen A., Stutzki J., Bensch F., 
  Falgarone E., Panis J.-F., 1999, Astr. Ges.: Rev. Mod. Astr. 12, 201
\bibitem[1971]{Hertel71} Hertel P., Thirring W., 1971, CERN preprint
  TH 1338, 309 
\bibitem[1878]{Hill1878} Hill G.W., 1878, Am. J. Math. 1, 5
\bibitem[1981]{Hockney81} Hockney R.W., Eastwood J.W., 1981, Computer
  Simulation Using Particles, McGraw-Hill, New York
\bibitem[2001]{Huber01b} Huber D., 2001, in The Promise of
  FIRST, eds. Pilbratt G.L., Cernicharo J., Heras A.M., Prusti T.,
  Harris R., ESA SP-460, in press, astro-ph/0101265 
\bibitem[1999]{Huber99} Huber D., Pfenniger D., 1999, in The Evolution 
  of Galaxies on Cosmological Timescales, eds. Beckman J.E. and
  Mahoney T.J., Astrophysics and Space Science, in press,
  astro-ph/9904209
\bibitem[2001]{Huber01a} Huber D., Pfenniger D., 2001, in The Promise of
  FIRST, eds. Pilbratt G.L., Cernicharo J., Heras A.M., Prusti T.,
  Harris R., ESA SP-460, in press, astro-ph/0101264
\bibitem[1999]{Joyce99} Joyce M., Montuori M., Pietronero L., Sylos
                        Labini F., 1999, ApJ 514, L5
\bibitem[1965]{Julian66} Julian W.H., Toomre A., 1966, ApJ 146, 810
\bibitem[1996]{Kuhlman96}Kuhlman B., Melott A.L., Shandarin S.F., ApJ
  470, L41
\bibitem[1978]{Larson79} Larson R.B., 1979, MNRAS 186, 479
\bibitem[1981]{Larson81} Larson R.B., 1981, MNRAS 194, 809
  Regular and Stochastic Motion, Springer, New York
\bibitem[1968]{Lynden68} Lynden-Bell D., Wood R., 1968, MNRAS 138, 495
\bibitem[1982]{Mandelbrot82} Mandelbrot B., 1982, The Fractal Geometry 
                             of Nature, Freeman, San Fransisco
\bibitem[1958]{McMillan58} McMillan W.D., 1958, The Theory of the
  Potential, Dover Publications, New York
\bibitem[1964]{Miller64} Miller R.H., 1964, ApJ 140, 250
\bibitem[1970]{Miller70} Miller R.H., Prendergast K.H., Quirk W.J., 1970, ApJ 
                        161, 903
\bibitem[1994]{Miller94} Miller R.H., 1994, in Ergodic Concepts in 
                            Stellar Dynamics, p. 137, eds. Gurzadyan
                            V.G., Pfenniger D., Springer
\bibitem[1983]{Myers88} Myers P.C., Goldman A.A., 1988, ApJ, 329, 392 
\bibitem[1990]{Pfenniger90} Pfenniger D., Norman C., 1990, ApJ 363, 391
\bibitem[1991]{Pfenniger91} Pfenniger D., Friedli D., 1991, A\&A 252, 75
\bibitem[1993]{Pfenniger93} Pfenniger D., Friedli D., 1993, A\&A 270,
  561
\bibitem[1994]{PfenComb94}  Pfenniger D., Combes F., 1994, A\&A 285, 94
\bibitem[1994]{Pfenniger94} Pfenniger D., 1994, in Ergodic Concepts in 
                            Stellar Dynamics, p. 111, eds. Gurzadyan
                            V.G., Pfenniger D., Springer
\bibitem[1996]{Pfenniger96} Pfenniger D., 1996, in Barred Galaxies,
                            p.273, eds. Buta R., Crocker D.A.,
                            Elmegreen B.G., A.S.P. Conf. Ser., Vol 91, 
                            San Francisco
\bibitem[1998]{Pfenniger98} Pfenniger D., 1998, in ${\rm H_2}$ in the Early
                            Universe , eds. Palla F., Corbelli E., 
                            Galli D., Memorie Della
                            Societa Astonomica Italiana, 429
\bibitem[1986]{Press86} Press W.H., Flannery B.P., Teukolsky S.A.,
  Vetterling W.T., 1986, Numerical Recipes, Cambridge University Press
\bibitem[1960]{Safronov60} Safronov V.S., 1960, Ann. d'Astrophysique
  23, 979 
\bibitem[1985]{Scalo85} Scalo J.M., 1985, in Protostars and Planets II,
                        p.201, eds. Black D.C. and Matthews M.S., Univ. of
                        Arizona Press
                          Astrophysics, p. 97, eds. Buchler J.R. and Eichhorn
                          H., New York Academy of Science, New York
\bibitem[1995]{Salo95} Salo H., 1995, ICAR 117, 287
\bibitem[1984]{Schwarz84} Schwarz M.P., 1984, MNARS, 209, 93
\bibitem[1985]{Sellwood85} Sellwood J.A., Carlberg R.G., 1984, ApJ
                           282, 61  
\bibitem[1999]{Semelin99} Semelin B., 1999, Ph.D. Thesis. Universit\'e
                         Paris 6
\bibitem[2000]{Semelin00} Semelin B., Combes F., 2000, to be published 
                          in A\&A, astro-ph/0007119
\bibitem[1999]{Stanimirovic99} Stanimirovi\'c S., Staveley-Smith L.,
                          Dickey J.M., Sault R.J., Snowden S.L., 1999, 
                          MNRAS 302, 417
\bibitem[2001]{Stanimirovic01} Stanimirovi\'c S., Lazarian A., 2001
  accepted for publication in the ApJ Let., astro-ph/0102191
\bibitem[2000]{Storzer00} St\"orzer H., Zielinsky M., Stutzki J.,
                          Sternberg A., 2000, A\& A 358, 682
\bibitem[1998]{Sylos98} Sylos Labini F., Montuori M., Pietronero L., 1998
                      Phys. Rep. 293, 66
\bibitem[1991]{Tauber91} Tauber J.A., Goldsmith P.F., Dickman R.L.,
                       1991, ApJ 375, 635
\bibitem[1964]{Toomre64} Toomre A., 1964, ApJ 139, 1217
\bibitem[1981]{Toomre81} Toomre A., 1981, in The Structure and Evolution of
                         Normal 
                         Galaxies, p. 111, eds. Fall S.M. and Lynden-Bell D., 
                         Cambridge Univ. Press
\bibitem[1990]{Toomre90} Toomre A., 1990, in Dynamics and Interactions
                         of Galaxies, p. 292, ed. Wielen R.,
                         Springer-Verlag 
\bibitem[1991]{Toomre91} Toomre A., Kalnajs A.J., 1991, in Dynamics of Disc
                         Galaxies, p. 341, ed. Sundelius B., G\"oteborg Univ.
                         Press
\bibitem[1994]{Vogelaar94} Vogelaar M.G.R., Wakker B.P., 1994, A\&A,
                           291, 557
\bibitem[1951]{Weizsacker51} von Weizs\"acker C.F., 1951, ApJ 114,
                             165
\bibitem[1999]{Westpfahl99} Westpfahl D.J., Coleman P.H., Alexander
                             J., Tongue T., 1999, AJ 117, 868
\bibitem[1988]{Wisdom88} Wisdom J., Tremaine S., 1988, Astron. J. 95, 925
\bibitem[1970]{Zeld70} Zeldovich Ya.B., 1970, A\&A 5, 84
\end{thebibliography}
\end{document}